\documentclass[final]{ws-procs9x6}
\usepackage{graphics}
\usepackage{axodraw}
\usepackage{natbib}

\usepackage{amsmath}
\usepackage{amssymb}
\usepackage{latexsym}
\usepackage{epsf}
\usepackage{psfrag}
\usepackage{feynarts}
\usepackage{graphicx}
\usepackage{color}

\newcommand{ \slashchar }[1]{\setbox0=\hbox{$#1$}   
   \dimen0=\wd0                                     
   \setbox1=\hbox{/} \dimen1=\wd1                   
   \ifdim\dimen0>\dimen1                            
      \rlap{\hbox to \dimen0{\hfil/\hfil}}          
      #1                                            
   \else                                            
      \rlap{\hbox to \dimen1{\hfil$#1$\hfil}}       
      /                                             
   \fi}                                             %

\newcommand{\gsim}{\gtrsim}
\newcommand{\lsim}{\lesssim}

\bibpunct{[}{]}{,}{n}{,}{,}

\begin{document}

\def\draftnote{}
\def\trimmarks{}
\setlength{\topmargin}{-0.01cm}

\pagestyle{plain}

\title{TASI 2014 Lectures: The Hunt for Dark Matter}

\author{Graciela B.  Gelmini}

\address{Department of Physics and Astronomy,\\ University of California, Los Angeles \\[3mm]
\textnormal{\texttt{gelmini@physics.ucla.edu}}}

\maketitle

{\vspace{-8.0cm}\begin{flushright}
NSF-KITP-15-005
\end{flushright}  }

\vspace{7.0cm}

\abstracts{These lectures, given at the 2014 Theoretical Advanced Study Institute (TASI), are an introduction to what we know at present about  dark matter and the major  current experimental and observational efforts  to identify what it consists of. They attempt to present the complexities of the subject,  making clear common simplifying assumptions, to better understand the reach of   dark matter searches.}

\section{Introduction}
Observations on all astrophysical and cosmological scales, from the scale of dwarf galaxies up to the largest scales, indicate that dark matter (DM) constitutes about 25\% of the content of the Universe. It cannot consist of atomic matter, which makes up stars, planets and ourselves. This known form of matter accounts for at most 5\% of the content of the Universe.  The remaining 70\% consists of dark energy, a component which, unlike matter, has repulsive gravitational interactions. 

  The nature of DM remains one of the most 
important open problems in science, and the efforts to resolve it are numerous.  The hunt for DM is multi-pronged and interdisciplinary, involving cosmology and astrophysics, particle physics, direct and indirect detection experiments and  searches at particle colliders. 
 
Many DM candidates have been proposed, many of them associated with beyond the Standard Model (SM) physics at the electroweak scale. Much has been said about the lightest supersymmetric particle, such as the lightest ``neutralino", or other Weakly Interacting Massive Particles (WIMPs) as ``natural" DM particle  candidates based on the   ``WIMP Miracle" argument (see e.g. \cite{Hooper:2009zm}) This  has lead to some degree of disillusionment about the prospect of discovering the nature of DM,  based on the lack of  any indication of new particle physics below energies of a few TeV at the Large Hadron Collider (LHC), and up to much higher energy scales in rare flavor physics processes. It is therefore important to avoid at this point oversimplifying assumptions in our presentation of what we know about the DM and what not having found it so far means.   

We will start by reviewing the evidence for DM and what we actually know about it. This is what defines the DM problem.  As we discuss the properties the DM must have, we will mention DM candidates which accommodate these properties in various different ways. Then we will concentrate on particle DM candidates, how to compute the relic abundance of WIMPs, and how they are searched for both in  direct and indirect detection experiments  and at the LHC.   We will discuss along the way recent DM hints in direct and indirect searches and DM particle candidates proposed to account for them. We will then mention briefly axions and sterile neutrinos.

\section{Evidence For Dark Matter}   

There is evidence for the existence of much more  matter than what can be assigned to visible matter at all scales from dwarf galaxies to the largest cosmological scales (see e.g.~\cite{Gorenstein:2014iba} for a  recent review). The excess is  what is called dark matter (DM). 

Galaxies are the building blocks of the present Universe. They range in mass from 10$^{9}M_\odot$ ($M_\odot$ is a solar mass) for dwarf galaxies to about 10$^{13}M_\odot$. Our own barred spiral galaxy, the Milky Way,  has a mass of about 10$^{12} M_\odot$. So far 25 satellite dwarf galaxies have been observed bound to the Milky Way, within a radius of several 100's of kpc from the galactic center (1pc (parsec) = 3.26 light-years). The closest of them is the Sagittarius Dwarf Galaxy, at 70 kpc from the center. Galaxies come in groups, clusters and superclusters. The Milky Way is part of the Local Group, together with two other large galaxies, Andromeda and Triangulum, and  many smaller satellite galaxies within a radius of several Mpc. The Local Group is in the outskirts of the nearest cluster, the Virgo Cluster, whose center  is  10 Mpc from the Milky Way. At the level of superclusters (scales of 10's  of Mpc)  the structure of the Universe is better described in terms of filaments, walls and voids.  The dominant component of all these structures is DM.

Evidence for the existence of DM was first  discovered by Fritz Zwicky in the 1930's. Measurements of the velocity dispersion of galaxies in the Coma Cluster led him to the conclusion that they could not be bound to the cluster by the gravitational attraction of the visible matter (stars, gas, dust) alone~\cite{zwicky}. The same argument  applied to the gas bound to clusters, as well as weak gravitational  lensing measurements of the total amount of mass in clusters, indicate that DM constitutes 6 times the mass in visible matter~\cite{clusters}. However, the seminal 1937 paper~\cite{zwicky} of Zwicky had only about ten citations in the first forty years.  

The existence of DM was rediscovered in the 1970's, this time at galactic scales,  by Vera Rubin and others. The flat rotation curves of disk galaxies, i.e. the  constant rotational speed $v$ as function of the  radius $r$ beyond the visible disk, indicates that the mass  continues to grow with the radius~\cite{rotationcurves}.  If all the mass of the galaxy $M$ would be concentrated at the disk,  an object of mass $m$ orbiting the galaxy beyond the disk would experience a gravitational force
$F={GMm}/{r^2} = mv^2/r $, and $v=(GM/r)^{1/2}$ would decrease as $r^{1/2}$. A constant $v$ requires the mass $M$ to grow with $r$. Most of the matter of galaxies resides in spheroidal ``dark haloes". This evidence, as well  as  weak~\cite{weaklensing} and strong~\cite{strong} gravitational lensing measurements, indicate that galaxies have more than 4 times the amount of mass in stars, gas and dust. Dwarf galaxies are the most DM-dominated systems known.

On cosmological scales, several observations are combined with General Relativity to determine the composition of the Universe.  The density fraction $\Omega_i \equiv \rho_i/\rho_c$, is the ratio  of the energy density of each component to the ``critical density" $\rho_c$, the value of the total energy density needed for the Universe to be spatially flat.  $\Omega = \Sigma_i \Omega_i$ has been measured to be very close to $\Omega=1$~\cite{Planck}. The latest observations by the Planck collaboration of the anisotropies in the Cosmic Microwave Background (CMB) combined with others (see Table 10 of~\cite{Planck}) have lead to  a composition of $\Omega_{DM} h^2 = 0.1187 \pm 0.0017$ of DM,  $\Omega_{b} h^2=0.02214 \pm 0.00024$ of ordinary matter (baryons, i.e. protons and neutrons) and $\Omega_{DE} =0.692 \pm 0.010$ of dark energy.  Here,  $h= 0.697 \pm 0.024$ is the present expansion rate of the Universe, the Hubble constant, in units of 100 km/Mpc s (we quote 68\% limits). Five independent measurements of the abundance of  atomic matter in the Universe show that it amounts to less than 5\% of the content of the  Universe: the X-ray emission from galaxy clusters, the relative height of the odd and even peaks in the angular power spectrum of CMB anisotropies,   the abundance  of light chemical elements generated in Big Bang Nucleosynthesis (BBN),  Baryon Acoustic Oscillations and absorption lines of the light of Quasars. 
   
 All measurements so far have confirmed the Big-Bang model of a hot early Universe expanding adiabatically for most of its lifetime of $t_U=$ 13.798 $\pm 0.037 \times 10^{9}$ y~\cite{Planck}. The CMB was emitted  379 ky after the Bang, when the temperature of the Universe was $T\simeq 3$ eV and atoms became stable for the first time. After this moment, called ``recombination", the Universe was populated by neutral constituents (the atoms) instead of plasma, and the mean free path of photons became very large on cosmological scales. Shortly before the emission of the CMB (when the Universe was about 100 ky old and $T\simeq 1$ eV) the Universe passed from being radiation-dominated to being matter-dominated. 
 
BBN is the earliest episode in the history of the Universe from which we have data, consisting of  the relative abundance of the light elements produced then: D, $^4$He and $^7$Li. It  took place between 3 and 20 minutes after the Bang, when the temperature of the Universe was $T\simeq$ MeV. In order for BBN and all the subsequent  history of the Universe to proceed as we know it, it is enough that the earliest and highest temperature of the radiation-dominated period in which BBN happens is just larger than  4 MeV~\cite{Hannestad:2004px}.
  
 Inflation is an early period of exponential expansion in the history of the Universe, advocated to produce
  through quantum fluctuations of a scalar field the small  density inhomogeneities which seed the large scale structure of the Universe  During inflation, the radiation and matter densities become zero and the Universe is repopulated during the ``reheating" period, after inflation. The earliest (highest) temperature of the radiation-dominated period after reheating is called the ``reheating temperature", $T_{RH}$. 
 
 The recent  BICEP2/Keck Array and Plank~\cite{Ade:2015fwj} upper limit on gravity waves produced during inflation (which replaced the  BICEP2 earlier claim of a measurement~\cite{Ade:2014xna}) determines an upper limit on $T_{RH}$. The upper limit on the ratio of tensor gravitational waves and scalar density perturbations in the primordial plasma tells us that the potential energy density during inflation was $ V \lsim (10^{16 }$GeV$)^4$. Under the assumption of an instantaneous reheating, the most efficient type of reheating, this energy goes completely into radiation,  $V\simeq T_{RH}^4$. If the reheating is not instantaneous, the energy density decreases with the expansion of the Universe and can be much smaller when transferred to radiation. Thus $T_{RH} \lsim 10^{16}$ GeV. How small $T_{RH}$ can be depends on the details of the reheating period~\cite{Dai:2014jja}. Thus, the lower  limit $T_{RH} \gsim 4$ MeV~\cite{Hannestad:2004px} imposed by BBN holds.  
 
 We do not know the thermal history of the Universe before its temperature was 4 MeV, and most DM particle candidates are produced during this period.

\section{What do we know about  Dark Matter?}

 We usually say that DM is neutral and stable, with very weak interactions. Let us review what we actually have learned about DM in the 80 years since we first found evidence of its existence. 

\vspace{0.1cm}

\noindent{\it{\bf -1- DM has attractive gravitational interactions and  is either stable or has a lifetime ${\bf{\gg t_U}}$.}}  
This is clear: DM is still present in the Universe and behaves as does regular matter for gravitational interactions. 

We have, in fact, no  evidence that DM has any other interaction but gravity. Thus, one can wonder if the many observational evidences for DM are instead  showing departures from the law of gravity itself.  This general idea  cannot be tested unless expressed in a particular model and the most successful model of this type proposed so far has been the ``Modified Newtonian Dynamics" (MOND)~\cite{mond}, and its covariant version, the ``Tensor-Vector-Scalar" (TeVeS) gravity model~\cite{Bekenstein:2004ne}. But they  cannot replace DM, as described below.

\vspace{0.1cm}

\noindent{\it{\bf -2- MOND with only visible matter is not enough at scales larger than galactic scales.}} 
In 1983 M. Milgrom~\cite{mond} proposed MOND as an alternative explanation for the flat rotation curves of galaxies. In MOND  $F=ma$ becomes $F=\mu(a) ma $, where $\mu(a)$ deviates from unity only for very small accelerations $a \ll a_0$, for which  $\mu=a/a_0$. Thus the gravitational force acting on a body of mass $m$ in an orbit of radius $r$ around a galaxy of mass $M$ leads to $F={GMm}/{r^2} = ma \mu$, which for large $r$, i.e. $a \ll a_0$, yields $a={\sqrt{GMa_0}}/{r}$.
Equating this with the centripetal acceleration $a= v^2/r$, one gets a constant orbital speed $v=(GMa_0)^{1/4}$.

MOND  is in good agreement with galaxy-scale observations for $a_0 \simeq 1.2 \times 10^{-10}$ m/s$^{2}$, without any need for DM~\cite{McGaugh:2014nsa} (and for this $a_0$, the effects of MOND are too small to be measurable in laboratory or solar-system scale experiments). But it fails at larger scales~\cite{McGaugh:2014nsa}, in particular in the ``Bullet Cluster"~\cite{Clowe:2006eq}, unless some form of DM is introduced~\cite{Angus:2007mn}. MOND is only a non-relativistic theory, thus it cannot be used where General Relativity is needed.  TeVeS~\cite{Bekenstein:2004ne} is a relativistically covariant theory
  in which the tensor field in Einstein's theory of gravity is replaced by scalar, vector and tensor fields, which interact in such a way to  yield MOND in the weak-field non-relativistic limit.
  
   In the ``Bullet Cluster"~\cite{Clowe:2006eq}, discovered in 2004, the baryonic matter (hot gas observed in X-rays) is at the center,   spatially segregated from the two lateral gravitational potential wells (measured via weak gravitational lensing).  The explanation of this system based on DM is that two galaxies collided, leaving behind the interacting gas, while the DM of both galaxies passed through (which yields an upper limit on the self interaction of the DM). To explain this system TeVeS  requires 2-3 times more matter than accounted for by the visible matter alone~\cite{McGaugh:2014nsa, Angus:2007mn}. TeVeS advocates propose the existence of  some ``cluster baryonic dark matter" (CBDM)~\cite{Milgrom:2008rv}, for example. But  once some kind of extra matter is necessary, it then becomes an issue of taste which type of DM one considers more likely.
   
   \vspace{0.1cm}

\noindent{\it{\bf -3- DM is not observed to interact with light}}. It is not observed to emit, reflect or absorb light of any frequency. This means that most of the DM  must have a small enough electromagnetic coupling or be very heavy. Upper limits are derived from background light at all frequencies~\cite{Overduin:2004sz}.

The DM could be neutral, maybe with a small  electric or  magnetic dipole moment~\cite{Pospelov:2000bq, Sigurdson:2004zp,  Barger:2010gv} or an anapole moment~\cite{Pospelov:2000bq, Ho:2012bg}.  It could  have a small effective electric charge, such as  that of ``Milli-Charged DM"~\cite{Feldman:2007wj}. Charged  DM particles  can also act as nearly collisionless if their mass is sufficiently large~\cite{CHAMPs, Jacobs:2014yca}).
 
   ``Millicharged DM" (see e.g. \cite{Feldman:2007wj, Cline:2012is}) can be part of a ``dark sector" which couples to SM particles only though the admixture of a ``dark photon"  of a ``dark U'(1)" gauge symmetry with  the usual photon (or the Z boson).  This could occur through a  ``kinetic mixing"~\cite{Holdom:1985ag,Burrage:2009yz} of the form $ \epsilon F_{\mu\nu} F'^{\mu\nu}$, where $F_{\mu\nu}$ and $F'^{\mu\nu}$ are respectively the field-strength tensors for the hypercharge and dark U(1) gauge groups.  After  diagonalizing the kinetic terms, the usual photon has an admixture $\epsilon$ of the dark photon. Thus if the DM particle couples with charge $Q'$ to the dark photon, the effective coupling with the usual photons is $\epsilon Q'$ and several limits imply that $\epsilon <10^{-3}$~\cite{Cline:2012is, McDermott:2010pa}.

``Millicharged DM" appears naturally as part of a complicated dark sector.  It could be ``Atomic DM", with dark protons and dark electrons forming dark atoms~\cite{Goldberg:1986nk, Feng:2009mn, Kaplan:2009de, Cline:2012is},  or  ``Mirror DM"~\cite{Foot:1991bp} whose Lagrangian is  a copy of that of the SM, but for the mirror particles (see e.g.~\cite{Kusenko:2013saa} and references therein). These are some possibilities in which the dark or ``secluded" sector has some of the richness of the visible sector, with hidden gauge interactions and flavor (e.g. \cite{Pospelov:2007mp} and~\cite{ArkaniHamed:2008qn}).

 Observational upper  limits on the cross section of elastic DM -photon interactions can be surprisingly large, e.g. a recent limit from simulations of Milky Way sub-haloes is  $\sigma^{\rm elastic}_{DM-\gamma} \leq  4 \times 10^{-33}$ cm$^2$ (m/GeV)~\cite{Boehm:2014vja}.
 
 An important consequence of the small interaction of DM with light is that the DM cannot cool by radiating photons during galaxy formation.
 
 \vspace{0.1cm}

\noindent{\bf{\it{-4- The bulk of the DM must be dissipationless, but part of it could be dissipative.}}} ``Dissipationless" mean that the DM cannot cool by radiating as baryons do to collapse in the center of  disk galaxies.  Otherwise,  their extended dark halos would not exist. Galaxies start as  structures made of the primordial admixture of dark and visible matter. Then, visible matter dissipates energy by emitting photons and falls into the potential well of the object.  Because this emission is isotropic, the visible matter  preserves any angular momentum it might initially have. Thus as it collapses to the center, it increases its angular speed until it becomes unstable towards the formation of a disk, which thus rotates much faster than the dark halo.

    While most of the DM must be nearly dissipationless, a small fraction of it could be dissipative. Part of the matter in the galaxy, the visible matter, is dissipative and its presence does not disrupt the stability of  dark haloes.  Thus a similar fraction, 5-10\%,  of the DM could also be dissipative and even form a ``Dark Disk". This is the idea behind ``Partially Interacting DM" (PIDM)  and ``Double Disk DM" (DDDM)~\cite{DDDM}. The dissipative  DM component could emit  ``dark photons" or other ``dark" particles.  
 
A Dark Disk was shown to arise  in some simulations of galaxy formation including baryonic matter besides the usual non-dissipative Cold DM~\cite{DD}, but with dissipative DM it should be a pervasive feature of all disk galaxies.

 \vspace{0.1cm} 
 
\noindent{\bf{\it{-5- The mass $m$ of the major component of the DM has only been constrained within some 80 orders of magnitude.}}}  There is a firm upper bound $ m \leq 2 \times  10^{-9}$ M$_\odot= 2 \times 10^{48}$ GeV  at the 95\% CL. It comes from unsuccessful searches for MACHOS (``Massive Astrophysical Compact Halo Objects") in the dark halo of our galaxy using gravitational microlensing with the Kepler satellite~\cite{Griest:2014cqa}, and from the ground-based MACHO and EROS surveys~\cite{Alcock:1998fx},    combined with bounds on the granularity of the DM for masses larger than 30 M$_\odot$~\cite{granularity}. Microlensing is a type of gravitational lensing in which the multiple images of the lensed star are superposed, producing a magnification of the star flux if an object passes near the line of sight to the star as it moves through the dark  halo. Above this limit, MACHOS  can account for only a small fraction of the dark halo of our galaxy.
 
 Among the best candidates for MACHOS are ``Primordial Black Holes" (PBH)~\cite{Carr:1974nx}. These are hypothetical black holes that could be created  in a primordial phase transition, maybe during inflation (see e.g.~\cite{Kusenko:2013saa,Green:2014faa}).  However, several limits apply to PBH which do not apply to other MACHOS and constrain the fraction of DM in PBH to be $<1$ for almost any $m_{\rm PBH}$ mass. PBH lighter than $10^{15}$g  would have decayed by Hawking radiation.  Not having observed this radiation  in $\gamma$-rays excludes $m_{\rm PHB}$$<$10$^{17}$g.  Accretion of PBH during star formation which would destroy neutron stars excludes $10^{16}$g$<$$m_{\rm PHB}$$<$10$^{22}$g.  Part of this last range, $10^{17}$g$<$$m_{\rm PHB}$$<$10$^{20}$g, is also excluded by the non-observation of ``femtolensing"  (in which the lensed images are separated by femto arc-seconds) of Gamma-Ray Bursts. The only narrow window remaining for PBH to make up all the DM, $10^{22}$g to $4 \times 10^{24}$g (i.e. $6 \times 10^{45}$ GeV to  $2 \times 10^{48}$ GeV), is being challenged by less-certain excluding arguments. For a current review on PBH and references to all these limits see~\cite{Green:2014faa}. PHB DM does not require physics beyond the SM (except some special inflation potentials).  
 
    There could be other DM candidates which might  arise within the SM, generically called  ``Macroscopic DM"~\cite{Jacobs:2014yca}. How to generate enough of them to account for all the observed DM is problematic. These include  ``Nuclearites", ``Strangelets", ``Strange Baryon Q-Balls", ``Baryonic Color Superconductors", ``Compact Composite Objects",  ``Strange Chiral Liquids Drops", and ``Compact Ultradense Objects". Macroscopic candidates could arise in physics beyond the SM too, e.g. ``Supersymmetric Q-balls". Various limits on them are presented in~\cite{Jacobs:2014yca} (see references therein for all these candidates).

 The limits just presented, and the fact that particle candidates can have the right relic abundance to be the DM,  constitute the only observational arguments we have in favor of elementary-particle candidates for DM.
 
  The lower limit on the DM particle mass is less well determined. It is at least 10$^{-31}$ GeV since there is a concrete particle candidate proposed with this mass,  ``Fuzzy DM". This is a boson with a de Broglie wavelength of 1 kpc~\cite{fuzzy-DM}.  For particles that reached thermal equilibrium in the early Universe the lower limit is 0.2-0.7 keV instead, due  the maximum occupation number these particles can have  in different structures in the Universe~\cite{Tremaine-Gunn}.

\vspace{0.1cm}

\noindent{\bf{\it{-6- DM has been mostly assumed to be collisionless, however the upper limit on DM self-interactions is very large.}}}$~$  The best limits on self-interactions are derived from the ``Bullet Cluster"~\cite{Clowe:2006eq} (where the DM of the two colliding galaxies had to pass through each other) and the non-sphericity of the halos of galaxies and  galaxy clusters. The upper limit  is huge:
 $\sigma_{\rm self}/m\leq$ 1cm$^2/$g $\simeq$ 2 barn/GeV $\simeq$ 2$ \times 10^{-24}$ cm$^2$/GeV. By comparison, the neutron capture cross section of uranium  is a few barns! The limit on the $\sigma_{\rm self}/m$ ratio comes from requiring that the self-interaction mean free path, $\lambda_{mfp} \simeq 1/ n \sigma_{\rm self} = m/ \rho \sigma_{\rm self}$, be long enough ($n = \rho/m$ is the DM number density, and the density $\rho$ of the system is measured). 
 
   Explaining the existence of DM cores in dwarf galaxies may constitute a problem for the usual collisionless Cold DM models.
   Collisions of the DM particles with themselves would erase small scale structure and turn cuspy central density profiles  into cored profiles  in dwarf galaxies, if the cross section is close to the mentioned upper limit~\cite{Zavala:2012us}.  DM with this large self-interaction is called ``Self-Interacting DM" (SIDM)~\cite{SIDM}. DM with less than an order of magnitude smaller $\sigma_{\rm self}$ is indistinguishable from collissionless.
   
     With SIDM the reduced central densities of dwarf galaxies imply reduced velocity dispersions, which can alleviate~\cite{Vogelsberger:2012ku} the ``too big to fail problem"~\cite{too-big-to-fail} of collisionless CDM (see below). 
    There are many particle models for SIDM, most with complicated ``dark sectors" (see e.g.~\cite{Kusenko:2013saa})
   
 \vspace{0.1cm}
 
\noindent{\bf{\it{-7- The bulk of the DM is Cold or Warm, thus particle  DM requires physics beyond the SM.}}} 
The DM is classified as hot, cold or warm according to how relativistic it is when galactic-size perturbations enter into the horizon (i.e. when these perturbations become encompassed by the growing horizon $\simeq ct$). This happens when $T\simeq$ keV. Hot DM (HDM) is relativistic, Cold DM (CMD) is non-relativistic and Warm DM (WDM) is becoming non-relativistic at this moment (see e.g.~\cite{Kolb:1990vq}).

The presence of DM in the early Universe is necessary for the formation of structure  in the Universe. Structures in baryons cannot grow until recombination, when atoms become stable. Before then the photon  pressure in the plasma prevents it. But at recombination baryons  must fall into already formed potential wells of DM, or there would not be enough time to form the structures  we observe now.  Thus, it is the DM which determines the major features of the large-scale structure of the Universe.

Perturbations in the DM survive at horizon crossing only if the DM is non-relativistic. With HDM, primordial galaxy-size density inhomogeneities would not survive,  superclusters would form first  and later galaxies through  fragmentation.  This does not reproduce the observed Universe. With CDM, inhomogeneities  much smaller than galaxy size survive, thus galaxies and clusters incorporate many smaller structures which form first. Some of them are not entirely tidally destroyed,
 thus, a large number of substructure is expected within CDM dark haloes. With WDM the smaller structures formed first are of the size of dwarf galaxy cores, thus there is much less substructure within large haloes than with CDM (see e.g.~\cite{Kolb:1990vq}).

Either CDM or WDM can account for all the large scale structure observations.  The difference between them is at the dwarf-galaxy scale, where observations and their interpretation are still not conclusive.

Very high resolution simulations of structure formation in the Universe  with only CDM (assumed to be collisionless) find within a halo similar to that of the Milky Way, of the order of 10 subhaloes so massive and dense that they seem ``too big to fail" to form lots of stars within them and be visible. But the Milky Way and  Andromeda  do not seem to have satellite galaxies with stars moving as fast as would be expected in these dense sub-haloes. This constitute the so called ``too big to fail" problem of CDM~\cite{too-big-to-fail}. Due to the large uncertainty in the mass of the Milky Way, this problem may be alleviated in our galaxy if its mass is sufficiently small, but the problem persists in Andromeda and in the Local Group~\cite{Kirby:2014sya}.  There are indications of the existence of a core  (a region of constant density) in the DM density profile of dwarf galaxies, instead of the cusp (with density growing towards the center) predicted by CDM-only simulations.

 Both these potential problems of (collissionless) CDM may disappear once the effect of visible matter  is fully taken into account, or with WDM or SIDM (Warm or Self-Interacting DM)~\cite{SIDM, Zavala:2012us, Vogelsberger:2012ku} instead of CDM (see also~\cite{Kusenko:2013saa} and references therein). With WDM  too few satellite galaxies might be produced in the Milky Way, if its mass is small enough~\cite{too-small-to-succeed}.

The only particles  in the SM  which are part of the DM  are neutrinos. They are lighter than 1 eV and remain in thermal equilibrium in the early Universe until temperatures of a few MeV (see below), thus they are HDM.  There are no CDM or WDM particle candidates in the SM but there are many  in extensions of the SM.  Candidates for CDM are ``axions", WIMPs, and many others. Sterile neutrinos and some non-thermal WIMPs (see below), among others, are  good WDM candidates.
 
 Because of spontaneous symmetry breaking arguments, completely independently of the DM issue, we do expect new physics beyond the SM to appear at the electroweak scale. Many extensions of the SM  were proposed because of this reason (supersymmetry, technicolor, large extra spatial dimensions,  the ``Little Higgs" model, the Inert Doublet model etc.), which provide both the main potential discoveries at the LHC, and also DM candidates.  However, the DM physics may be entirely different from that at the electroweak scale. Many models have been proposed in recent years just to account for DM hints in direct and indirect DM searches.
   We already mentioned some of them, and will mention more later in the lectures.
  
   When computing the properties of DM candidates we should keep in mind our assumptions, among them our cosmological assumptions.
 
  \vspace{0.1cm}

\noindent{\bf{\it{-8- Most DM candidates are relics from the pre-BBN era,  from which we have no data.}}}  The computation of the relic abundance and primordial velocity distribution of  particle DM candidates produced  before the BBN  temperature  limit of 4 MeV depends on assumptions made regarding the thermal history of the Universe. 
  With different viable cosmological  assumptions (see e.g.~\cite{Gelmini:2010zh}), the relic density and velocity distribution of the DM candidates may change considerably (see section  4.1.4).

As already pointed out early on (e.g. by J. Barrow in 1982~\cite{Barrow:1982ei}), if we ever discover a DM particle from the pre-BBN era,   we would use its properties as a cosmological probe to gain information about this epoch.  

\section{The relic abundance of Dark Matter particle candidates}

We usually characterize DM particle candidates according to how they are produced as ``thermal" or ``non-thermal" relics (for a review see e.g.~\cite{Gelmini:2010zh}).  ``Thermal" relics are produced via interactions with the thermal bath, reach equilibrium with the bath
 and then ``decouple" or ``freeze-out" when their interactions cannot keep up with the expansion of the Universe. 
 Chemical equilibrium is achieved when reactions that change the number of particles are faster than the expansion rate of the Universe $H$ (or the reaction time is shorter than the lifetime of the Universe $\simeq H^{-1}$). After chemical decoupling or freeze-out  the number of particles per comoving volume remains constant. After kinetic decoupling, the exchange of momentum with the radiation bath ceases to be effective.  ``Non-thermal" DM particles are all those not produced in this way. For example, they could be produced via the decay of other particles, which themselves may or may not have a thermal abundance.
 
\subsection{Thermal relics}

 The evolution of the number density $n$ of thermal DM particles  $\chi$ and antiparticles $\bar{\chi}$  interacting with particles in a thermal bath  is given by the Bolzmann Transport Equation (see e.g.~\cite{Kolb:1990vq}),  
\begin{equation}
\frac{dn}{dt}=-3Hn-\left<\sigma_{A} v\right>(n^2-n_{\rm EQ}^2),
\label{BE}
\end{equation}
assuming  for simplicity that $\chi$ and $\bar{\chi}$ can only annihilate and be created in  $\chi\bar{\chi}$ pairs and that there is no particle-antiparticle asymmetry, $n_\chi =n_{\bar{\chi}}=n$.
Both of these assumptions are fulfilled for the neutralino and many other candidates that coincide with their own antiparticles. Here $t$ is time, $H$ is the expansion rate of the Universe (the Hubble parameter, $H\equiv{\dot{a}}/{a}$, where $a$ is the scale factor of the Universe),  $\sigma_{A}$ is the $\chi\bar{\chi}$ annihilation cross section, and $v$ is the relative speed of the annihilating DM particles. The angle brackets $\left<~\right>$ denote an average over the thermal distribution of momenta of the DM particles (an average over initial states and sum over all final states, see e.g.~\cite{Gondolo:1990dk}), and  $n_{\rm EQ}$ is  the equilibrium number density  at the temperature $T$ of the thermal bath.

The three terms on the right-hand side of Eq.~\eqref{BE} account respectively for the expansion of the Universe, particle annihilation and  creation (see e.g.~\cite{Kolb:1990vq}).  If particles were neither created nor destroyed, the solution $n\sim a^{-3}$ would only be diluted by the  increase of $a$, with ${dn}/{dt}= -3 ({\dot{a}}/{a}) n=-3Hn$.
 With only annihilation,  $n\sim e^{-t/t_A}$ and ${dn}/{dt}= -n/t_{A}$,  where the typical annihilation time $t_A$ is proportional to the mean free path $t_A \simeq \lambda_{mfp}/v=1/ (\sigma_A n v)$. Thus the term proportional to $n^2$ accounts for the annihilation.   Include the  creation, stop the expansion when the thermal bath temperature is $T$  and wait until equilibrium is achieved, in which case $n= n_{\rm EQ}$  and ${dn}/{dt}=0$, thus the creation term must cancel the annihilation term with $n= n_{\rm EQ}$. 

Consider the entropy density $s=({2 \pi^2}/{45}) g_{\rm s-eff}(T) ~T^3$, where $g_{\rm s-eff}(T)$ is the effective entropy number of degrees of freedom~\cite{Kolb:1990vq, BBC-pdg},   and $T$ the photon temperature.
Eq.~\eqref{BE} and the conservation of entropy per comoving volume, $S = s a^3$=constant, which implies ${ds}/{dt}=-3Hs$, can be combined into a single equation for the dimensionless variables
 $Y\equiv n/s$ and  $x \equiv m/T$,
\begin{equation}
\frac{dY}{dx}=\frac{1}{3H}\frac{ds}{dx}\left<\sigma_A v\right>(Y^2-Y_{EQ}^2).
\label{BE-Y}
\end{equation}
Notice that with $S=s a^3$ constant, $Y$  is  proportional to the number of particles per comoving volume $n a^3$. When $g_{\rm s-eff}(T)$ is approximately constant (thus $a \sim T^{-1}$),  a good approximation  most of the time, Eq.~\eqref{BE-Y} becomes
\begin{equation}
\frac{x}{Y_{EQ}} \frac{dY}{dx}= - \frac{\Gamma_A}{H}\left[\left(\frac{Y^2}{Y_{EQ}^2}\right) -1 \right],
\label{BE-Y-2}
\end{equation}
where $\Gamma_A = n_{\rm EQ} \left<\sigma v\right>$ is the equilibrium annihilation rate.   Eq.~\eqref{BE-Y-2} clearly shows that the number of particles per comoving volume becomes constant (${dY}/{dx} = 0$) when ${\Gamma_A}/{H} << 1$. At high $T$,  ${\Gamma_A} >{H} $ for particles that are in equilibrium, but ${\Gamma_A}$ decreases with decreasing $T$ faster than $H$, and  crosses $H$ at  the chemical decoupling or freeze-out temperature $T=T_{fo}$,
\begin{equation}
\Gamma_A (T_{fo})= \left< \sigma_A v \right>_{T=T_{fo}} n_{EQ}(T_{fo}) \simeq H(T_{fo}).
\label{fo}
\end{equation}

$H$ is related to the total  energy density of the Universe $\rho$ by Friedmann's equation $H  =\sqrt{{8} \pi G \rho/3}$~\cite{Kolb:1990vq, BBC-pdg}. $G$ is the gravitational constant and $G= M_P^{-2}$ defines the Planck mass  $M_P \simeq 10^{19}$ GeV. For a radiation-dominated Universe, $\rho \sim T^4$ and thus $H \simeq T^2/ M_P$.

Let us estimate $T_{fo}$ for particles that are relativistic  (e.g. active neutrinos) or non-relativistic (e.g. WIMPs) at decoupling, using estimates of the annihilation cross section  into light SM fermions of mass $m_f <<T$ via the exchange of a mediator of mass $M$  and coupling $g$.

 For relativistic particles, $m <T $, thus $T$ is the dominant quantity of mass dimension which can appear in the numerator of the cross section to compensate the  $1/ M^4$ propagator factor (the  mass dimension of $\sigma$ is $-2$)
\begin{equation}
\sigma^{\rm R}_A \simeq \frac{g^4}{M^4} T^2.
\label{sigmaR}
\end{equation}
For weak interactions $g^4/M^4\simeq G_F^2$ ($G_F\simeq 10^{-5}/$GeV$^2$
 is the Fermi constant).
For non-relativistic DM particles there are two possible situations. If $M > m >T$,  $m$ and not $T$ appears in the cross section, namely
\begin{equation}
\sigma^{\rm NR}_A \simeq \frac{g^4}{M^4} m^2.
\label{sigmaNR-light}
\end{equation}
 If instead the DM mass is much larger than any other quantity with mass dimension, including $M$, only $m$ appears in the cross section, 
\begin{equation}
\sigma^{\rm NR}_A \simeq \frac{g^4}{m^2}.
\label{sigmaNR-heavy}
\end{equation}

\subsubsection{Particles relativistic at decoupling - active neutrinos}

Active neutrinos interact via weak interactions, and Eq.~\eqref{fo}  becomes
\begin{equation}
\Gamma_A (T_{fo})\simeq n_{EQ}(T_{fo}) \sigma^{\rm R}_A(T_{fo})  \simeq G_F^2  T_{fo}^5 \simeq H(T_{fo}) \simeq \frac{T_{fo}^2}{M_{P}}.
\label{fo-R}
\end{equation}
We used that for relativistic particles  $n_{\rm EQ} \simeq T^3$, and that Eq.~\eqref{sigmaR} gives $\sigma^{\rm R}_A \simeq G_F^2  T^2$.  We have also used the form of $H$ valid for a radiation-dominated Universe.  Substituting in the numerical values for $G_F$ and  $M_{P}$ we get $T_{fo} \simeq$ few MeV,  the same result as from a full calculation.  Since the active neutrinos have masses smaller than a few eV, they are relativistic at decoupling. These thermal neutrinos constitute the Cosmic Neutrino Background and are the most abundant particles in the Universe after the CMB photons. Their temperature is smaller than the CMB photon temperature, $T_\nu=(4/11)^{1/3}T$, because $e^+e^-$ pairs annihilate after neutrinos decouple,  so that their entropy goes almost entirely to photons.

\subsubsection{Particles non-relativistic at decoupling- the ``WIMP Miracle"}
 
For particles with $m>>T_{fo}$,  $n_{EQ} = g_\chi \left({mT}/{2\pi}\right)^{3/2} e^{-m/T}$
where $g_\chi$ is the $\chi$ number of degrees of freedom (normalized so that $g_\gamma=2$ for a photon). Thus Eq.~\eqref{fo} becomes
\begin{equation}
 \Gamma (T_{fo}) \simeq \sigma^{\rm NR}_A~ n_{\rm EQ} \simeq  \sigma^{\rm NR}_A(T_{fo}) \left(\frac{m T_{fo}}{2\pi}\right)^{3/2} e^{-m/T_{fo}}  \simeq H(T_{fo}) \simeq \frac{T^2_{fo}}{M_{P}}.
 \label{fo-NR}
\end{equation}
The crossing of the terms proportional to $e^{-m/T_{fo}}$ and $T_{fo}^2$ depends mostly on the exponential factor: it will happen when the argument of the exponential is neither very small nor very large, i.e.  $m/T_{fo} \simeq 1$ (actually one finds $m/T_{fo}$ of O(10)). Using  $T_{fo} \simeq m$ in Eq.~\eqref{fo-NR} we get
\begin{equation}
n_{\rm EQ}(T_{fo}) \simeq  \frac{T^2_{fo}}{\sigma^{\rm NR}_A(T_{fo}) } \simeq \frac{m^2}{\sigma^{\rm NR}_A(T_{fo}) }.
 \label{neq-NR}
\end{equation}
After decoupling, at $T < T_{fo}$,  the number density $n(T)$  only decreases as $T^{3}$ due to the expansion of the Universe, $n(T) = n_{\rm EQ}(T_{fo} ) (T^3/T_{fo}^3)$. Thus, using Eq.~\eqref{neq-NR}, the DM density at $T < T_{fo}$ is
\begin{equation}
\rho(T) = m~n(T) \simeq \frac{m^3 }{\sigma^{\rm NR}_A (T_{fo})} \frac{T^3} {m^3}=\frac{T^3 }{\sigma^{\rm NR}_A (T_{fo})},
 \label{rho-NR}
\end{equation}
With this simple argument~\cite{Baltz:2004tj}, we obtain, the crucial result that the relic density is inversely proportional to the annihilation cross section at $T_{fo}$.

\begin{figure}[t]
\centerline{\includegraphics[width=0.57\hsize]{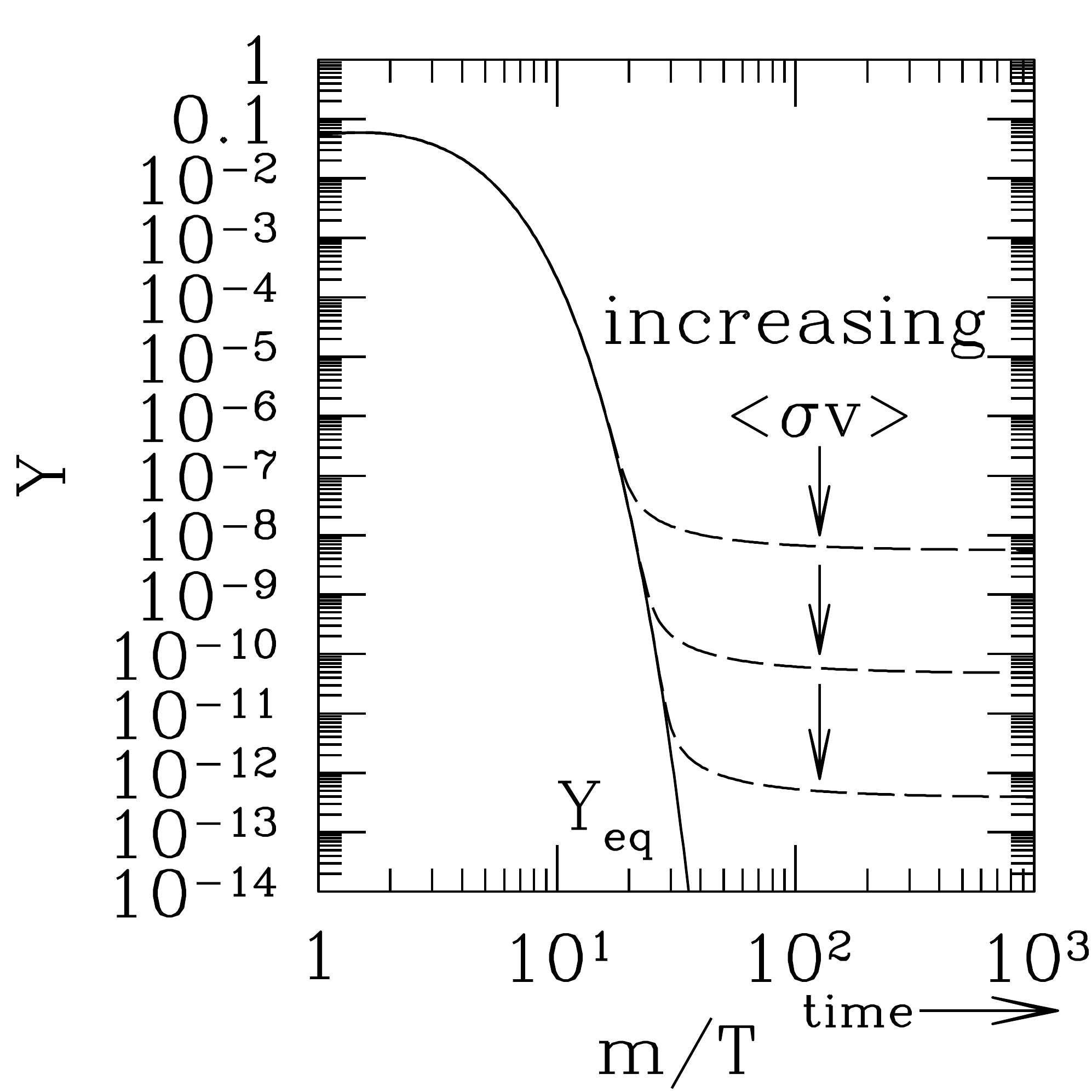}}
\caption{Typical evolution of the WIMP number density per comoving volume $Y$ in the early Universe assuming no
 particle-antiparticle asymmetry as function of $x=m/T$. Here $Y_{eq}$ is the equilibrium  value of $Y$. Fig. from~\cite{Gelmini:2010zh}.}
\label{yyy}
\end{figure}

The numerical solution of the Boltzmann equation, assuming the Universe is radiation dominated and there is conservation of entropy in matter and radiation,  is illustrated in Fig.~\ref{yyy}.  At high temperatures $Y$ closely tracks its equilibrium value 
 until at freeze-out, when annihilation and creation of WIMPs become ineffective.  Then, $Y$ becomes constant.
  The freeze-out for weak-strength interactions occurs at  $x_{fo}\equiv m/T_{fo} \simeq 20$, when the typical WIMP speed  is $v_{fo}= (3T_{fo}/2m)^{1/2} \simeq 0.27c$,  and the relic density is
\begin{equation}
\Omega h^2 \simeq 0.1 \left(\frac{x_{fo}}{20}\right) \left(\frac{60}{g_{\rm eff}}\right)^{1/2}  \frac{ 3 \times 10^{-26} {\rm ~cm^3/s} }{a + 3 b/ x_{fo}},  
\label{omegawimp}
\end{equation}
where $\left< \sigma^{\rm NR}_{A} v \right> \simeq a + b \langle v^2 \rangle + O(v^4)$ ($a$  and $bv^2$ correspond to s-wave and   p-wave annihilation, respectively). $g_{\rm eff}$ is the effective number of degrees of freedom in the radiation bath at  freeze-out  (which coincides with the number $g_{\rm s-eff}$ in the entropy density except at very low temperatures). Within the SM, $g_{\rm eff}$ is 10.75 at  1MeV $ < T <$ 100 MeV, is about 60 above the QCD phase transition and up to $T \simeq$ GeV, becomes about 90 at  $T \simeq$100 GeV, and  reaches 106 above the top quark mass~\cite{Kolb:1990vq, Gondolo:1990dk, BBC-pdg}.

Smaller annihilation cross sections lead to larger relic densities (``the weaker prevail''). 
 WIMPs with stronger interactions remain in equilibrium  longer, hence decouple when  their density is further suppressed by a smaller Boltzmann factor (e.g. for strong interactions,
   $x_{fo} \simeq 45$ instead of 20).

\begin{figure}[t]
\centerline{\includegraphics[width=0.75\textwidth]{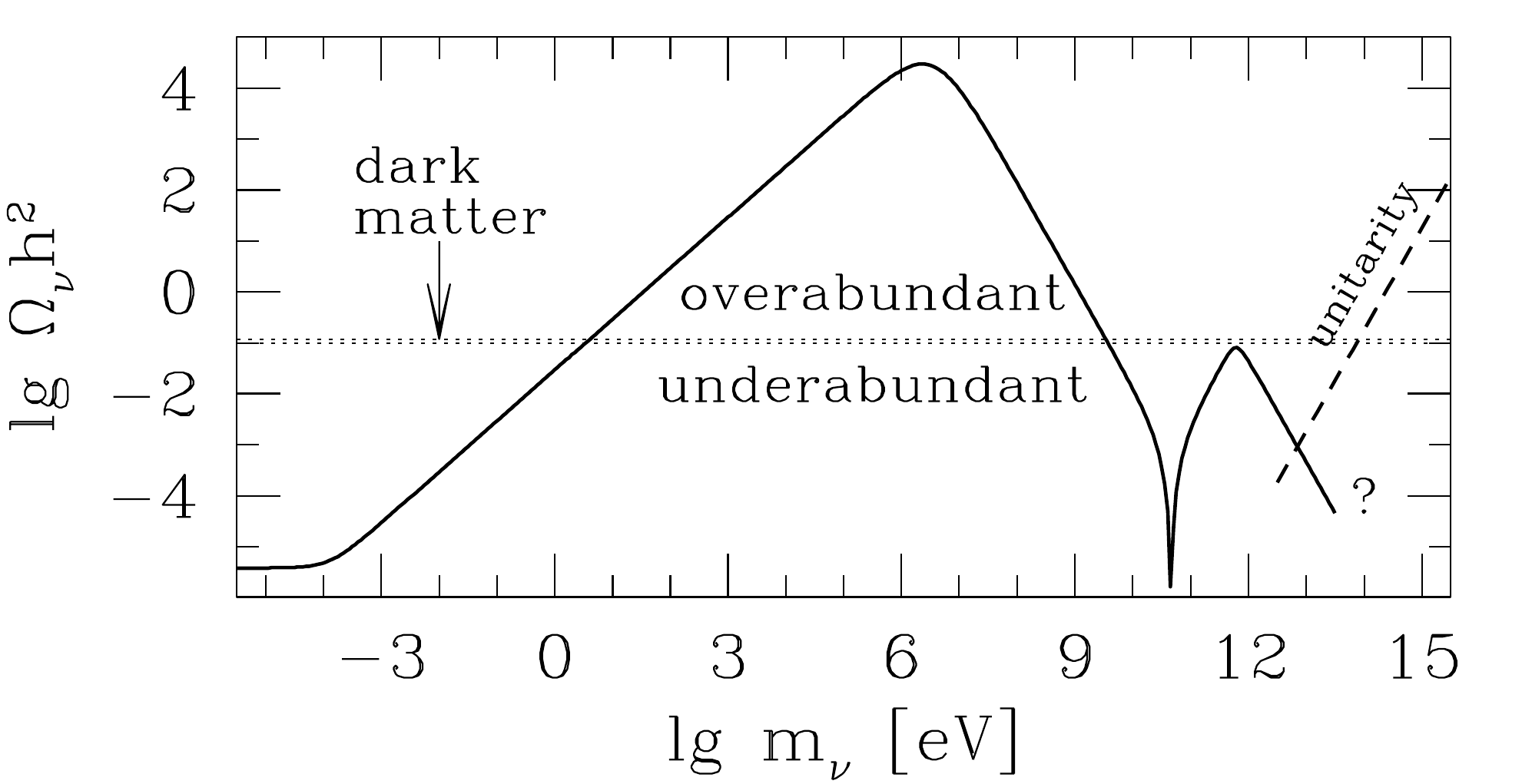}}
\caption{Relic density $\Omega_\nu h^2$ of a 4th generation  thermal Dirac neutrino  with SM interactions as a function of the neutrino mass $m_\nu$ (solid line). Fig. from~\cite{Gelmini:2010zh}.
}
\label{Omega-nu}
\end{figure}

The ``WIMP miracle" refers to the fact that for typical cross sections of weak order of magnitude,  Eq.~\eqref{omegawimp} gives the right order of magnitude of the DM density $\Omega h^2 \simeq 0.1$ for WIMP masses in the GeVs-to-TeV  range.  This is shown in Fig.~\ref{Omega-nu} for
the original WIMP,  studied in 1977 by  Lee and Weinberg~\cite{Lee:1977ua}, an active Dirac neutrino of a 4th generation with no lepton asymmetry. Although this candidate has been rejected by a combination of LEP and direct detection limits,
it is still interesting as a pedagogical example. Fig.~\ref{Omega-nu} shows its relic density $\Omega_\nu h^2$  as function of its mass $m_\nu$.  For $m_\nu <$ MeV these neutrinos are relativistic at decoupling, thus  $n_\nu$ is the same for any mass and $\Omega_\nu h^2 \sim m_\nu$. For $m_\nu >$ MeV
they are non-relativistic at decoupling and  Eq.~\eqref{omegawimp} applies.   Their annihilation is mediated by the Z-boson, thus the  Z-resonance in the cross section  at $m_\nu \simeq  M_Z/2$ gives rise to the characteristic V shape in the $\Omega_\nu h^2$ curve.  For $m_\nu < M_Z$, 
 $\left<\sigma_{annih}v\right >  \simeq G_F^2 m_\nu^2 \simeq (m_\nu / $GeV$)^2$ $10^{-26} {\rm ~cm^3/s}$ (Eq.~\eqref{sigmaNR-light})  and for $m_\nu > M_Z$, 
$\left<\sigma_{annih}v \right> \simeq \alpha/ m_\nu^2 \simeq  (m_\nu/ $TeV$)^{-2}$ $10^{-26} {\rm ~cm^3/s}$ (Eq.~\eqref{sigmaNR-heavy}) . Here, $\alpha$ is the electromagnetic coupling constant.  This shows that GeV- and TeV-mass WIMPs have the relic abundance of the DM 

 Above $m_\nu$ $\simeq$ 100 GeV, new annihilation channels into Z or W boson pairs open up, and when they become dominant, the annihilation cross section increases.  This decreases $\Omega_\nu h^2$. For larger $m_\nu$, a general unitarity argument~\cite{Griest:1989wd}  limits $\Omega_\nu h^2$  to be below the dashed line  in Fig.~\ref{Omega-nu}, and then the neutrino Yukawa coupling becomes too large for perturbative calculations to be reliable (indicated by a question mark in Fig.~\ref{Omega-nu}). 

The relic density of other WIMP candidates exhibits features similar to that of the Dirac neutrino just discussed.
State-of-the-art calculations of WIMP relic densities strive to achieve a precision comparable to that of observations, around a few percent, given the assumptions made.  Notice the assumptions that go into the calculations presented above. 1) The DM is in thermal equilibrium and then decouples while the Universe is radiation dominated. 2) There is no entropy change in radiation plus matter, either during or after decoupling.  3) There is no DM particle-antiparticle asymmetry. 4) The DM particles are stable.  A change in any of these assumptions could lead to a  very different relic abundance:
the DM could be asymmetric,  the Universe may not be radiation dominated at decoupling,  the entropy per comoving volume of radiation plus matter could change, DM WIMPs could be produced in the decay of other particles, thermal or not,  or WIMPs could be unstable and decay into the DM after they decouple.

 This last possibility is the  ``SuperWIMP'' scenario~\cite{Feng:2003xh}, in which the DM particles could have interactions much weaker than weak (super-weak)  because they inherit the correct DM density from unstable thermal WIMPs which decay to them at late times.
   
\subsubsection{Asymmetric Dark Matter (ADM)} 

We owe our very existence to a particle-antiparticle asymmetry in baryons so, why not also the DM? Baryons $B$ (protons and neutrons) decouple while they are non-relativistic.  If the number of  $B$  and $\bar{B}$ would be the same before and during their decoupling, then after decoupling $n_B/s = n_{\bar{B}}/s \simeq 10^{-19}$, and thus $\Omega_B = \Omega_{\bar{B}}  \simeq 10^{-11}$ (see e.g.~\cite{Kolb:1990vq}).
What we observe instead is $\Omega_B \simeq 0.05$ and $ \Omega_{\bar{B}}=0$.  Thus an early Baryon Asymmetry 
 $A_B= ({n_B - n_{\bar{B}}})/s \simeq 10^{-10}$  must exist, so that $n_B$ cannot be smaller than $s A_B $, as $n_{\bar{B}}$ goes to 0.

   The idea of ADM is almost as old as the ``WIMP Miracle". It was proposed first in 1985~\cite{Nussinov:1985xr} for candidates with mass of 100's of GeV in the context of technicolor models, and in 1986~\cite{Gelmini:1986zz}  for  particles of a few GeV of mass (in the context of models for ``Cosmions", DM particles which could cool the central region of the Sun). The idea has been pursued ever since, and became very popular for 1-10 GeV mass WIMPs in recent years (see e.g.~\cite{Petraki:2013wwa} and references therein). The simplest models~\cite{Nussinov:1985xr, Gelmini:1986zz} assume a similar asymmetry of the DM and baryons, $A_{DM} \simeq A_B$, thus their relic number densities are similar,  $n_{DM}\simeq n_B$ and
\begin{equation}
\frac{\Omega_{DM}}{\Omega_B} \simeq \frac{n_{DM} m}{n_{B} m_{B}} \simeq \frac{m}{m_B} \simeq \frac{m}{1 {\rm GeV}}.
\label{ADM}
\end{equation}
If $m \simeq 5$ GeV we obtain  the right ratio ${\Omega_{DM}}/{\Omega_B} \simeq 5$. Thus ADM explains  the present
  ${\Omega_{DM}}/{\Omega_B}$ ratio, which is otherwise a coincidence.

\begin{figure}[t]
\centerline{\includegraphics[width=0.60\textwidth]{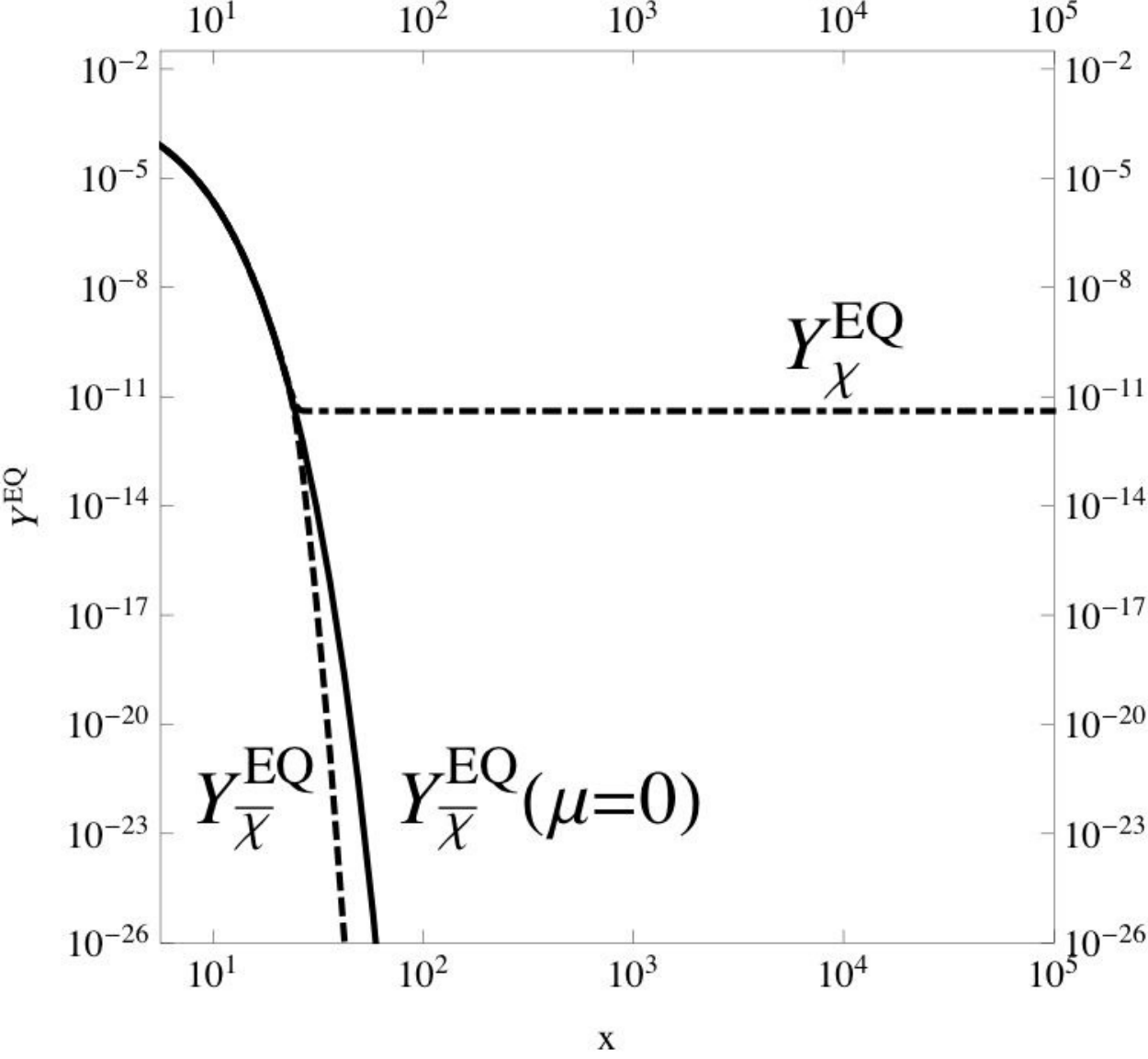}}
\caption{Typical evolution of the equilibrium number density per comoving volume in the early Universe of particles $\chi$ and antiparticles $\bar{\chi}$,  $Y^{\rm EQ}_{\chi}$ and $Y^{\rm EQ}_{\bar{\chi}}$, in the presence of an asymmetry $Y- \bar{Y} =A> 0$, compared with $Y^{\rm EQ}$ for no asymmetry (chemical potential $\mu=0$), as function of $x=m/T$. Fig. from~\cite{Gelmini:2013awa}.
}
\label{Y-ADM}
\end{figure}
 
 Fig.~\ref{Y-ADM} shows the evolution of $Y_{\chi}= n_\chi/s$ and $Y_{\bar{\chi}}= n_{\bar{\chi}}/s$ while particles $\chi$ and antiparticles ${\bar{\chi}}$ are in thermal  equilibrium in the presence of an asymmetry $Y- \bar{Y} =A> 0$. The equilibrium number densities  of $\chi$ and $\bar{\chi}$ 
 in this case differ by the chemical potential  $\mu_{\chi}$  (in equilibrium $\mu_{\chi}=-\mu_{\bar{\chi}}$):
\begin{equation}
n_{\chi}^{EQ}=g_{\chi}\left(\frac{mT}{2\pi}\right)^{3/2}e^{(-m+\mu_{\chi})/T},~n_{\bar{\chi}}^{EQ}=g_{{\chi}}\left(\frac{mT}{2\pi}\right)^{3/2}e^{(-m-\mu_{\chi})/T}.
\label{nxeq}
\end{equation}
Therefore $n_{\chi}^{EQ}(\mu_{\chi}=0)\left[\exp{(\mu_\chi /T)}- \exp{(-\mu_\chi /T)}\right] =As$.  Notice that
$Y^{\rm EQ}_{\chi}$ cannot become smaller than  $A$ (it approaches $A$ as $Y^{\rm EQ}_{\bar{\chi}}$ goes to zero). $Y^{\rm EQ}_{\bar{\chi}}$ decreases exponentially until ${\bar{\chi}}$ freezes-out, when $\Gamma_{\bar{\chi}}^{\rm EQ}  \simeq H$, where $\Gamma^{\rm EQ}_{\bar{\chi}} \sim n^{\rm EQ}_{\chi} \sim A T^3$. The previous  Boltzmann Eqs.~\eqref{BE} and \eqref{BE-Y} become (if the DM particles annihilate and are created only  in $\chi \bar{\chi}$ pairs)
\begin{equation}
\frac{dn_{\chi}}{dt}+3Hn_{\chi}=-\left<\sigma_{\chi{\bar{\chi}}} v\right>(n_{\chi}n_{\bar{\chi}}-n^{EQ}_{\chi}n^{EQ}_{\bar{\chi}}).
\label{dndt}
\end{equation}
and 
\begin{equation}
\frac{x}{Y^{EQ}_{\chi}} \frac{dY_{\chi}}{dx}=\frac{\Gamma_\chi}{H} \left(\frac{Y_{\chi}Y_{\bar{\chi}}}{Y^{EQ}_{\chi}Y^{EQ}_{\bar{\chi}}} -1 \right),
\label{dYdx}
\end{equation}
where $\Gamma_\chi= \left<\sigma_{\chi{\bar{\chi}}} v\right> n^{EQ}_{\bar{\chi}}$  is the annihilation rate of particles $\chi$. The corresponding equations for $\bar{\chi}$ are obtained by exchanging $\chi$ and $\bar{\chi}$. Annihilations cease once the minority component (here $\bar{\chi}$) decouple, when $\Gamma_{\bar{\chi}} <H$. 

In the standard cosmology (i.e. assuming radiation domination at decoupling and entropy conservation) the relic density of the minority component~\cite{Griest:1986yu, Iminniyaz:2011cd} is exponentially small with respect to the majority component density, which means that there is no DM annihilation after decoupling.  This is one of the main characteristics ADM is assumed to have.  However, this is a pre-BBN cosmology dependent feature. In some non-standard cosmologies, the present ADM annihilation rate can be very large (see e.g.~\cite{Gelmini:2013awa}).

\subsubsection{Non-Standard pre-BBN cosmologies} 

The relic density and  velocity distribution of many WIMPs and other DM candidates, e.g thermal WIMPs of mass  $m>$ 80 MeV,  heavy sterile neutrinos and axions, depend on the unknown characteristics of the Universe before $T\simeq$ 4 MeV~\cite{Hannestad:2004px}, when the content and expansion history of the Universe may differ from the standard assumptions.  In non-standard  pre-BBN cosmological models, the WIMP relic abundance may be higher or lower than the standard abundance. The density may be decreased by reducing the rate of  thermal production (through a low $T_{RH} < T_{fo}$) or by producing radiation after freeze-out (entropy dilution). The density may also be increased by creating WIMPs from decays of particles or extended objects (non-thermal production) or by increasing the expansion rate of the Universe at the time of freeze-out (see e.g.~\cite{Gelmini:2010zh, Gelmini:2009yh}). 

Not only the relic density of WIMPs but their relic velocity distribution can differ from the standard predictions. WIMPs could be warmer and even WDM~\cite{Lin:2000qq}, if they are produced in the decay of a heavy particle and do not exchange momentum with the thermal bath after, or could be colder~\cite{Gelmini:2008sh}.

Non-standard cosmological scenarios are more complicated than the standard scenario (e.g. to generate the baryon number asymmetry).  They contain additional parameters that can be adjusted to modify
the DM relic density. However these are due to physics at a high energy scale, and do not change the particle physics model at the electroweak scale or below. 

\section{WIMP DM searches}

WIMP's are actively searched for in direct and indirect DM detection experiments, and at colliders. 
 
 Direct searches look for energy deposited within a detector by the DM particles in the dark halo of the Milky Way. They are  sensitive  to even a very subdominant  WIMP component of the DM, if the scattering cross section $\sigma_{S}$  is large enough.  For thermal WIMPs ($\rho \sim 1/\sigma_{A}$), the event rate $R\sim \rho \sigma_{S} \sim \sigma_{S}/ \sigma_{A}$.  Because of the crossing symmetry relating  annihilation and scattering amplitudes, if the annihilation cross section $ \sigma_{A}$ is large so is  $\sigma_{S}$. Therefore the rate may remain large even  for WIMPs with a very small relic density (see e.g. \cite{Duda:2001ae}).  Direct searches are subject to uncertainties in the local dark halo characteristics, and are relatively  insensitive to DM that couples to leptons only (``leptophillic"). They would fail if the WIMP is so light its signal is below threshold, or if $\sigma_{S}$ is too small. 
     
    Indirect searches look for annihilation (or decay) products of WIMPs in the dark haloes of the Milky Way and other galaxies, as well as WIMPs accumulated within the Sun or Earth.  These searches are sensitive to interactions with all SM particles and directly probe the  annihilation process in the ``WIMP miracle".  The caveat to this type of search is that the DM may not annihilate (e.g. if it is asymmetric and consists exclusively of particles and no antiparticles) or decay.   Experimental  sensitivities  to several types of signal are expected to improve greatly in
  the coming decade, but the discovery of the DM through these searches requires
  understanding astrophysical backgrounds and the expected signal is  subject to uncertainties in  dark halo characteristics.
   
At colliders, in particular the LHC, WIMPs are searched for as missing transverse energy.
 The caveats to this type of search is that the DM particle may be too heavy to be produced (above  a few TeV at the LHC)   or  its signal may be hidden by backgrounds. Also, a signal produced by a particle escaping the detectors with lifetime $\simeq 100$ ns cannot be distinguished from one with lifetime $> 10^{17}$ s as required for  DM particles.
 Hadron colliders are relatively insensitive to DM that interacts only with leptons. 

All three types of searches are independent and complementary to each other. Even if the LHC finds  a good DM candidate, to prove that it is the DM (and that it did not decay on cosmological time scales or was not produced in large enough amounts in the early Universe) we will need to find it in the dark haloes of our galaxy and other galaxies.

\subsection{Direct Detection of WIMPs}

The flux of  WIMPs passing through a detector on Earth can be large, $n v=\rho_{DM} v/m \simeq 10^7$(GeV/$m$)/cm$^2$s. However the expected energies deposited  and interaction rates are very small, keV to 10's of keV and  less than an  event per 100 kg per day.
 Measuring these small energies and rates requires a constant fight against backgrounds. The experiments need to be underground,  in mines or tunnels under mountains,  to shield them from cosmic rays. The signal should consist of single hits and be  uniform throughout the volume of  the detector (this is a powerful way of discriminating agains neutrons, which tend to interact multiple times in a detector and closer to its surface). 

Most direct searches are non-directional, but some, still in the stage of development, attempt to measure the recoil direction beside the energy (see e.g.~\cite{Cushman:2013zza}). In a directional search, 10's of events would be enough to determine the direction of arrival and thus the DM origin of the signal. Without directionality, the unmistakable signature of DM is a few percent annual modulation of the rate, due to the variation in the velocity of the ``WIMP-wind" on Earth as Earth rotates around the  Sun~\cite{Drukier:1986tm}. 

\subsubsection{The dark halo model}

 In order to compare the results of different experiments usually the Standard Halo Model (SHM) is used. In the SHM the local DM density is $\rho_{DM} \simeq $ 0.3 GeV/cm$^3$  (its error is $\pm$0.1GeV/cm$^3$, see e.g.~\cite{Read:2014qva}).  The local DM velocity distribution in the galactic reference frame $f_\text{G}(\vec{v}_G)$  is a Maxwell-Boltzmann distribution with zero average  and dispersion $ v_0$, truncated at the local escape speed $v_{\rm esc}$ from our galaxy  (see e.g. \cite{Savage:2008er}).
The WIMP velocity with respect to the Galaxy is $\vec{v}_G = \vec{v} + \vec{v}_\odot + \vec{v}_{E}(t)$ where $\vec{v}$ is the WIMP velocity  with respect to Earth, $\vec{v}_\odot$  is the velocity of the Sun with respect to the galaxy, and $ \vec{v}_{E}(t)$ is the velocity of Earth with respect to the Sun. Thus the velocity distribution with respect to the  Earth is
\begin{equation}
f(\vec{v}, t) = f_\text{G}(\vec{v} + \vec{v}_\odot + \vec{v}_{E}(t)).
\label{f}
\end{equation}
 There are uncertainties in the parameters of the SHM (see e.g.~\cite{Bozorgnia:2012eg}). Some usual values are $v_\odot=232$ km/s~\cite{Savage:2008er}, ${v}_0=220$ km/s, and ${v}_{\rm esc}=544$ km/s~\cite{Smith:2006ym}.
 The most recent Radial Velocity Experiment (RAVE)  2013 results~\cite{Piffl:2013mla} give  ${v}_{\rm esc}=533^{+54}_{-41}$ km/s but values  of  $v_{\rm esc}$ between 500  and 650 km/s can be found in the literature.  In the SHM the maximum average velocity of WIMPs with respect to Earth happens between the end of May and the beginning of June (see e.g.~\cite{Bozorgnia:2012eg}).  The minimum occurs six months later (although not exactly, because of the small ellipticity of Earth's orbit).
 
 We expect the actual halo to deviate  from this simplistic model. The local density and velocity distribution could actually be very different if Earth is within a DM clump, which is unlikely~\cite{Vogelsberger}, or in a DM stream, or if there is a ``Dark DiskÓ~\cite{DD, DDDM} (see section 2)  in our galaxy.  The DM of the Sagittarius Stream, tidally stripped from the Sagittarius Dwarf Galaxy, could be passing through the Solar system, perpendicularly to the galactic 
 disk~\cite{stream}. A large amount of DM clumps are expected to remain within the dark halo of our 
 galaxy~\cite{Vogelsberger, Kuhlen:2008aw}, because haloes grow hierarchically, incorporating lumps and tidal streams from earlier phases of structure formation. However, clumps are more effectively destroyed by tidal effects near the center of the galaxy, thus most of them are far from the Sun. The chance that a random point close to the Sun  is lying within a clump is smaller than 10$^{-4}$~\cite{Vogelsberger}. ``Debris flows", which are spatially homogeneous structures in velocity, are expected from this complicated merger history, and they would also modify the velocity distribution primarily in the high velocity tail~\cite{debris-flows}. 

 \subsubsection{Recoil energies and rates}

The maximum recoil energy of a target of mass $M_T$  in an elastic collision with a WIMP of mass $m$ is
\begin{equation}
E_{max} = 2 \mu_T^2 v^2/ M_T= q^2_{max}/ 2 M_T,
\label{Emax}
\end{equation}
 where $\mu_T = {m M_T}/{(m+M_T)}$ is the  reduced mass.  For light WIMPs with  $m \ll M_T$, $\mu_T \simeq m$, 
and typically $E_{max} \simeq 2{\rm keV}\left({m}/{\rm GeV}\right)^2\left({10~ {\rm GeV}}/ {M_T}\right)$, since $v \simeq 10^{-3}$c. The threshold recoil energy in most detectors at present is O(keV), thus detectable WIMPs must have $m\gsim$  GeV.
For heavy WIMPs with $m \gg M_T$, 
$\mu = M_T$.  The energy is large enough, $E_{max} = 2A_T$ keV (we use $M_T \simeq A_T$ GeV if  the nuclear mass number is $A_T$) but  the limits die out  because the WIMP flux decreases as $1/m$.   A DM candidate with sub-GeV mass, called ``Light DM" (LDM),  with  $m \simeq$ MeV to GeV, could deposit enough energy through interactions with electrons  (between 1 to 10 eV) to  be detected via electron ionization or  excitation, or molecular dissociation~\cite{LDM, Angle:2011th}. 

 The typical momentum transfer in an elastic collision is $q \simeq \mu_T v \simeq$ O(MeV), which becomes $q \simeq$
 MeV$\left({m}/{{\rm GeV}}\right)$ for $m\ll M_T$,  and $q \simeq A_T$MeV for  $m\gg M_T$. We note that
\begin{equation}
q  <{1}/{R_{\rm Nucleus}}  \simeq {\rm MeV} \left({160}/{A_T^{1/3}}\right),
\label{q}
\end{equation}
 thus WIMPs interact coherently with nuclei. $R_{\rm Nucleus} = 1.25~ {\rm fm} ~A_T^{1/3}$ is the radius  of a target nucleus  (we recall that fm=$10^{-15}$meters=(197 MeV)$^{-1}$) and $A_T^{1/3}$ is a number of about 3 to 5 for most nuclei. If  $1/q >> R_{\rm Nucleus}$ the nucleus interacts like a point-like particle.
For larger $q$, the loss of complete coherence is taken into account by a  nuclear form factor.  For spin-independent (SI) interactions (see below) the form factor is the Fourier transform of the density of nucleons in the nucleus.
A usual form  for it is the Helm form factor for the distribution of charge (protons), assuming that the distributions of neutrons and protons are similar~\cite{Helm:1956zz}.

The recoil rate expected (in non-directional detectors) is given by the local WIMP flux $n v$ 
times the number of target nuclei in the detector times the scattering cross section, integrated over the WIMP velocity distribution. 
In units of events/(unit mass of detector)/(keV of recoil energy)/day  the expected differential recoil rate is
\begin{equation}
\frac{dR}{dE_R} =  \sum_T \frac{dR_{T}}{dE_R}=\sum_T \int_{v>v_{\rm min}}  \frac{C_T}{M_T}
\times \frac{d\sigma_T}{dE_R}  \times n  v f(\vec{v},t) d^3v.
\label{dRdE}
\end{equation}
Here, $E_R$ is the recoil energy, $T$ denotes each target nuclide (elements and isotopes), $C_T$ is the mass fraction of nuclide $T$ in the detector, $C_T/M_T$ is the number of targets $T$ per unit mass of detector, $d\sigma_T/dE_R$ is the differential WIMP-nucleus scattering cross section and $n= \rho/m$. The local WIMP density  $\rho$ is the local DM density,  $\rho$=$\rho_{DM}$, if the WIMP in question constitutes all the DM. If instead  {\it R}=$\Omega_{\rm WIMP}/ \Omega_{DM}$$<$1, then $\rho$={\it R}$\rho_{DM}$, assuming that the DM close to Earth has  the same composition as the DM as a whole.
The  local  DM density and velocity distribution  with respect to Earth, $f(\vec{v},t)$, depend on the dark halo model adopted.  

The minimum speed $v_{\rm min}$ a WIMP must have  to communicate to the target $T$ a recoil energy $E_R$ depends on the collision being elastic or inelastic. In some particle models a DM particle of mass $m$ may collide inelastically, producing a different state with mass $m' = m + \delta$, while the elastic scattering is either forbidden or suppressed~\cite{TuckerSmith:2001hy}).  $\delta$ can either be positive (``inelastic DM", iDM, with ``endothermic" scattering)~\cite{TuckerSmith:2001hy} or negative (``exothermic DM", exoDM, with``exothermic" scattering)~\cite{Graham:2010ca}. These types of interactions enhance the potential signal in some targets and suppress it in some others:  ``endothermic" scattering favors heavier targets (so the iodine in DAMA/LIBRA is preferred over the germanium in CDMS) while ``exothermic" scattering behaves in the opposite fashion (the silicon in CDMS is preferred over the xenon in LUX or Xenon100). For elastic collisions $\delta=0$.   For $\left| \delta \right| \ll m$,
\begin{equation}
v_{\rm min}=\ \left|  \sqrt{ \frac{M_T E_R}{2\mu_T^2}}  + \frac{\delta}{\sqrt{2M_T E_R}} \right|.
\label{vmin}
\end{equation}

The differential recoil rate, Eq.~\eqref{dRdE},  is not directly experimentally accessible because of energy-dependent experimental efficiencies and  energy resolution functions, and because what is often measured is a part $E'$ of the recoil energy $E_R$. The observable differential rate is
\begin{equation}
\label{Obs-rate}
\frac{dR}{dE'} = \epsilon(E') \, \int_0^\infty dE_R \, \sum_{T}  G_{T}(E_R,E') \, \frac{dR_{T}}{dE_R}.
\end{equation}
Here $E'$ is the detected energy, often quoted in keVee (keV electron-equivalent) or in photoelectrons.  $\epsilon(E')$ is a counting efficiency or cut acceptance. $G_T(E_R, E')$  is a (target nuclide and detector dependent) effective energy resolution function that gives the probability that a recoil energy $E_R$ is measured as $E'$.  It incorporates the mean value $\langle E' \rangle = Q_T E_R$, which depends on  an energy dependent ``quenching factor" $Q_T(E_R)$, and the energy resolution $\sigma_{E_R}(E')$. These functions must be measured (although sometimes the energy resolution is obtained from computation).
 
 WIMP interactions in crystals produce mostly phonons. Only a fraction $Q_T$ of the recoil energy goes on average into ionization or scintillation.  For example, $Q_{Ge} \simeq$ 0.3, $Q_{Si} \simeq$ 0.25, $Q_{Na} \simeq$ 0.3, and $Q_I \simeq$ 0.09. In noble gases such as Xe,  a similar factor $L_{eff}$ measures the scintillation efficiency of a WIMP relative to a photon. There are large experimental uncertainties in the determination of these parameters at low energies.
 
We can see in Eqs.~\eqref{Obs-rate} and \eqref{dRdE} that three elements enter into the observed rate in  direct detection experiments:  the detector response, the local dark halo model and, finally, the particle physics input, given through the cross section and mass of the DM candidate.

 \subsubsection{The scattering cross section}

For contact interactions in the non-relativistic limit $v \to 0$ the differential cross section has the form 
\begin{equation}
\frac{d\sigma_T}{dE_R}=  \frac{\sigma_T(E_R)}{E_{max}} = \sigma_T(E_R)~\frac{M_T}{ 2\mu_T^2 v^2}.
\label{dsigmadE}
\end{equation}
for momentum transfer $q$ and velocity-independent interaction operators. There are only two types of interactions of this kind,  SI and SD.

\begin{figure}[t]
\centerline{\includegraphics[width=1.0\textwidth]{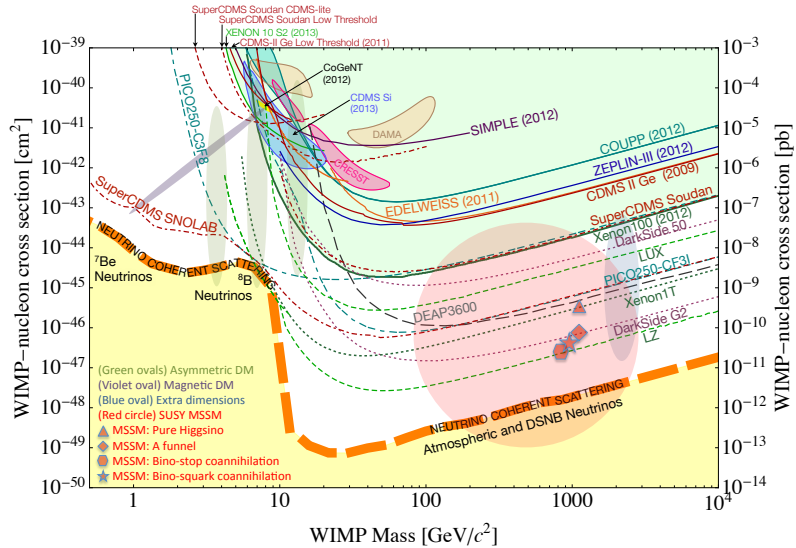}}

\vspace{-0.9cm}

\caption{Overview of 90\%CL existing direct detection limits (solid lines) and future sensitivity regions for a WIMP with 
SI interactions and $f_p= f_n$ constituting the whole of the DM, assuming the SHM, plotted in the $(m, \sigma_{\rm p})$ 
plane~\cite{Cushman:2013zza}. DAMA/LIBRA (light brown), CoGeNT (yellow), CRESST-II (pink) and CDMS-II-Si (light blue) signal regions for this WIMP, are shown. The  level at which neutrinos of different sources would constitute a background to a DM signal, the ``neutrino floor"(thick dashed orange lines) and regions of interest for particular DM candidates are also shown (see~\cite{Cushman:2013zza} for details).}
\label{SI-overview}
\end{figure}

 ``Spin-independent" (SI) contact  interactions are due to scalar or vector couplings. The DM couples to the nuclear density.  We can write $\sigma^{SI}_T(E_R)$=$\sigma_{T0} F^2(E_R)$, where $F^2(E_R)$ is the nuclear form factor, $F^2(0)$=1, (see e.g.~\cite{Jungman:1995df}), and 
\begin{equation}
\sigma_{T0} = \bigl[ Z + (A-Z) (f_n/f_p) \bigr]^2
(\mu_T^2/\mu_{\rm p}^2)\sigma_{\rm p}.
\label{sigma-SI}
\end{equation}
Here, $\sigma_{\rm p}$ is the WIMP-proton cross section while $f_n$,$f_p$ are the WIMP couplings to $n$,$p$. For $(f_n/f_ p)$=1 (isospin-conserving coupling), $\sigma_{T0} =A^2 (\mu_T^2/\mu_{\rm p}^2) \sigma_{\rm p}$.  Heavier nuclei are more neutron rich than lighter nuclei. Thus, the ratio  $(f_n/f_p) = -Z_T/ (A_T -Z_T)$ cancels the coupling of the DM with the particular nuclide with atomic and mass numbers $Z_T$ and $A_T$, and changes the couplings to all other nuclei too (no choice can make  the coupling  with an  element zero because of its isotopic composition).  ``Isospin-Violating" DM~\cite{IsospinViolating} with  $f_n/f_p = -0.7$ minimizes the coupling with xenon (thus weakening the limits of Xenon10, Xenon100 and LUX, some of the strongest  at present).  $f_n/f_p = -0.8$ instead reduces maximally the coupling to germanium~\cite{Gelmini:2014psa} (thus weakening preferentially the limits of CDMS and SuperCDMS).

\begin{figure}[t]
\centerline{\includegraphics[width=0.58\textwidth]{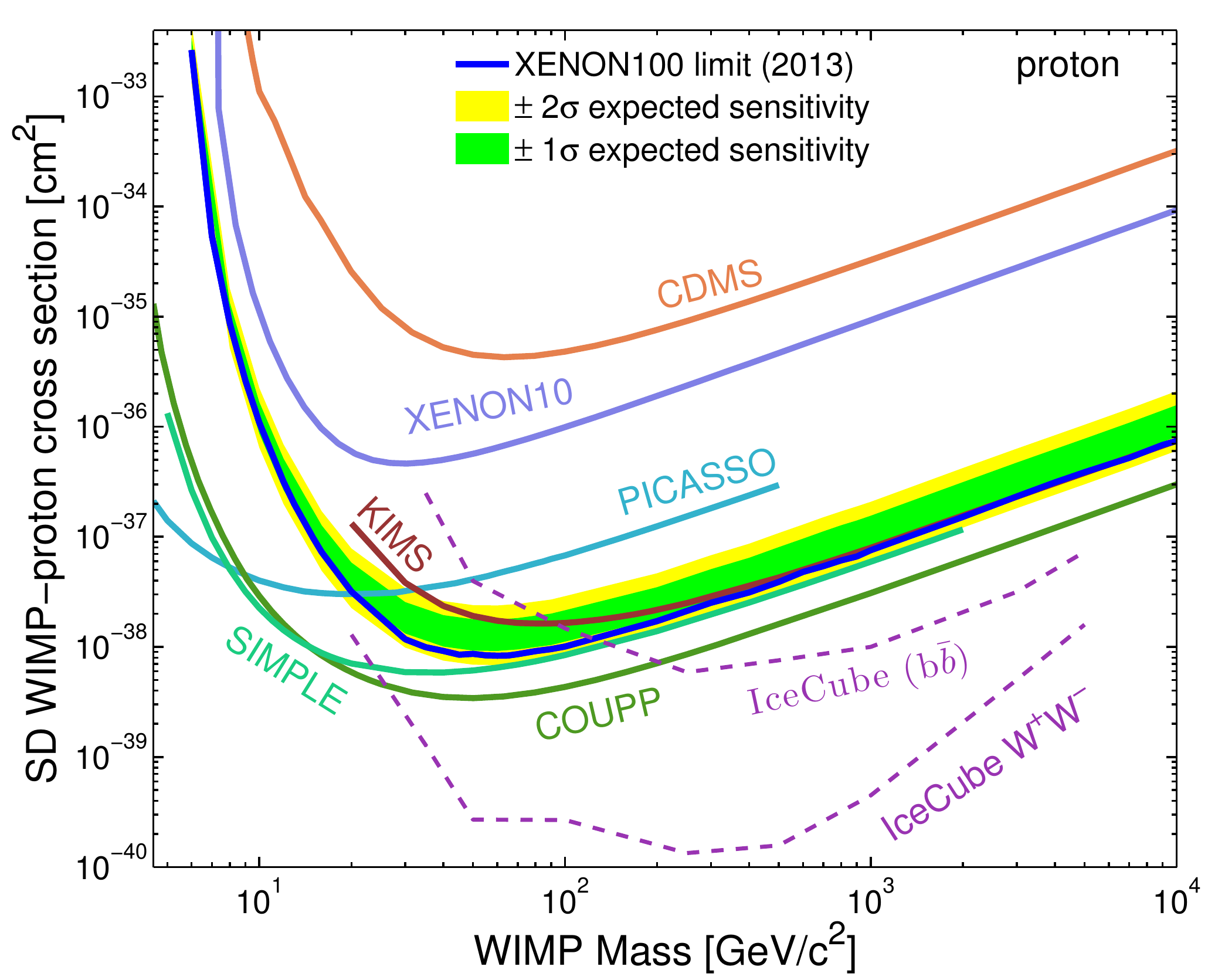}}
\caption{Best bounds on WIMPs with spin-dependent  (SD) interaction coupled only to protons, assuming the SHM. (Fig. from \cite{Cushman:2013zza}). The IceCube limits come from the non-observation of neutrinos from the Sun (indirect detection), assuming annihilation into the mentioned channels. The others  are direct detection limits. See~\cite{Cushman:2013zza} for details.}
\label{SD-proton}
\end{figure}

``Spin-dependent" (SD) interactions result from an axial-vector coupling. The DM couples to the nuclear spin density, leading to 
(see e.g.~\cite{Jungman:1995df})
\begin{equation}
\sigma^{SD}_T(E_R)= 32\mu^2G_F^2 \left[(J_T+1)/J_T\right] \bigl[ \langle S_{\rm p}\rangle 
a_{\rm p} + \langle S_{\rm n}\rangle 
a_{\rm n} \bigr]^2 F^2_{SD}(E_R).
\label{sigma-SD}
\end{equation}
Here, $J_T$ is the nuclear spin,  $a_{\rm p,n}$ are the WIMP couplings to p and n and $F^2_{SD}(E_R)$ is the  nuclear form factor, with
$F^2_{SD}(0)$=1. $\langle S_{\rm p,n}\rangle$, the expectation values of the p and n spin content in the target nucleus, are numbers $\lsim$ O(1) that can differ easily by factors of 2 or more in different nuclear models (see for example Ref.~\cite{Bednyakov:2004xq}). Since also the nuclear spins are of O(1), SD cross sections are a  factor  $A_T^2$ smaller than SI cross sections. The bounds on the second are therefore typically better than the bounds on the first (Figs.~\ref{SI-overview}, \ref{SD-proton}).

There are many other types of possible DM-nucleus interactions besides the two mentioned, and many of them have been considered in recent years to try to accommodate different hints of a DM signal in direct or indirect searches. All other interation operators contain extra powers of the momentum transfer $q$ or WIMP velocity. Refs.~\cite{interactions,  Barello:2014uda}
list all possible operators in the non-relativistic limit up to  $O(q^2)$. For example,  a pseudo-scalar mediator yields a coupling  
 $(\vec{S}_{DM} \cdot \vec{q})(\vec{S}_N \cdot \vec{q})$,  with two extra powers of $q$, dependent on the spin of the DM particle, $\vec{S}_{DM}$, and the spin of the interacting nucleon, $\vec{S}_N$. The mediators can be either heavy or light
with respect to the momentum transfer.

Starting from the fundamental interactions with which the DM particles couple to quarks, there are uncertainties on how to pass from quarks to protons and neutrons, and then to nuclei. Each type of interaction requires its own nuclear form factor, most of which are poorly known. 

\subsubsection{The average inverse speed function $\eta(v_{\rm min}, t)$}

With the contact differential cross-section of ~Eq.\eqref{dsigmadE},  
Eq.~\eqref{dRdE} becomes
\begin{equation}
\frac{dR}{dE_R} = \sum_T \frac{ \sigma_T(E_R) }{2 m \mu_T^2}  \rho \left[\int_{v>v_{\rm min}}
\frac{f(\vec{v},t)}{v} d^3v \right] = \sum_T \frac{\sigma_T(E_R)}{2 m \mu_T^2}  \rho \eta(v_{\rm min}, t),
\label{dRdE-2}
\end{equation}
where the last equality defines the velocity integral  $\eta(v_{\rm min}, t)$. Due to the revolution of the Earth around the Sun,  
 ${\eta}({\rm v}_{min},t)$ has an annual modulation generally well approximated by the first two terms of a harmonic series,
\begin{equation}
\eta(v_{\rm min}, t)= \eta_{0}(v_{\rm min}) + \eta_{1}(v_{\rm min}) \cos(\omega (t - t_0),
\label{eta} 
\end{equation}
where $t_0$ is the time of the maximum of the signal and $\omega = 2 \pi/$yr.  In the SHM,   $\eta(v_{\rm min}, t)$ is maximum when the average WIMP velocity with respect to Earth is maximum for $v_{\rm min} >$ 200 km/s. For lower $v_{\rm min}$ values,  $\eta(v_{\rm min}, t)$ is instead a minimum then. The time average $\eta_{0}$ and the modulated component  of ${\eta}$ enter respectively into  the average and modulated parts of the rate in Eq.~\eqref{dRdE-2}.  Notice that the factor $\rho \eta(v_{\rm min}, t)$ includes all the dependence of the rate  on the dark halo model for any detector.

 \subsubsection{Hints and limits: Halo-Dependent and Independent analyses}

There are many direct DM detection experiments that are either running, in construction or in the stage of research and development (see e.g.~\cite{Cushman:2013zza} and references therein).  They use different target materials and  detection strategies (see Fig.~\ref{SI-overview}). Three direct detection experiments, DAMA/LIBRA~\cite{Bernabei:2010mq} (containing NaI), 
CoGeNT (Ge)~\cite{Aalseth:2010vx} and CDMS-II-Si~\cite{Agnese:2013rvf} 
have at present claims of  having observed potential signals of DM. DAMA/LIBRA finds in 14 years  of data an annual modulation at the 9.3$\sigma$ C.L. with a phase compatible with that expected from DM in the SHM.  A few years ago CoGeNT found both an unexplained  rate excess and an annual modulation (at  the $2\sigma$  initially, and later with a smaller C.L.), both attributable to WIMP interactions. CDMS-II (had Ge and Si) found 3 unexplained low-energy events in their Si component which could be due to DM. CRESST-II with an upgraded detector no longer finds an unexplained excess in their 
rate~\cite{Angloher:2014myn}, as they had found in their previous  2010 results~\cite{Angloher:2011uu}. All other direct detection searches, including LUX (Xe), XENON100 (Xe), XENON10 (Xe), CDMS-II-Ge, CDMSlite (Ge) and SuperCDMS (Ge), have produced only upper bounds on the interaction rate and annual modulation amplitude of a potential  WIMP signal (see e.g~\cite{DelNobile:2014sja} and references therein). 

It is thus essential to compare these data to decide if the potential DM signals are compatible with each other and with the upper bounds set by searches with negative results for any particular DM candidate. There are two ways of presenting and comparing direct detection data: ``Halo-Dependent" and  ``Halo-Independent".

 As we have seen, the rate observed in a particular direct-detection detector due to DM particles in the dark halo of our galaxy depends on three main elements: 1) the detector response to potential WIMP collisions within it; 2) the WIMP-nucleus cross section and WIMP mass; 3) the local density $\rho$ and velocity distribution $f(\vec{\rm v}, t)$ of WIMPs passing through the detector. All three elements have considerable uncertainties.
 
\begin{figure}[t]
{\includegraphics[width=0.48\textwidth]{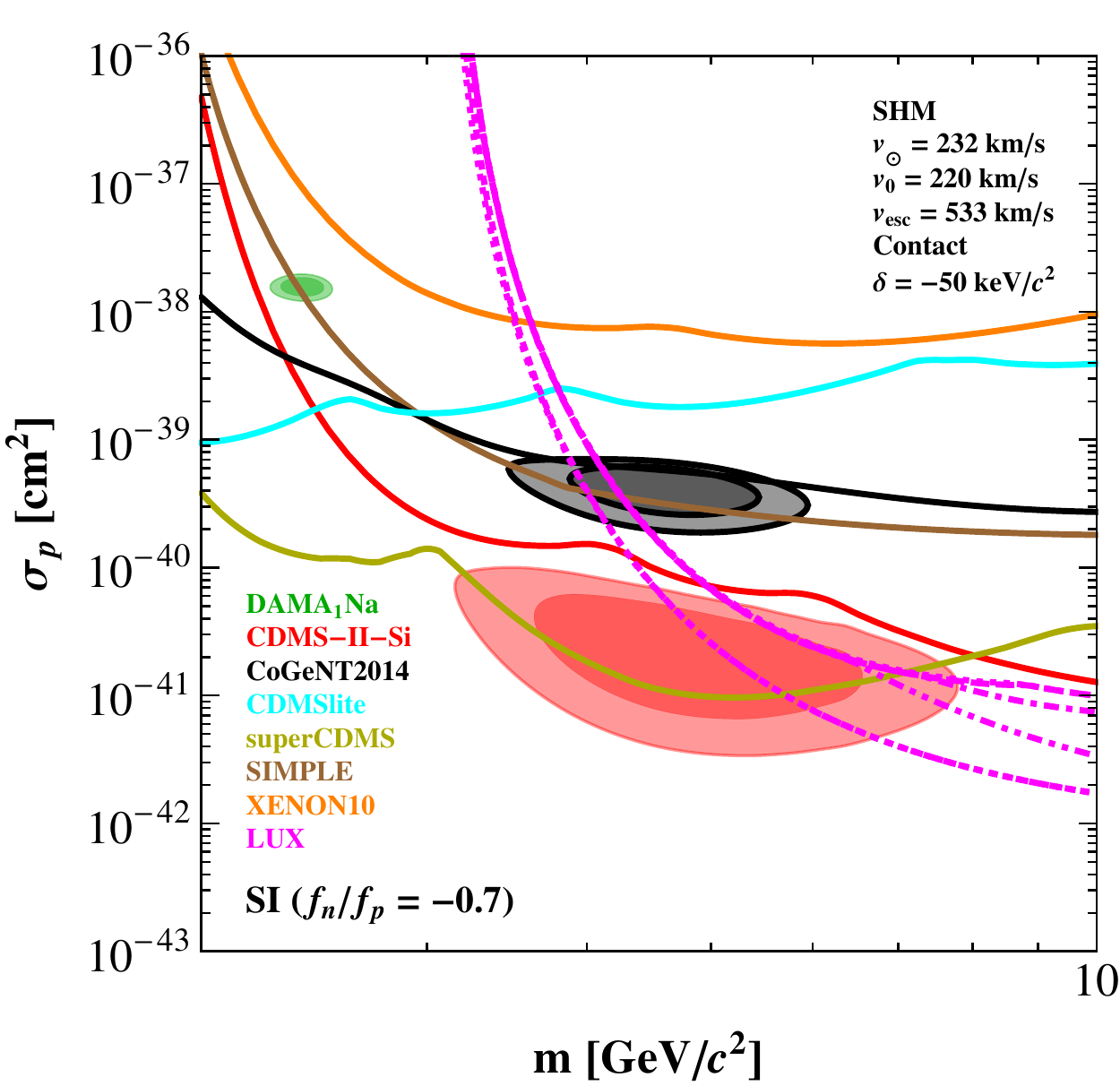}}
{\includegraphics[width=0.50\textwidth]{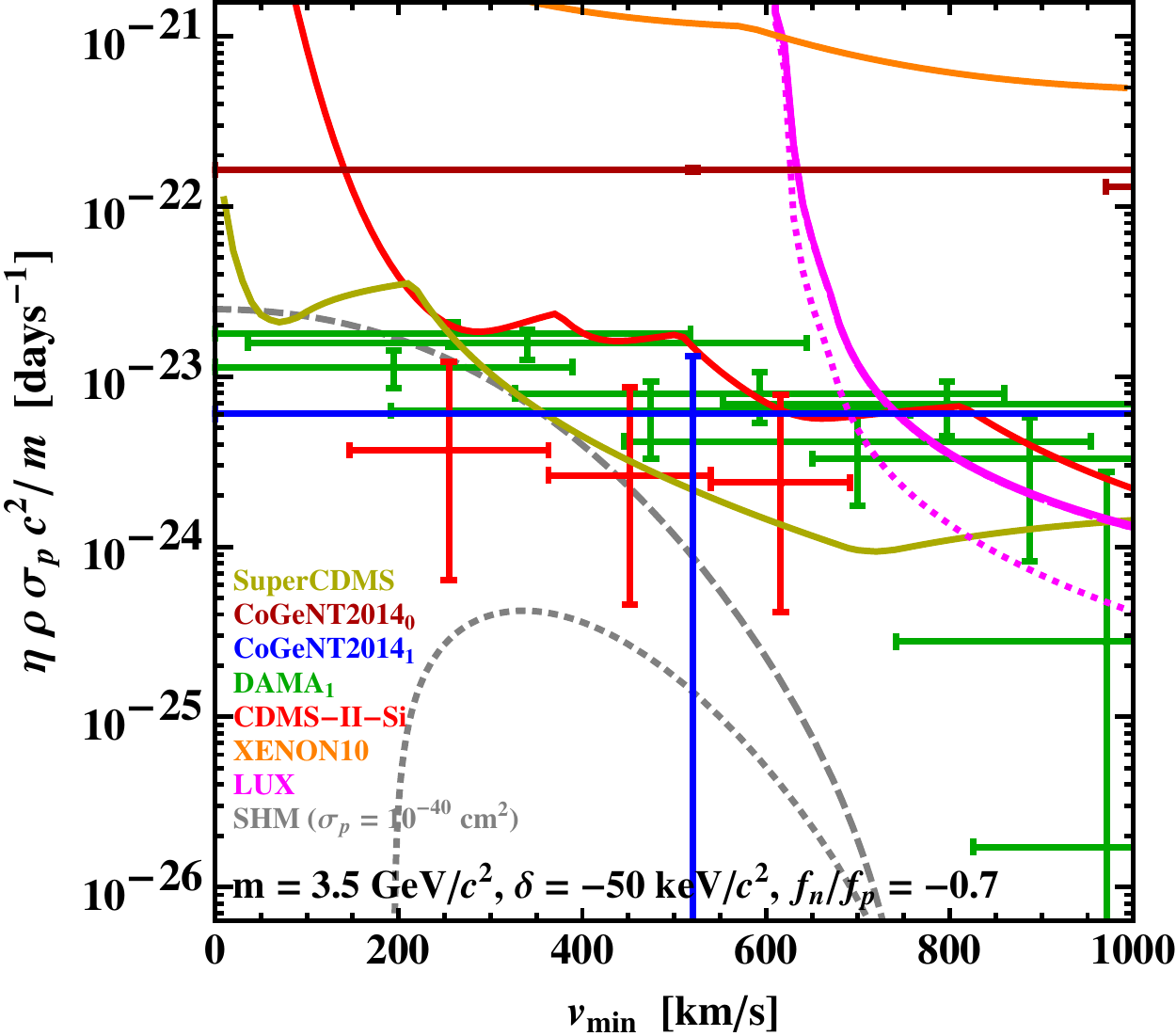}}
\caption{Halo-Dependent direct detection data comparison in the $(m, \sigma_p)$ plane, in the SHM (left)  and Halo-Independent  data comparison in the $(v_{\rm min}$, $\rho \sigma_{p} \eta({ v}_{min})/m)$ plane with $m=$ 3.5 GeV (right)  for an exothermic WIMP ($\delta=-50$ keV) with Isospin-Violating ($f_n/f_p = - 0.7$) SI coupling, for which the CDMS-II-Si signal (in pink-red) is compatible with all limits. 
 The dashed gray lines in the right panel show the shape of  the average  (upper line) and  the annual modulation amplitude
  of  $\sigma_{p}\rho\eta/m$ (lower line) in the SHM (with $\sigma_{p} = 1 \times 10^{- 40}$ cm$^2$). Fig. from~\cite{Gelmini:2014psa} (see this reference for details).
}
\label{Halo-Dep-Indep}
\end{figure}

 The   ``Halo-Dependent" data comparison  method, used since the inception of direct detection~\cite{Ahlen:1987mn, Smith:1988kw}, fixes the three aforementioned  elements of the rate, usually assuming the SHM for the galactic halo, except for the WIMP mass $m$ and a reference  cross section parameter $\sigma_{\rm ref}$ extracted from the cross section ($\sigma_{\rm ref}=\sigma_p$ for SI interactions). Data are plotted in the $(m, \sigma_{\rm ref})$ plane (as in Figs.~\ref{SI-overview},  \ref{SD-proton} and the left panel of Fig.~\ref{Halo-Dep-Indep}). All experimental collaborations use this method and assume WIMPs with SI with $f_n=f_p$ to present their results. Fig.~\ref{SI-overview} shows that with these assumptions, the DAMA, CoGeNT and CDMS-II-Si regions almost overlap for ``Light WIMPs" with $m \simeq10$ GeV,  but they are all rejected by upper limits.  One caveat is that in the ``Halo-Dependent" plots it is usually assumed that the WIMP constitutes the whole of the DM. If it constitutes a fraction ${\it R}$ of the DM, the vertical axis is ${\it R} \sigma_{\rm ref}$ and not just $\sigma_{\rm ref}$.

 In the  ``Halo-Independent" data comparison method one fixes the elements 1) and 2) of the rate, except again for a reference cross section parameter $\sigma_{\rm ref}$, but does not make any assumption about the element 3), circumventing in this  manner the uncertainties in our knowledge of the local characteristics of the dark halo of our galaxy~\cite{Fox:2010bz,
 Frandsen:2011gi, Gondolo:2012rs}  (see e.g.~\cite{Gelmini:2014boa} and references therein). The main idea  is that for a particular DM candidate the interaction rate at one particular recoil energy depends for any experiment on one and the same  function $\sigma_{\rm ref}\rho\eta( {\rm v}_{min}, t)/m \equiv \tilde{\eta}({\rm v}_{min}, t)$.  Thus, all rate measurements  and bounds can be mapped onto measurements of and bounds on the unique function $\tilde{\eta}({\rm v}_{min}, t)$  for a fixed WIMP mass $m$ and plotted in the $({\rm v}_{min}, \tilde{\eta})$ plane.  To be compatible, experiments must measure the same  $\tilde{\eta}$ function. This method was initially developed for a SI  WIMP-nucleus interaction~\cite{Fox:2010bz, Frandsen:2011gi, Gondolo:2012rs} and only in~\cite{DelNobile:2013cva} extended to any other type of WIMP-nucleus interaction. In this case the rate in an observed energy interval $[E'_1,E'_2]$ is written as $R_{[E'_1,E'_2]}= \int_{E'_1}^{E'_2} dE'~ {dR}/{dE'} =\int_0^\infty dv_{\rm min}  \mathcal{R}_{[E'_1,E'_2]}(v_{\rm min})\tilde\eta(v_{\rm min}, t)$ with a DM candidate and detector dependent response function $\mathcal{R}$~\cite{Gondolo:2012rs, DelNobile:2013cva}. 

 The right panel of Fig.~\ref{Halo-Dep-Indep} shows the Halo-Independent comparison for the same candidate whose Halo-Dependent analysis assuming  the SHM is shown in the left panel of the same figure: a WIMP with Isospin-Violating ($f_n/f_p = - 0.7$) SI interactions and exothermic scattering ($\delta=-50$ keV)  (see~\cite{Gelmini:2014psa}  for details). The crosses in this panel represent potential rate and modulation amplitude measurements and we see that part of the three red crosses corresponding to the three events observed by CDMS-II-Si escape all bounds. This is consistent with the analysis in the left panel, where part of the CDMS-II-Si region is also seen to escape all bounds.

So far the results have indicated that one of the potential direct DM detection  signals at a time could be compatible with all negative results for particular DM candidates~\cite{Barello:2014uda, Arina:2014yna, Gelmini:2014psa}, but not all of them. For example, a signal in DAMA/LIBRA is favored by a magnetic dipole-moment coupling~\cite{Barello:2014uda} or a spin-dependent coupling to protons, such as $(\vec{S}_{DM} \cdot \vec{q})(\vec{S}_p \cdot \vec{q})$~\cite{Arina:2014yna}. Both favor couplings to Na and I and disfavor couplings with Xe and Ge.  ``Magnetic Inelastic DM"~\cite{Barello:2014uda} may work as a candidate for DAMA/LIBRA because inelasticity further disfavors lighter targets (better I than Ge). 

The situation will be clarified with more data and very fast advances are expected in direct detection in the next decade. 
Fig.~\ref{SI-overview} presents an overview of existing direct detection limits (solid lines) and future sensitivity regions. The next generation of multi-ton experiments will reach the rate level at which neutrinos of different sources (atmospheric, the Sun, the diffuse background from supernovas or DSNB)  will constitute a background to a DM signal, called the ``neutrino floor" 
(see~\cite{Cushman:2013zza}, for details and references).

\subsection{Indirect Detection of WIMPs}

The DM annihilation or decay products searched for  are high energy neutrinos produced by WIMPs captured inside the Sun or the Earth, photons  and anomalous cosmic rays, such as positrons and antiprotons, which do not come from astrophysical sources. Indirect searches assume that DM particles annihilate or decay. Here we assume that they annihilate (e.g. because  they are self-conjugated Majorana fermions or  bosons). 

\subsubsection{Indirect detection through neutrinos from the Sun}

As the Sun and Earth move through the dark halo of our galaxy,  WIMPs can occasionally scatter with the baryonic matter in one of these bodies and loose enough energy to become gravitationally bound to it. The capture rate $\Gamma_C$ is proportional to the scattering cross section and the local DM number density, $\Gamma_C \sim \sigma_S n$. As the number of  captured WIMPs $N= (\Gamma_C ~t)$  increases with time,  the annihilation rate increases too as $\Gamma_A \sim \sigma_A N^2$, and the number of WIMPs inside the Sun or Earth changes as $dN/dt=\Gamma_C- 2 \Gamma_A$. The capture rate is constant, since the captured number of DM particles  is too small to affect the DM density and distribution in the halo. Given enough time, the annihilation rate  grows enough to compensate the capture, $\Gamma_A = \Gamma_C/2$, so  $dN/dt= 0$ and a constant equilibrium number is reached. For typical WIMPs, equilibrium is reached in the Sun within the lifetime of the Sun ($4 \times 10^9$ y) but not in the Earth~\cite{Gould:1987ir, Jungman:1995df}.  

From the annihilation products, only neutrinos can escape from the Sun or Earth and travel to large-volume underwater or ice ``neutrino telescopes", where they could be detected. These are neutrinos with energies much larger than those otherwise produced in these bodies. Besides, a signal from the Sun would follow its position, thus eliminating all possible backgrounds.  

The best limits on neutrinos from the Sun come from the IceCube telescope~\cite{Aartsen:2012kia}, an instrumented kilometer cube of ice in Antarctica (with a more densely instrumented core, called DeepCore). A smaller underwater neutrino telescope in the Mediterranean, Antares~\cite{Zornoza:2014dma}, is a prototype for a possible ``KiloMeter-cube Neutrino Telescope", KM3NeT,  to be built somewhere in the Mediterranean.  In the limit of equilibrated capture and annihilation, the flux of neutrinos only depends on the rate at which WIMPs are captured. This  rate depends on the scattering cross section.  Since direct detection depends on the same quantity both types of bounds can be compared, as in Fig.~\ref{SD-proton}.   
Because the Sun contains primarily hydrogen, the limits on the spin-independent (SI) scattering cross section, Eq.~\eqref{sigma-SI}, obtained by IceCube are not competitive with those obtained in direct detection experiments. Spin-dependent  (SD) WIMP-nucleus cross sections, Eq.~\eqref{sigma-SD}, are of the same order of magnitude for hydrogen and heavy nuclei, and therefore the Ice Cube limits on these cross sections are very important, as shown in Fig.~\ref{SD-proton}.

 The indirect limits due to neutrinos in the Sun are particularly important for Self-Interacting DM (defined in section 3) because the capture rate is enhanced by the scattering of halo DM with captured DM, once the number of captured particles becomes large enough~\cite{Zentner:2009is}.  

\subsubsection{Indirect detection through photons}

Gamma-ray astronomy is done with ground and space instruments. The Fermi Space Telescope was launched in 2008. Its main instrument is the Large Area Telescope (LAT) which detects photons between 20 MeV and 300 GeV. Photons with energy above 20 GeV and up to several TeV are detected by  ground-based Air Cherenkov Telescopes (ACT): HESS in Namibia, MAGIC in Las Palmas and Veritas in the US.  On the planning stage is a large array of ACTs, the CTA (Large Telescope Array), which could detect photons from 10's of GeV to above 100 TeV.

Photons reveal the spatial distribution of their sources. The Universe is totally transparent to photons below 100 GeV. At higher energies photons interact with infrared and optical backgrounds, but below 10's of TeV still arrive to us from cosmological distances (at higher energies they are absorbed by the CMB; only above $10^{10}$ GeV, the range of Ultra-High Energy Cosmic Rays, can they again reach us from cosmological distances). 

Monochromatic photons  can be  produced through $\chi \chi \to\gamma \gamma$ (or $\gamma Z$) with energy equal (or close) to the annihilating WIMP $\chi$ mass $m$. Detection of this monoenergetic  spectral line would be a ``smoking gun" signature. However, usually  (but not always) these processes happen only at the one-loop order and  they are suppressed with branching ratios 10$^{-3}-10^{-4}$. Secondary photons, in particular from  pion $\pi^0 \to \gamma \gamma$ decay,  would be produced with a spectrum whose cutoff at high energies is $m$. So a cutoff in the observed spectrum would be a signature of DM annihilation, but not exclusively. The spectrum of astrophysical sources also has a cutoff, because  there is a limit to the energy with which emitted particles can be produced. 

The $\gamma$-ray flux (number per unit area, time and energy) expected from the annihilation of DM particles coming from  a particular direction is
\begin{equation}
\Phi_{\gamma}(E_{\gamma}) \simeq 
 \left<\sigma_{A} v\right>  \frac{dN_{\gamma}}{dE_{\gamma}} \int_{\rm {line ~of ~sight}} 
 \frac{\rho^2(r)}{m^2_\chi} dl(\theta) d\theta ,
 \label{photon-flux}
\end{equation}
where $\left<\sigma_{A} v\right>$ is the annihilation cross section times relative speed  at the source, ${dN_{\gamma}}/{dE_{\gamma}}$ is the $\gamma$-ray spectrum per  annihilation (for example, it would be a delta function for $\chi \chi \to \gamma\gamma$).  The integration of the DM number density squared  as function of distance,  $(\rho(r)/m)^2$,  is along the line of sight and over the angular aperture or resolution of the detector.

Since the annihilation rate depends on the square of the  DM density, high density regions such as the galactic center (GC), DM clumps, dwarf galaxies and other galaxies and galaxy clusters would boost the rate.  Thus a signal is expected from them. 

\begin{figure}[t]
\centerline{\includegraphics[width=0.80\textwidth]{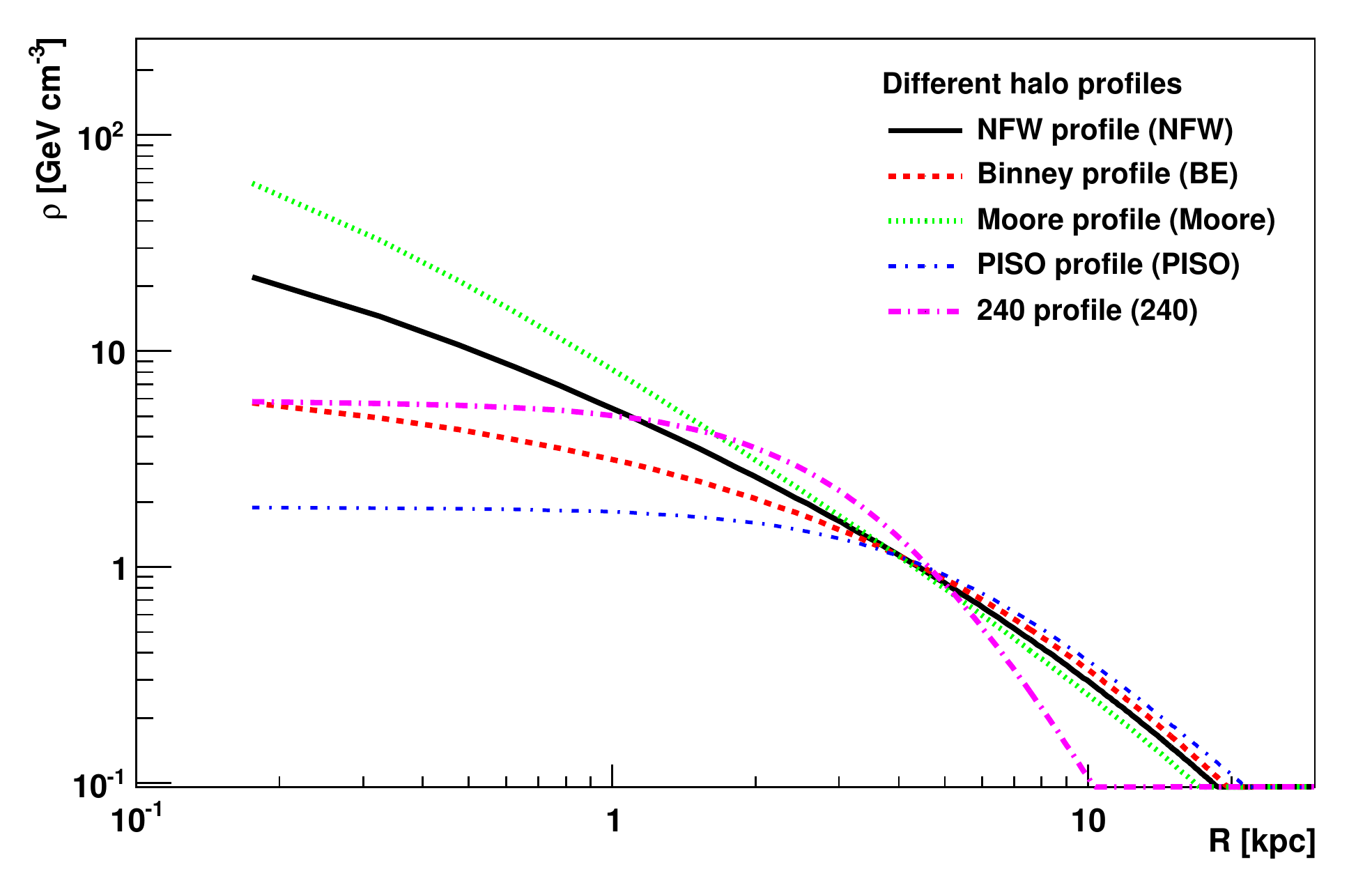}}
\caption{Radial DM density distribution for different galactic halo models. The  central density value is fixed by the requirement of the rotation speed of the Sun~\cite{Weber:2009pt}
}
\label{profiles}
\end{figure}

The amount of DM in the GC is uncertain (see Fig.~\ref{profiles}). In fits  to numerical simulations of structure formation containing CDM only, Navarro, Frenk and White (NWF) found a universal density profile~\cite{Navarro:1996gj}, at all scales  (large clusters, galaxies and dwarf galaxies) which goes as $1/r^3$ at large distances from the GC and as $1/r$ close it. This is called a ``cuspy profile."  It is expected to be modified close to the center by the effect of baryons, which are dissipative, and thus fall to the center and dominate over the DM.  Fits to observed rotation curves of galaxies find instead ``cored profiles", in which the DM density becomes constant close to the GC.  A commonly used halo profile is the ``generalized NFR", 
\begin{equation}
\rho(r)= \rho_0 \left({r}/{r_s}\right)^{-\gamma} \left[ 1+ \left({r}/{r_s}\right) \right]^{\gamma - 3},
\label{profile}
\end{equation}
where $r_s \simeq 20$ kpc is  called the ``scale radius" and the central density value $\rho_0$ is fixed by imposing that the DM density at the distance of the Sun from the GC is equal to the value inferred by observations ($\simeq$ 0.3 GeV/cm$^3$). For $r \ll r_s$, $\rho \sim 1/r^\gamma$, while for $r \gg r_s$, $\rho \sim 1/r^3$. The NFW profile has $\gamma = 1$ and a cored profile has $\gamma \simeq 0$. When considering the region of the GC, the most important feature of the halo profile is the inner slope, $\gamma$. As shown in Fig.~\ref{profiles}~\cite{Weber:2009pt},  in some halo profiles, this slope can be considerably steeper than in the NFW profile. E.g. the Moore profile has $\gamma=1.5$~\cite{moore}.

Several hints for DM have been found at the GC.  We review these potential signals below.

\subsubsection{The INTEGRAL 511 keV signal}

The satellite INTEGRAL, launched in 2002,  confirmed the emission of 511 keV photons from the GC, a 40 year old signal first observed by balloon-born $\gamma$-ray detectors. It is clearly due to a non-relativistic $e^+$ annihilating with an $e^-$ almost at rest. Initially the emitting region seemed spherically symmetric, as expected for a DM signal, but in 2008 INTEGRAL~\cite{Weidenspointner:2008zz} revealed that is it not,  and  found evidence of a population of binary stars consistent with being the  $e^+$ source. These observations decreased the motivation to consider DM as the origin of the signal. Special DM candidates were proposed to explain  this signal:  ``Light DM", LDM, with MeV mass, annihilating as $\chi \chi \to e^+e^-$ into  $e^+e^-$ almost at rest ~\cite{Boehm:2003bt}; ``eXciting DM", XDM, in which a DM particle  $\chi'$ decays into a lighter one $ \chi$ with mass difference  $m' -m = \delta \simeq$ MeV and an $e^+e^-$ almost at rest, $\chi' \to  \chi e^+e^-$~\cite{Finkbeiner:2007kk}.

\subsubsection{The ``WMAP Haze"}

 This is an excess of microwave emission in the inner 20 degrees, about 1 kpc,  around the GC.
  It was discovered by D. Finkbeiner in 2003~\cite{Finkbeiner:2003im}. It can be explained as synchrotron radiation from $e^-$ and $e^+$ (accelerated in magnetic fields) produced in astrophysical sources or by the annihilation of DM particles. Almost any WIMP annihilating into $e^-$ or $e^+$  could produce it. However, it is now considered part of the ``Fermi Bubbles" (discovered in 2009 also by Finkbeiner and collaborators)~\cite{Dobler:2009xz}. These are two  large structures of $\gamma$-ray emission of  8 kpc in diameter each, extending to both sides of the galactic plane. They could be due to an early  period  of strong jet-like activity of the black hole at the GC, which is now dormant. 

\begin{figure}[t]
\centerline{\includegraphics[width=0.75\textwidth]{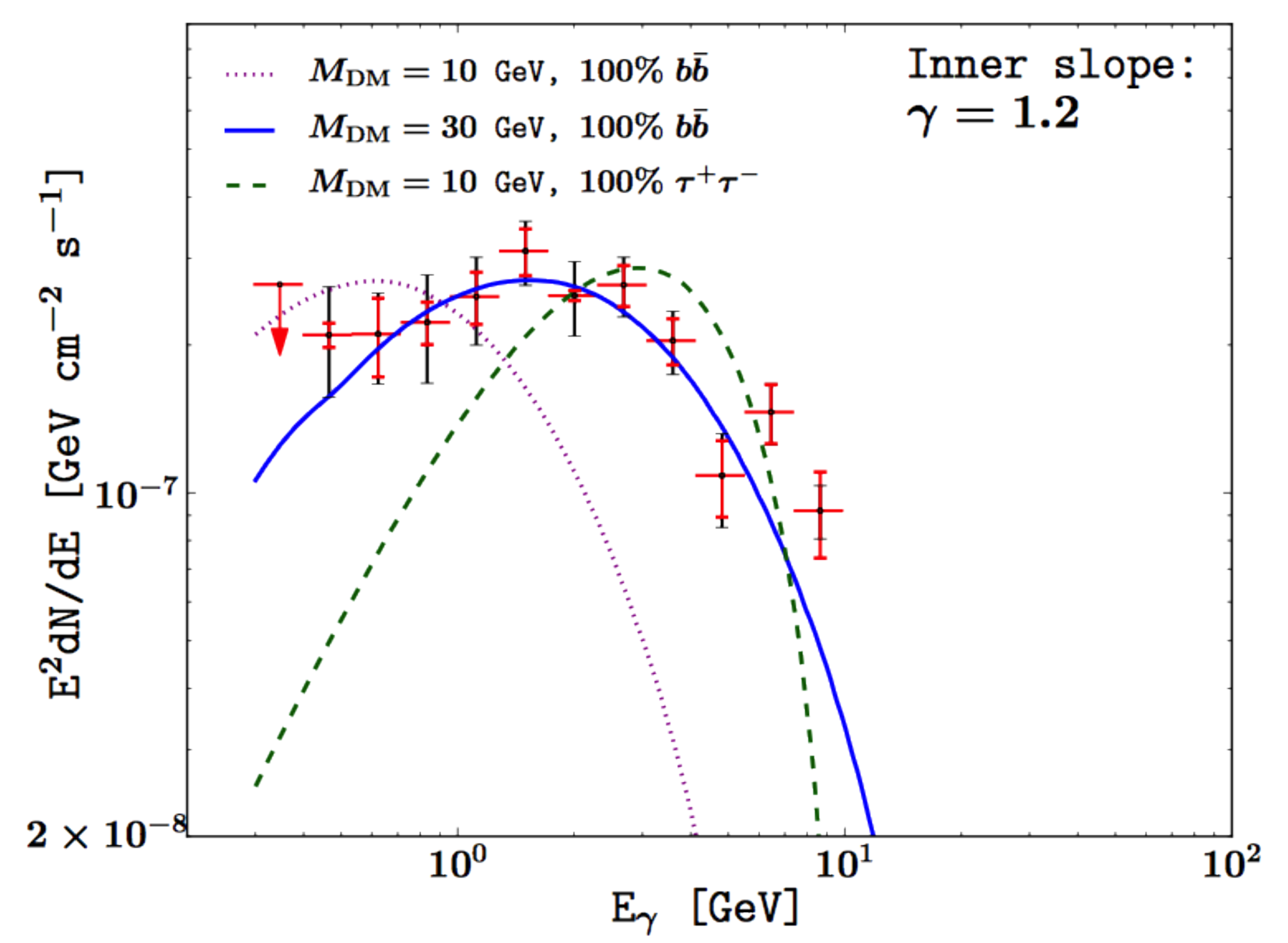}}
\caption{Spectrum of the extended GeV excess from the GC (red crosses) and several fits with DM annihilating into $b \bar{b}$ and  $\tau^+ \tau^-$. Fig. from~\cite{Gordon:2013vta}. Notice the required inner slope of the galactic density profile $\gamma =1.2$ (a cuspy profile) necessary to fit the data.}
\label{GeV-spectrum}
\end{figure}

\subsubsection{The extended GeV excess} 

In 2011, Hooper and Linden~\cite{Hooper:2011ti} found by  subtracting from the Fermi data all known contributions, an unexplained extended excess of GeV photons, peaking at 2 -3 GeV, coming from the GC (see Fig.~\ref{GeV-spectrum}).   The  existence of  the excess was confirmed by several other 
groups~\cite{Abazajian:2012pn, Gordon:2013vta}  and later also found in the inner galaxy (up to 10 degrees from the GC)~\cite{Daylan:2014rsa}.  

\begin{figure}[t]
\centerline{\includegraphics[width=0.68\textwidth]{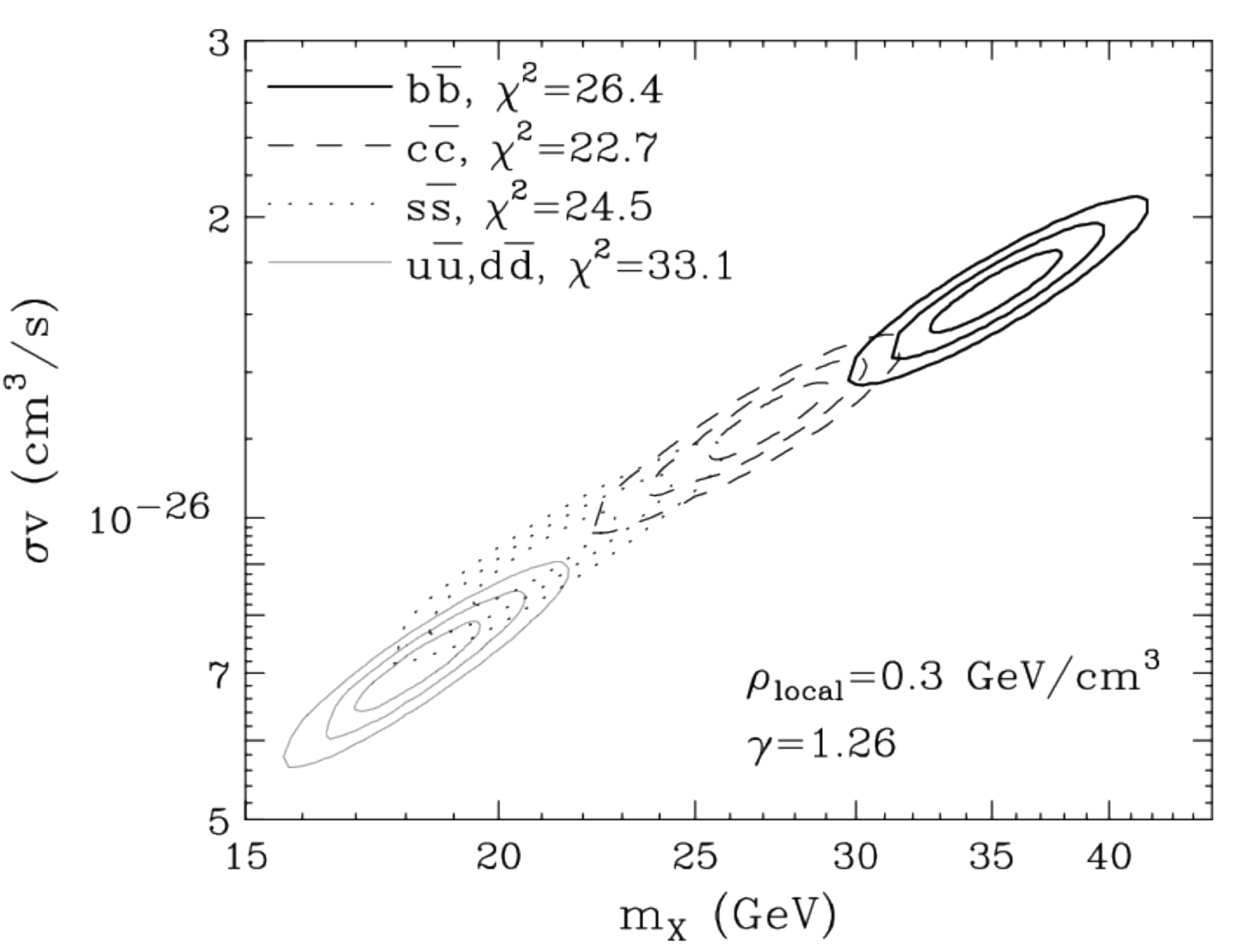}}
\caption{DM mass and cross section required to explain the extended GeV emission from the GC if the DM annihilates into the  shown quark-antiquark modes. Fig. from~\cite{Daylan:2014rsa}. Notice the required inner slope of the galactic density profile, $\gamma =1.26$.}
\label{Hooperon}
\end{figure}

This signal, shown in Fig.~\ref{GeV-spectrum}, can be interpreted as possible evidence of DM particles  with mass of 7-12 GeV annihilating to $\tau^+ \tau^-$ (among other leptons) or with mass of  22-45 GeV annihilating to quarks, with an annihilation cross section  close to  the value $\simeq10^{-26}$ cm$^3$/s required by thermal WIMPs at decoupling 
 (see Fig.~\ref{Hooperon}).  In both cases to fit the observed signal the galactic profile must be ``cuspy" with an inner slope $\gamma \simeq 1.2 $ to 1.4.  A cuspy DM halo profile enhances the rate from the GC considerably with respect to a core profile. Fig.~\ref{profiles} shows that the ratio of the central densities of the ``cuspy" Moore ($\gamma=1.5$) to the ``cored" $\gamma=0$ profiles is about 30, which translates into a factor of 900 in the rate predicted from the GC. 

The GeV excess could also be explained by unresolved millisecond 
pulsars~\cite{Abazajian:2010zy, Abazajian:2012pn, Gordon:2013vta, Abazajian:2014fta}. These are rotating neutron stars which have been spun-up, through accretion from a companion star with which it forms a close binary system, to a period of 1-10 milliseconds.  They emit X-rays and $\gamma$-rays, possibly from the matter being accreted. Their distribution in or near the GC is not known, as they cannot be observed.  There are also uncertainties in the photon spectrum they emit. Several assumptions are needed to fit the observed GeV excess.

\subsubsection{The 130 GeV line} 

A line at this energy with 3.5$\sigma$ significance was found in 2012  in the FermiLAT data by C. Weniger and collaborators~\cite{Bringmann:2012vr} coming from the GC. This monochromatic line could have been produced by DM annihilation into $\gamma \gamma$.  A hint of a second line was found at 111 GeV~\cite{Su:2012ft} which could have corresponded to  annihilation of the same particles into $Z^0 \gamma$. However, no evidence of the line was found elsewhere where it was expected if due to DM annihilation, e.g. in dwarf galaxies~\cite{GeringerSameth:2012sr}.  The line was also found where it could not be if due to DM annihilation: in Earth's Limb~\cite{Finkbeiner:2012ez} and in the vicinity of the Sun~\cite{Whiteson:2013cs} (Earth's Limb is the edge of the disk of the Earth, where $\gamma$-rays can be produced by cosmic rays). The last two are clear indications of a detector effect. The FermiLAT paper~\cite{Ackermann:2013uma} about the line finds a hint of it at the 1.6$\sigma$ level from the GC, and also  at 2$\sigma$ in the Limb. Another troubling indication for a DM interpretation is that the accumulated significance of a real annihilation line should increase with time but it is decreasing for this signal~\cite{Weniger:2013dya}. Thus, the line seems spurious, and possibly due to an experimental effect, although the issue is not entirely resolved yet~\cite{Weniger:2013tza}.

\begin{figure}[t]
\centerline{\includegraphics[width=0.83\textwidth]{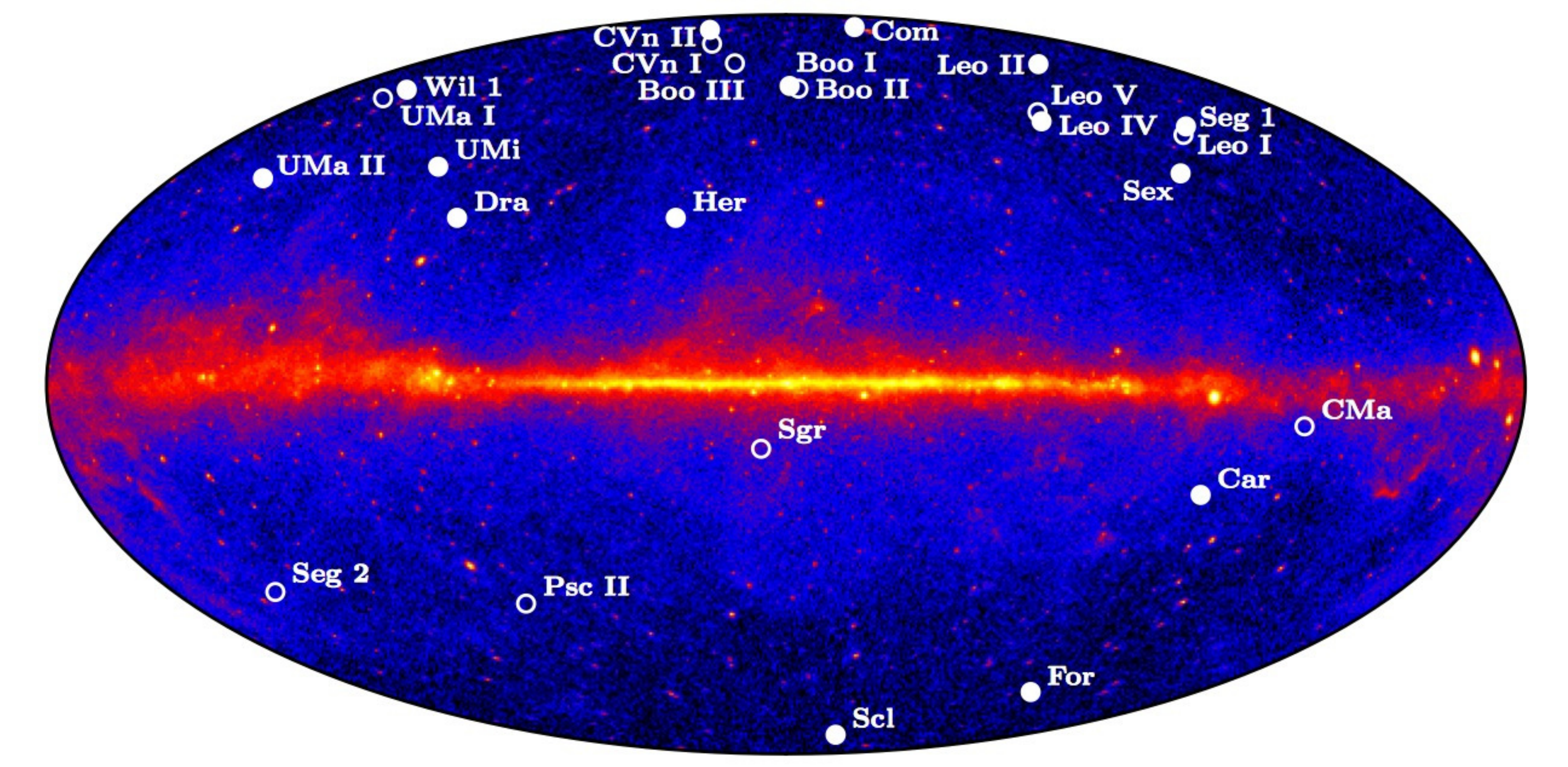}}
\caption{Map of known dwarf galaxies overlaid on  Fermi data. Fig. from~\cite{Ackermann:2013yva}.
}
\label{dwarfs-map}
\end{figure}

\subsubsection{FermiLAT limit on thermal WIMPs from dwarf galaxies} 

The GC is a complicated place, with large uncertainties in the DM profile, and other powerful photon sources.  Other
 overdense regions might provide a cleaner signal.
A large amount of DM clumps are expected to remain within the dark halo of our galaxy. 
 CDM simulations of structure formation (the Via Lactea II~\cite{Kuhlen:2008aw} and the Aquarius Project~\cite{Vogelsberger}) suggest that $O(10)$ clumps could be discovered by Fermi, although none has been observed so far. Dwarf galaxies are simpler sources because they are the most DM-dominated structures observed so far. A map of the known dwarf galaxies overlaid with the Fermi gamma-ray map is shown in Fig.~\ref{dwarfs-map}.
 
\begin{figure}[t]
\centerline{\includegraphics[width=0.85\textwidth]{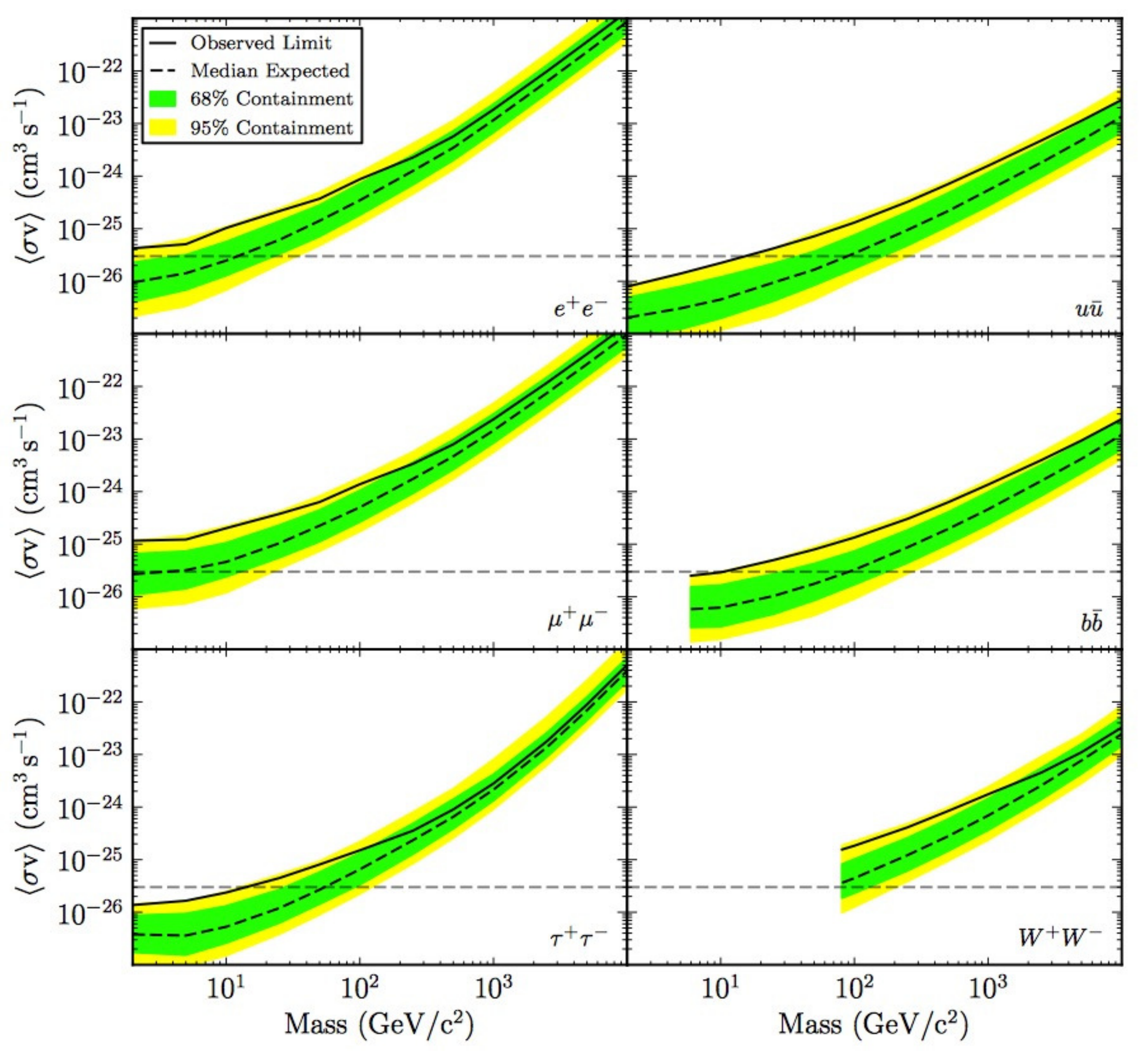}}
\caption{Fermi upper limits from 15 stacked dwarf galaxy images on the DM on $\left<\sigma_A v \right>$, assuming exclusive annihilation in each of the modes shown~\cite{Ackermann:2013yva}. The lower limit on $\left<\sigma_A v \right>$ at freeze-out  for thermal WIMPs is also shown (horizontal  dashed line).
}
\label{stacked-dwarfs}
\end{figure}

The best  limit on thermal WIMPs from the Fermi-LAT collaboration comes from 15 stacked dwarf galaxy
 images~\cite{Ackermann:2013yva}.   When observations on each of several very similar objects are not good enough, stacking their images increases the statistics and allows one to obtain better limits. Fig.~\ref{stacked-dwarfs} shows the upper limits on $\left<\sigma_A v \right>$  assuming exclusive annihilation into the pairs of leptons or quarks indicated in each panel. To have an acceptable density $\Omega \leq \Omega_{DM}$, thermal WIMPs must have 
  $\left<\sigma_A v \right> \geq 3 \times 10^{-26}$ cm$^3$/s at decoupling.  This requires the limit to be above the horizontal dashed line in Fig.~\ref{stacked-dwarfs}. Both limits together impose $m > 20$ GeV for thermal WIMPs (from the $\tau^+ \tau^-$ and $u \bar{u}$ modes) and a limit $m >$ O(100) GeV is expected in 4 y~\cite{Ackermann:2013yva}.   
 
 The caveat to this limit is that $\left<\sigma_A v \right>$ at decoupling and now may not be the same,  because the characteristic WIMP speed at decoupling is $v \simeq c/3$ and in the galaxy is $v \simeq 10^{-3} c$. For  s-wave annihilation $\left<\sigma v\right>$ is independent of  $v$ and the limit applies, but for p-wave annihilation $\left<\sigma v\right> \sim v^2$, and the limit does not apply.  Also, there could be a ``Sommerfeld enhancement" of the present annihilation which would invalidate the limit. This enhancement is due to  a modification of the wave function of annihilating  low velocity particles due to attractive long-range forces. The particles form an almost bound state which enhances the annihilation cross section.

 \subsubsection{Anomalous cosmic rays}
 
 Positrons and antiprotons would be produced in DM annihilations in equal numbers as electrons and protons. They are an interesting potential signal of WIMP annihilation because there is not much antimatter in the Universe. Unlike photons, which travel in straight lines and are not absorbed  for energies $<100$ TeV,
 $e^+$ and $e^-$ interact with the magnetic fields of the galaxy.  They rapidly (within a few kpc) loose energy through the emission of synchrotron radiation and  inverse Compton scattering interactions with photons (in Compton scattering a high energy 
 $\gamma$ interacts with an $e$ almost at rest; in inverse Compton  a high energy $e$ interacts with a $\gamma$ with much smaller energy and produces a lower energy $e$ and a higher energy $\gamma$). $p$ and  $\bar{p}$ suffer convective mixing and spallation.  They propagate further than electrons, but still only from a fraction of the size of the galaxy.

 Balloon-born experiments detecting positrons have found since the 1980's a possible excess over secondary cosmic ray fluxes. It was called the ``HEAT excess" and
  is now called the  ``PAMELA excess".  PAMELA (Payload for Antimatter Matter Exploration and Light-nuclei Astrophysics), a satellite carrying a magnetic spectrometer that was in operation from 2006 to 2011,
reported  in 2008 an excess in the positron fraction $e^+/(e^+ + e^-)$  in the 10 to 100 GeV energy range~\cite{Adriani:2008zr} compatible with the HEAT excess. The result was confirmed by FermiLAT~\cite{FermiLAT:2011ab} and more recently by  AMS-02~\cite{Aguilar:2013qda} (Alpha Magnetic Spectrometer), a cosmic ray research module mounted to the exterior of the International Space Station, in operation since 2011. Fermi does not have a magnet to distinguish positively  and negatively charged particles, but they cleverly used the magnetic field of the Earth~\cite{FermiLAT:2011ab}.

\begin{figure}[t]
\centerline{\includegraphics[width=0.74\textwidth]{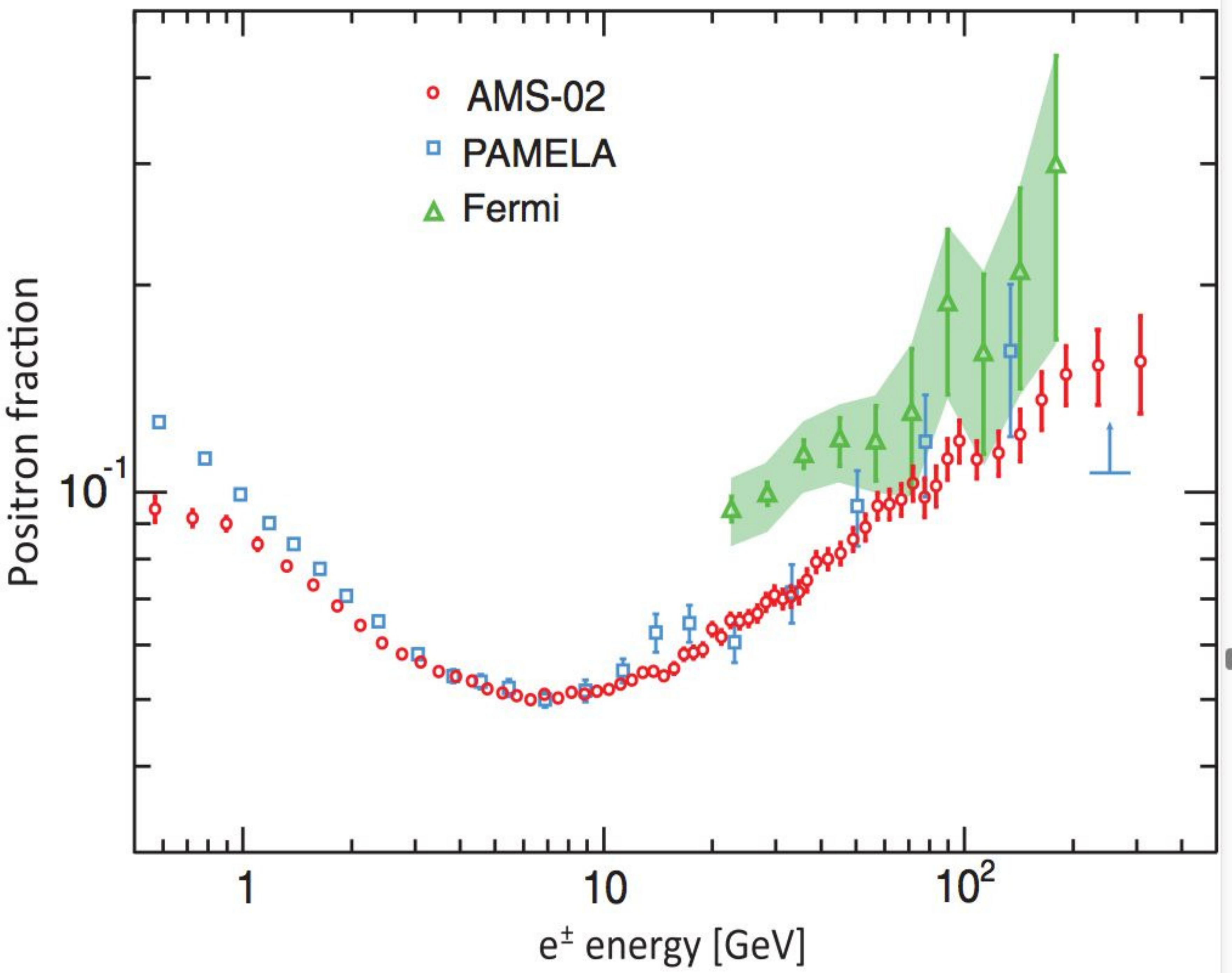}}
\caption{The positron fraction $e^+/(e^+ + e^-)$ measured by AMS (red), PAMELA (blue)  and Fermi-LAT (green). Fig. from
\cite{Aguilar:2013qda}. }
\label{AMS}
\end{figure}

The rapid rise shown in Fig.~\ref{AMS} in the cosmic ray $e^+$ fraction above 10 GeV measured by PAMELA and AMS indicates the existence of nearby primary sources of high energy positrons, such as pulsars or annihilating/decaying DM. The spectrum of secondary positrons produced through the collisions of cosmic rays in the interstellar medium is predicted to fall rapidly with energy, and thus is unable to account for the observed rise. 

It had been proposed that secondary positrons could be produced and then accelerated in nearby supernova remnants, potentially explaining the observed rise without the need of primary sources. If so, other secondary cosmic ray species (such as boron nuclei and $\bar{p}$) should also be accelerated, leading to rises in the boron-to-carbon and $\bar{p}/p$ ratios. Recent measurements  show no sign of such a rise, what disproves this mechanism~\cite{Cholis:2013lwa}.

It is very important for the DM interpretation that no excess in $\bar{p}$ has been found~\cite{Adriani:2010rc}.  This means that the DM particles should be  ``leptophilic", i.e. only annihilate (or decay) into leptons and not quarks, either because the DM carries lepton number or because of kinematics.

 No endpoint of the $e^+$ excess was observed by PAMELA or AMS, which would be an indication of the mass $m$ of the DM particle.   HESS measured the electron plus positron  (which they call generically electrons) spectrum showing a soft decrease with energy from 1 to 10 TeV~\cite{Aharonian:2008aa}.  Together these observations require DM with $m \simeq$ few TeV annihilating not directly into $e^+ e^-$ pairs (which would generate a sharp spectrum) but into leptons of the 2nd or 3rd generation: $\tau^+\tau^-$, 4$\mu$, 4$\tau$ or into pions~\cite{Cholis:2013psa}. This could be explained if the DM annihilates into a pair of
 bosons  $\phi$ with $m_\phi <$ 1 GeV,  $\chi \chi  \to \phi \phi$, and then $\phi$ decays primarily into  $\mu^+ \mu^-$   or pions because of kinematic  reasons~\cite{ArkaniHamed:2008qn}. Due to the high multiplicity of such processes, the resulting $e^+$ energy distribution at injection is soft, as needed. 
 
  Since $\phi$ is so light, it must  be  very weakly coupled to ordinary matter to have escaped detection. The paper of Arkani-Hamed et al. titled ``A Theory of Dark Matter"~\cite{ArkaniHamed:2008qn} containing this idea 
  proposed the existence of a complicated dark sector in which the particles $\phi$ are the light gauge bosons  (``dark photons") of a hidden gauge symmetry. In this model WIMPs with mass $m \simeq $500 to 800 GeV have exited states  with small mass differences $\delta$ between 0.1 to 1 MeV, which can be ``inelastic DM" (see section 5.1 and Eq.~\eqref{vmin}) and  ``eXiting DM" (see the INTEGRAL signal  above).
   
 As if all these necessary properties the DM should have to account for the PAMELA excess are not complicated enough, the annihilation rate must be larger than expected for thermal WIMPs by a boost  factor $B\simeq$ 10 to 10$^3$.  The boost factor $B$ is simply a factor that must multiply the spectrum obtained from the annihilation of thermal WIMPs and that must somehow be explained. Astrophysical enhancements due to nearby overdense structures of DM cannot be larger than a few. Most DM clumps  are expected to be far from the Sun~\cite{Vogelsberger}. A possibility is that the annihilation cross section is larger in the dark halo at present, but that it had the value necessary for thermal WIMPs at freeze-out.  We already mentioned this idea of a Sommerfeld enhancement of the present annihilation. The exchange of a light boson $\phi$ may produce a Yukawa potential and as the annihilating particles approach at very low relative velocity they almost form a bound state, which enhances the cross section~\cite{ArkaniHamed:2008qn}.  No boost factor at all is needed if WIMPs have a large annihilation cross section in both the early Universe and  the dark galactic halo near Earth, which would produce a too-small relic abundance for thermal WIMPs but could be fine if the pre-BBN cosmology is non-standard~\cite{kane}.  
  
Besides all these requirements on the type  of DM, constraints imposed by its annihilation  in the GC are only compatible with  halo models that predict a relatively small amount of DM in the GC (cored  profiles)~\cite{Meade}.  Decaying DM has also been considered (see e.g.~\cite{Meade, Cholis:2013psa}). It  must decay mostly into leptons of the 2nd or 3rd generation, have multi-TeV mass and a very long lifetime, $\tau \simeq 10^{26}$s.

Upcoming AMS data may help to settle the origin of the positron excess not only by increasing statistics and extending studies to higher energies, but also by further  constraining any anisotropy in the positron and electron flux. If the origin of the positrons is one of the pulsars nearby, there should be an anisotropy at some level.

 \subsection{WIMP searches at the LHC}
 
\begin{figure}[t]
{\includegraphics[width=0.25\textwidth]{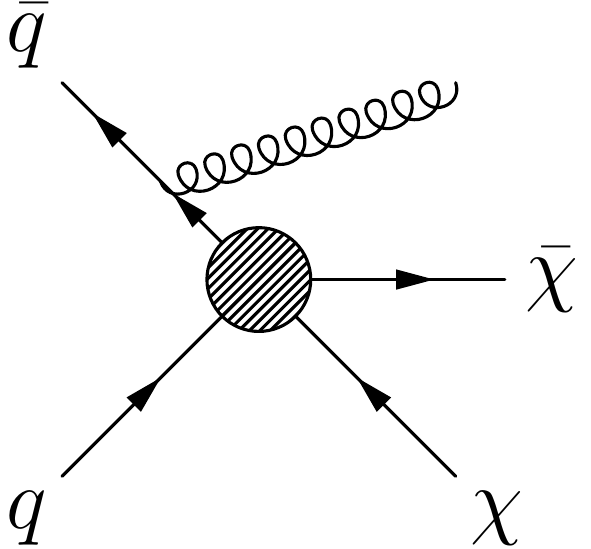}}
{\includegraphics[width=0.74\textwidth]{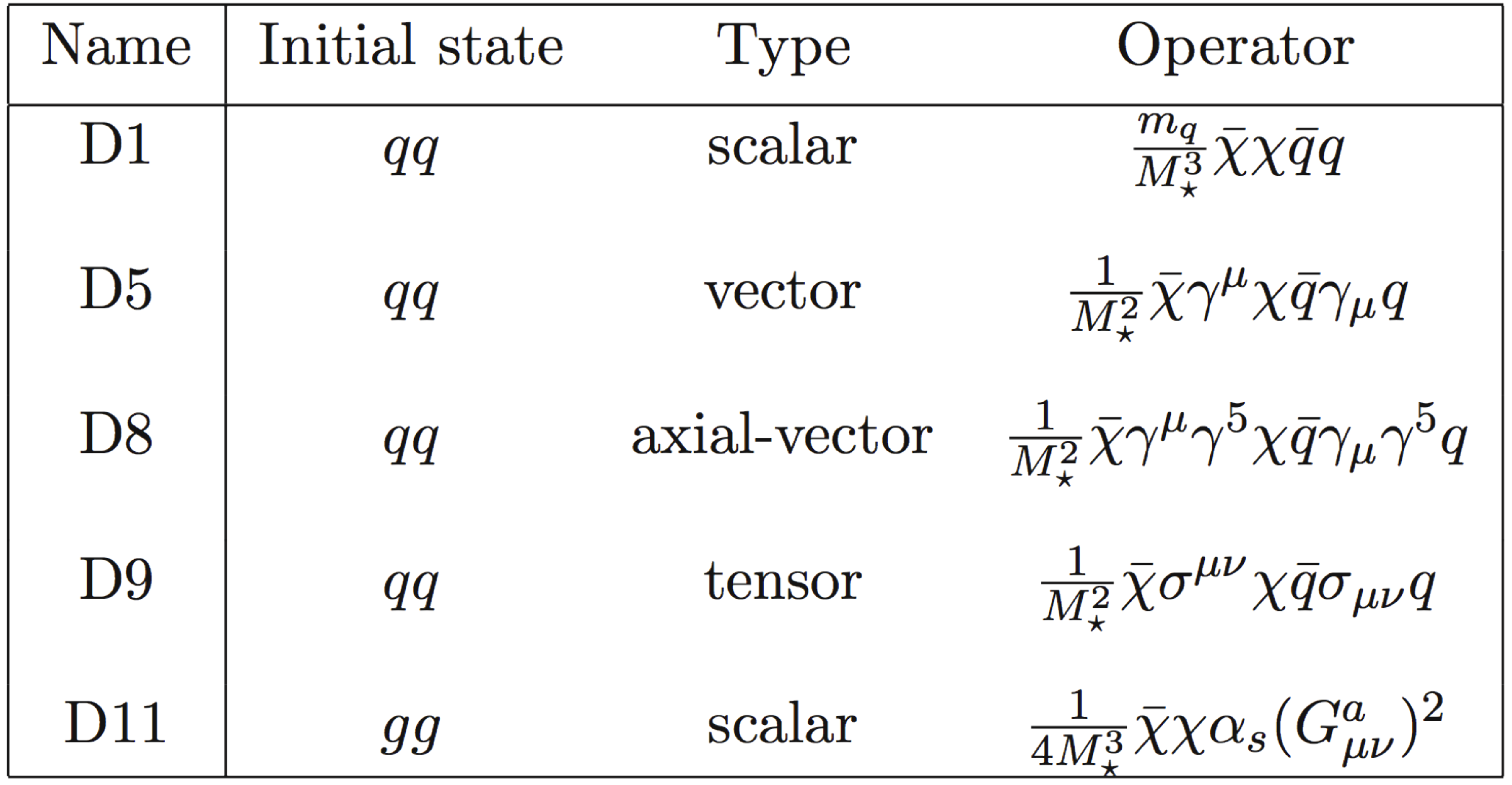}}
\caption{Monojet diagram (from~\cite{Kopp:2011eu}). Table 1. Some Effective Field Theory operators~\cite{ATLAS:2012ky} coupling a WIMP $\chi$ to quarks $q$ or gluons $g$, valid when the energy of the process is smaller than $M_*$. If the process is due to a mediator of mass $M$ and coupling $\lambda$, the operator represents a contact interaction with propagator  $\lambda^2/ M^2 = 1/ M_*^2$.}
\label{Mono}
\end{figure}

	DM particles escape  detection at colliders, thus they are characterized by missing transverse energy in collider events. One way to search for DM particles at colliders is to know the particle spectrum and how the DM  couples to other charged and/or strongly interacting particles that can be directly observed in the detectors. This is how complicated decay chains predicted in supersymmetric models are searched for.  Another way is to search for the production of a pair $\chi\bar{\chi}$ of WIMPs  and one visible particle emitted either by the initial or the intermediate SM particles to detect the event~\cite{Beltran:2010ww, Goodman:2010ku, Fox:2012ee}. If the one observable particle emitted is a photon, it is a ``monophoton" event; if it is a gluon, it is a ``monojet" event  (see the diagram in Fig.~\ref{Mono}). Other mono-particle events such as mono-W's (leptons), mono-Z's (dileptons), or even mono-Higgses, have  been  considered~\cite{Bai:2012xg, Askew:2014kqa}.

So far the limits obtained using monophoton, monojet, or other mono-particle events have been derived assuming effective couplings. Non-renormalizable operators from effective field-theory analysis such as those in the Table 1~\cite{ATLAS:2012ky} have been used to couple the DM particle $\chi$ to quarks or gluons. These effective couplings are valid when the energy of the process is small compared with the scale $M_*$. If a process is due to the exchange of a mediator of mass $M$ and coupling $\lambda$, the mediator is assumed to be heavy with respect to  the  momentum transfer  $q$ of the LHC partonic reaction, $M >>q$,  so that its propagator reduces to a constant $\lambda^2/ M^2 = 1/ M_*^2$. This approach is limited because it does not include possible interference between different operators, and it does not deal with lighter mediators.

\begin{figure}[t]
\centerline{\includegraphics[width=0.90\textwidth]{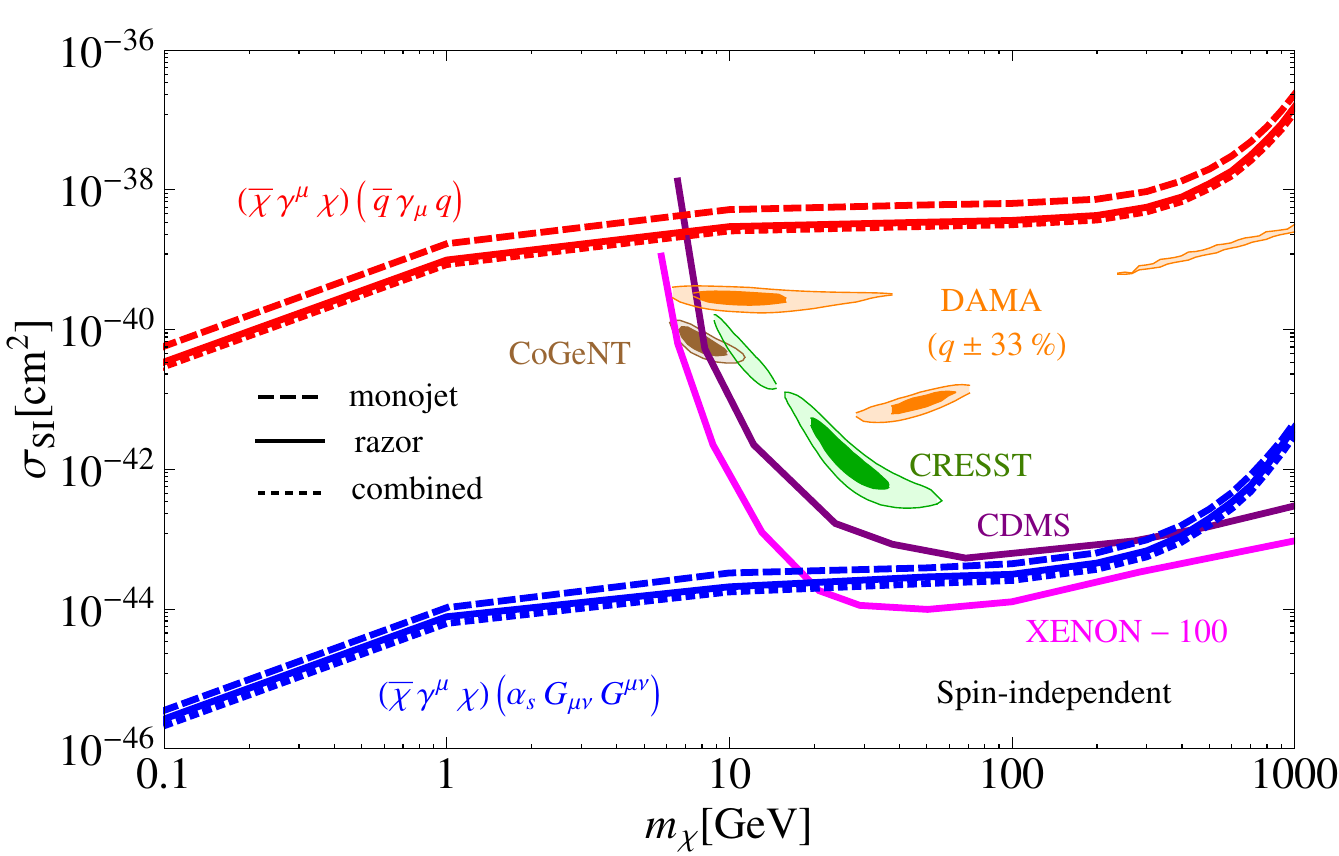}}
\caption{LHC bounds (monojet limits and others) on  SI DM-nucleon scattering compared to direct detection limits~\cite{Fox:2012ee}. For  DAMA and CoGeNT 90\% and $3\sigma$ contours  and for CRESST $1\sigma$ and $2\sigma$ contours are also shown.  See~\cite{Fox:2012ee} for details.}
\label{LHC-DD}
\end{figure}

 The same effective couplings in Table 1 can be used to compute the WIMP interaction with nuclei in direct searches, and the results can be compared in plots like the one shown in Fig.~\ref{LHC-DD} (from~\cite{Fox:2012ee}; see this reference for details). This type of plot must be understood with care. If the WIMP in question constitutes a fraction {\it R} of the local DM density, the direct detection limits must be multiplied by  $1/${\it R}, while the LHC limits do not change. More importantly,
 while it is valid to include the direct detection limits when presenting the LHC limits derived from contact interactions, the reverse is not correct. The reason is that mediators that are heavy with respect to typical LHC partonic energies, $M \gsim $100's GeV,  are also heavy in comparison with the typical MeV momentum transfer in direct detection experiments.   The opposite is not true: what is a contact interaction in a direct detection experiment may very well not be such at the LHC. 
If the mediator is light enough,  it could itself  be produced at the LHC or other colliders,  and the analysis of the collider data would be different. 

Once the mediator is light,  its couplings with different SM particles must be specified to study it. The problem is then that many theories will show a common low energy behavior when the mediating particles are heavy compared to the energies involved.  Each  effective contact interaction corresponds to many different possible particle models for the mediator. Without going to  a  complete theory, like a complete supersymmetric model, an intermediate step is to try to study simplified models for DM at the LHC. These incorporate all the known constraints on different interactions, and classify mediators according to whether they propagate in the s-channel or the t-channel, or by the way the DM relic density occurs~\cite{deSimone:2014pda, Abdallah:2014hon}.  Significant work remains to be done in this direction.

\subsection{Complementarity of WIMP Searches}

Direct, indirect and collider DM searches are independent and complementary. They differ in essential characteristics and  rely on different DM properties to see a signal. If a compelling DM signal is discovered, 
complementary experiments will be necessary to verify the initial discovery, and to determine the actual abundance in the Universe and properties of the particle in question.

\begin{figure}[t]
\centerline{\includegraphics[width=0.50\textwidth]{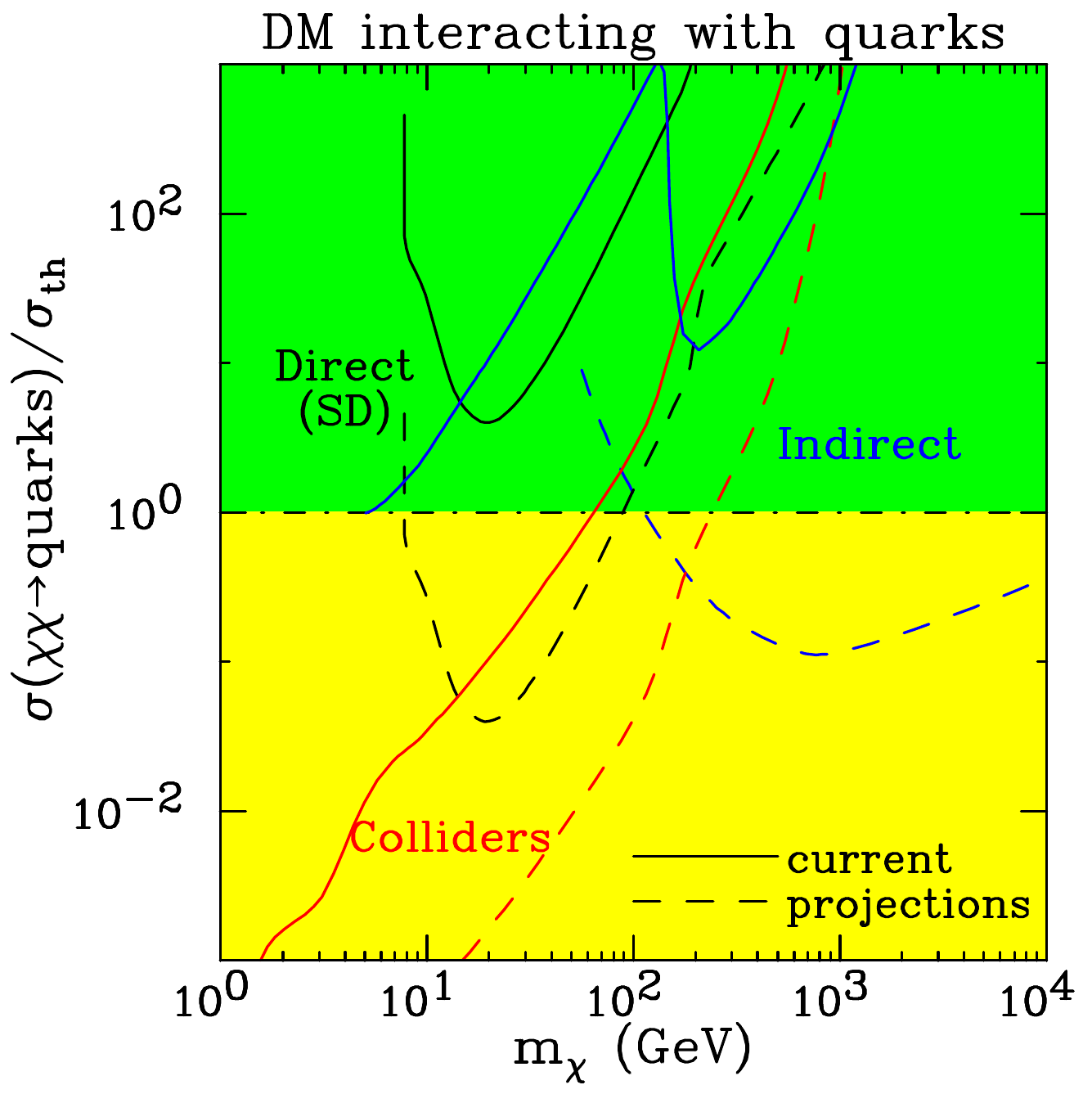}}
\caption{DM discovery prospects for current and future direct detection, indirect detection and particle colliders on the cross section $\sigma(\chi \chi \to$ quarks)$/\sigma_{th}$ as function of the WIMP mass $m_\chi$ for WIMPs with the simple contact interaction in Eq.~\eqref{lagrangian} (see~\cite{Bauer:2013ihz} for details). Here $\sigma_{th} = 3 \times 10 ^{-26}$cm$^3$/s  taken as the  reference value for the annihilation cross section, is the value required  for a thermal WIMP to account for all of the DM.}
\label{complementarity-1}
\end{figure}

Attention must be paid to the assumptions that go into the quantitative comparison of  limits coming from the three types of searches, which is necessarily model dependent. For example, as part of the Snowmass 2013 study a simple contact  interaction model was assumed~\cite{Bauer:2013ihz}:
\begin{equation}
\frac{1}{M_q^2} ~\bar{\chi} \gamma^\mu\gamma_5 \chi \sum_q \bar{q} \gamma_\mu\gamma_5  q
+ \frac{\alpha_S}{M_g^3}~ \bar{\chi}  \chi G^{a \mu \nu} G^a_{\mu \nu}
+ \frac{1}{M_\ell^2} ~
\bar{\chi} \gamma^\mu  \chi \sum_\ell \bar{\ell} \gamma_\mu  \ell~,
\label{lagrangian}
\end{equation}
where the interactions with quarks mediate spin-dependent direct signals, and those with gluons mediate spin-independent direct signals. The coefficients $M_q$, $M_g$, and $M_\ell$ characterize the strength
of the interaction with the respective SM particle, and were chosen so that the combined
annihilation cross section into all three channels provides the
correct relic density for the thermal WIMP $\chi$ to constitute the whole of the DM. With this simple model it is possible to compare the limits imposed by direct, indirect and collider searches as well as their reach (see 
Fig.~\ref{complementarity-1} from~\cite{Bauer:2013ihz}).

The limits  in Fig.~\ref{complementarity-1} still depend on the particular halo model and the details of the detector response assumed (see~\cite{Bauer:2013ihz} and references therein for details). Besides, the limits would become very different if the particular DM candidate studied would constitute only a fraction {\it R} of the DM.  In this case,  the upper limits from direct detection on the scattering cross section $\sigma_{S}$ become larger by a factor  of $1/${\it R} (since the rate is  
  $\sim${\it R}$\sigma_{S}$), the indirect detection limits  on the annihilation  $\sigma_{A}$ would change by this factor squared, $1/${\it R}$^2$, and the LHC limits would remain unchanged.

\begin{figure}[t]
\centerline{\includegraphics[width=0.99\textwidth]{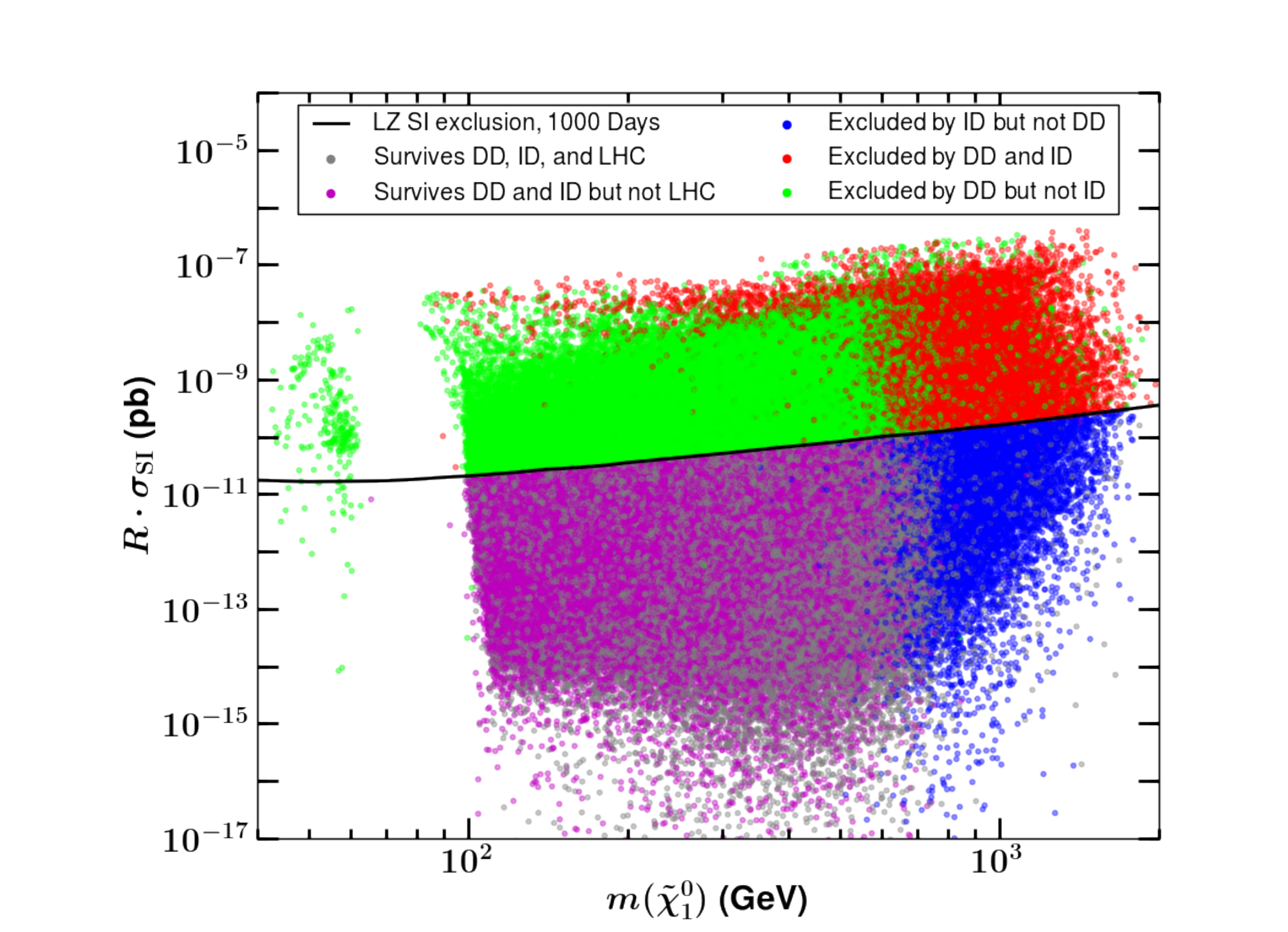}}
\caption{Results from a  scan of the 19 pMSSM parameters  plotted in the mass $m_\chi$ vs. {\it R}$\sigma_{SI}$ plane. {\it R}=$ \Omega_\chi / \Omega_{DM}$ is the density fraction of neutralinos $\chi$ in the DM, and $\sigma_{SI}$  is the WIMP-proton cross section  for SI WIMP nucleus interactions. Each point is a model: those in the reach of direct detection (the future LZ, whose discovery limit is shown with a black line) are green,  those within the reach of indirect detection (FermiLAT and the future CTA) are blue, those in the reach of both are red. The magenta points are models  tested by LHC searches which escape detection in direct or indirect detection. The gray points escape all searches in the near future. 
See~\cite{Cahill-Rowley:2014boa} for details.
 }
\label{complementarity-2}
\end{figure}

The complementarity of searches has also been studied in  a complete model, the pMSSM (phenomenological Minimal  Supersymmetric Standard Model), scanning over 19 parameters for DM mass values of 50 GeV to 
4 TeV (see Fig.~\ref{complementarity-2} from~\cite{Cahill-Rowley:2014boa}). The DM candidate is the lightest neutralino $\tilde\chi^0_1$, which is its own anti-particle.

In Fig.~\ref{complementarity-2} each point represents one particular
supersymmetric model; there are 200,000 points. Within each model, the DM interactions
are completely specified and  all relevant DM signals can be computed.  Models which produce a subdominant component of neutralino DM have also been included. The vertical axis of Fig.~\ref{complementarity-2}  shows the product of the computed density fraction of neutralino DM, {\it R}=$\Omega_\chi / \Omega_{DM}$, times the cross section. The models are divided into categories indicated by the color of each point, depending on whether the DM candidate is within the reach of future direct detection experiments (green points) such as LZ (the future upgrade of the LUX experiment, whose reach is indicated with a black line), or  within the reach of   indirect detection experiments (blue points), in particular FermiLAT and the future CTA,   or both (red points). The magenta points represent models that escape detection in the direct and indirect DM experiments just mentioned, and are tested at the LHC. Fig.~\ref{complementarity-2} shows clearly that the three different
DM probes are necessary  to test most  of the supersymmetry models in this scan.  It is interesting to see that the DM candidates in the gray models  escape all searches in the near future (see~\cite{Cahill-Rowley:2014boa} for details.)

\section{Axions as DM candidates}

In the SM there is a source of CP violation besides the phase of the Kobayashi-Maskawa mixing matrix. It is associated with the so called $\theta$ parameter, which provides the magnitude of a CP-violating
interaction among gluons allowed by all general principles.  Experimental bounds on electric dipole moments constrain this parameter to be very small, $\theta  \lsim 10^{-10}$, while a priori we would expect $\theta \simeq 1$. This unexplained smallness of the $\theta$ parameter constitutes  the so-called ``strong CP problem". The only known viable solution of this problem is the Peccei-Quinn mechanism. It consists of augmenting the SM (e.g. by the addition of a second doublet Higgs field, but there are other ways) so that the resulting Lagrangian has a new global U(1)$_{PQ}$ chiral symmetry~\cite{Peccei:1977hh}.  This global symmetry must be spontaneously broken, and this generates a Goldstone boson, the
 ``axion"~\cite{Weinberg:1977ma}.  The Goldstone mode is the component  $a$ of the scalar field along the degenerate orbit of minima of the characteristic inverted mexican hat potential.
 
\begin{figure}[t]
\centerline{\includegraphics[width=0.86\textwidth]{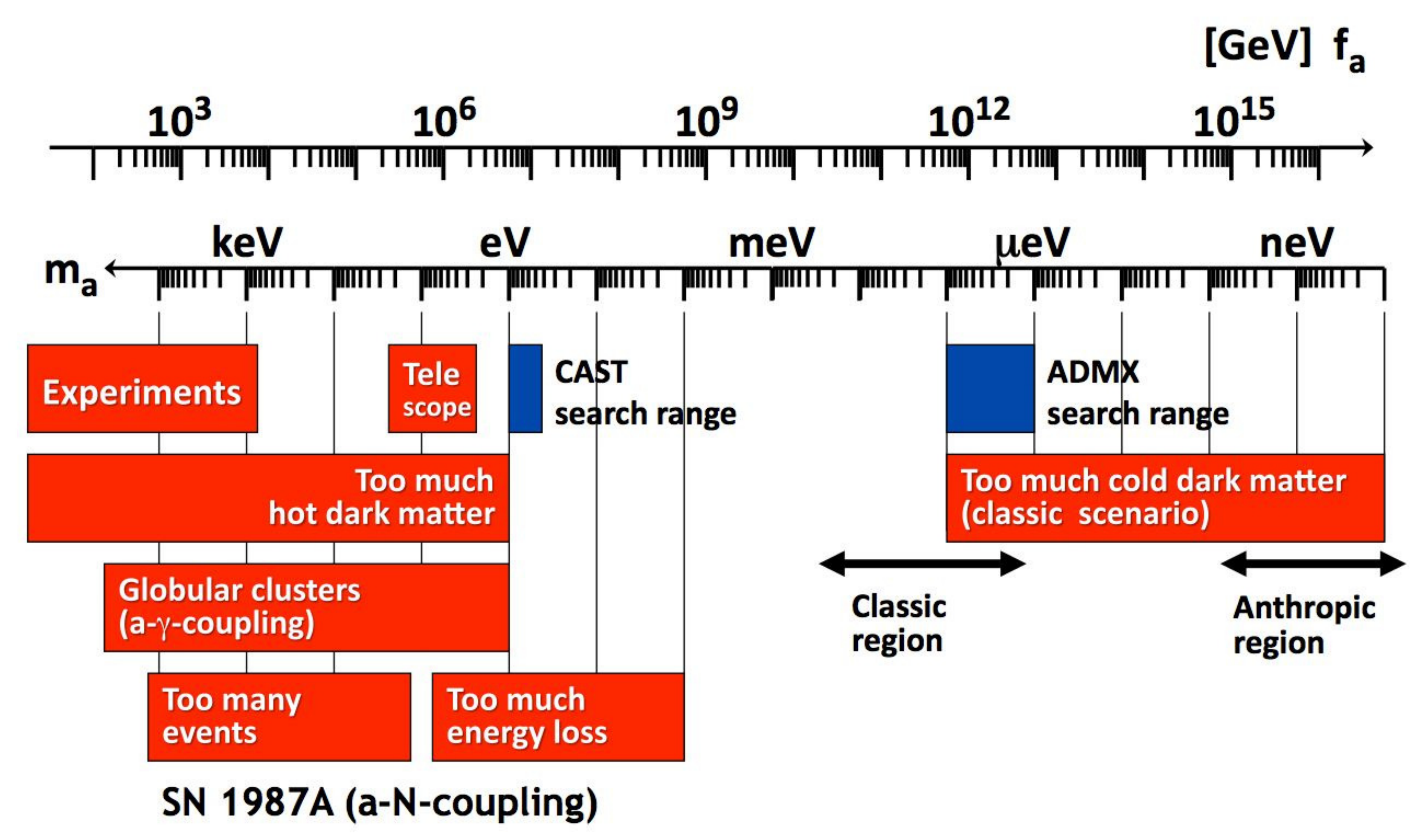}}
\caption{Excluded ranges and regions of interest  in terms of axion mass $m_a$ or PQ  spontaneous symmetry breaking scale $f_a \propto 1/m_a$. The mass range $\mu$eV to meV correspond to the so called ``classic region" of CDM axions  and the  mass range below (larger values of $f_a$)  is the ``anthropic region" of CDM axions. CAST and ADMX are experiments that search for axions. Fig taken from~\cite{Beringer:1900zz}. See also~\cite{Raffelt:2006cw}.}
\label{axion-limits}
\end{figure}

 In order for axions to be phenomenologically viable the  spontaneous symmetry breaking (SSB) scale, $f_a$, must be large, $f_a \gg 100$ GeV. QCD effects violate explicitly the PQ symmetry and generate a potential $ m_a a^2/2$ for the axion field $a= \theta f_a$ with $m_a \simeq \Lambda_{QCD}^2/ f_a$,  whose minimum is at $\theta=0$. The small explicit breaking produces a tilt in the orbit of the minima, which has then a minimum value and  a non-zero curvature close to it.  The PQ symmetry therefore solves the strong CP problem by transforming the $\theta$ parameter into a  field that has a minimum at $\theta$=0, simultaneously giving a small mass to the axion, which becomes a pseudo-Goldstone boson. For good reviews on axions with complete reference lists  see~\cite{Peccei:2006as, Raffelt:2006cw}.

An initial displacement $a_i= f_a \theta_i$ of the axion field from its minimum will result in  coherent oscillations of this field with frequency $m_a c^2/ \hbar$. The present energy density  in these oscillations is~\cite{Bae:2008ue} (recall $m_a \sim 1/ f_a$)
\begin{equation}
 \Omega_a h^2= 0.195~ \theta_i^2 \left(\frac{f_a}{10^{12}~ {\rm{GeV}}}\right)^{1.184} = 0.105~ \theta_i^2  \left(\frac{10~ \mu{\rm{ eV}}}{m_a} \right)^{1.184}   
\label{axion-density}
\end{equation}
and behaves as CDM.  Other types of light bosons, called ``Axion-Like Particles" (ALPs), are pseudo-Goldstone bosons of other broken global symmetries which do not couple to QCD, and can also be good DM candidates. They can acquire masses through their interactions with strongly-coupled hidden sectors or explicit breaking of the associated global symmetry~\cite{Beringer:1900zz}. ALPs together with very light  ``hidden" gauge bosons (``dark photons")  are generically called ``Weakly Interacting Slim Particles" (WISPs)~\cite{Ringwald:2012hr}.

  Axions can also be HDM for ``large" masses $m_a \simeq$ eV (when produced thermally via a  coupling with pions such as $a \pi \pi \pi$). The CERN Axion Solar Telescope (CAST) experiment  looks for this type of axion, which could be emitted by the Sun. Axions as CDM must have much smaller masses. They are searched for with resonant cavities  in the Axion DM eXperiment (ADMX) through the (model dependent) axion coupling with photons.  Fig.~\ref{axion-limits}~\cite{Beringer:1900zz}. shows the excluded ranges and  regions of interest in  axion mass $m_a$ and  PQ SSB scale $f_a$ (in particular the ranges within reach of CAST and  ADMX).

\begin{figure}[t]
\centerline{\includegraphics[width=0.75\textwidth]{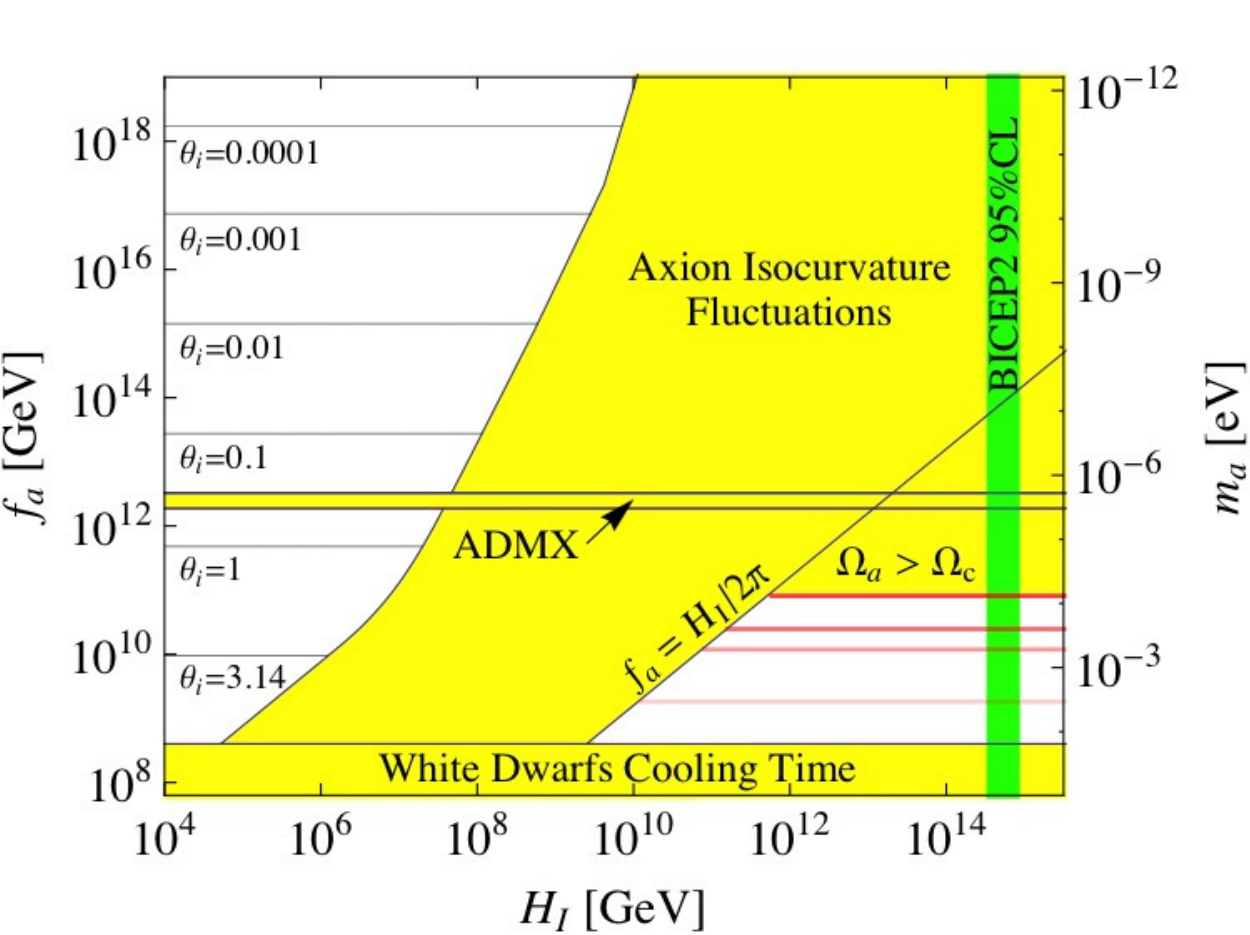}}
\caption{CDM axion parameter space, using the standard computations of the axion relic abundance~\cite{Visinelli:2014twa}. The yellow region is excluded by astrophysical and CMB constraints. The now rejected BICEP2 claim would have implied a Hubble scale during inflation indicated by the green vertical band.  
Fig. from~\cite{Visinelli:2014twa}. }
\label{Visinelli-Gondolo}
\end{figure}

  The mechanism of production of CDM axions depends on when the PQ SSB occurs relative to inflation: after or before. 
When the PQ SSB happens, the position along the orbit of minima where the field ends up  is different in causally disconnected volumes in the Universe.  This is the initial value $a_i= f_a \theta_i$ of the field oscillations which happen much later,  after the explicit PQ symmetry breaking becomes important.  

 If inflation happens after the PQ SSB, our whole Universe comes from only one causally connected volume before inflation.  The so-called ``misalignment"  $\theta_i$ is then the same in the whole visible Universe. For this to happen the condition is $f_a > H_I/2\pi$, where the Hubble expansion $ H_I$ is the only dimensional relevant quantity during inflation (see e.g.~\cite{Visinelli:2014twa}). In this case, if $\theta_i \simeq$ O(1), axions would have the DM density for $m_a\simeq 10\mu$eV and $f_a \simeq 10^{12}$ GeV (see Eq.~\eqref{axion-density}). This is called the ``classic region"  (see Fig.~\ref{axion-limits}). If instead
  $\theta_i \ll 1$,  Eq.~\eqref{axion-density}  shows that axions should have a much smaller mass (and larger $f_a$) to be the DM. This is  called  the ``anthropic region" (see Fig.~\ref{axion-limits}). 
  
With the standard axion production mechanism, the possibility of  $f_a > H_I/2\pi$ would have been excluded~\cite{Visinelli:2014twa} by the recent BICEP2~\cite{Ade:2014xna} claim (now rejected~\cite{Ade:2015fwj}) of a measurement  of gravity waves produced during inflation, as shown in Fig.~\ref{Visinelli-Gondolo}.
The BICEP2 claimed measurement would have fixed  the Hubble expansion during inflation to be $H_I \simeq 6 \times 10^{14}$ GeV,
 for which  limits  on axion isocurvature fluctuations in the Universe
  forbid  $f_a > H_I/2\pi$. Isocurvature fluctuations have their origin in quantum fluctuations of the value $a_i$ of the field during inflation, which translate into fluctuations in the number density of axions after the coherent oscillations of the axion field start. 

 If the PQ SSB happens after inflation, i.e. if $f_a < H_I/2\pi$, the present Universe
contains many different domains that were causally disconnected when the SSB happened and thus have randomly distributed values of $\theta_i$. In this case a network of cosmic strings is generated at the moment of  SSB of the global U$_{PQ}$(1) via the Kibble mechanism.  This mechanism consists of a  correlation of the field values after the SSB of a global U(1) symmetry around an axis in space and along the orbit of degenerate minima  in internal space.  This  prevents SSB along the axis itself, with the consequent formation of a topological defect along the axis, i.e. a string (see e.g.~\cite{Hindmarsh:1994re}). These strings may become connected with walls after the explicit PQ symmetric breaking, and then the whole network of strings (and possibly walls) decays into axions (see e.g.~\cite{Hiramatsu:2012gg}).  Understanding the evolution of this network is very complicated  and different calculations  of the axion density they produce differ by up to a factor of a 1000. The red horizontal bands in Fig.~\ref{Visinelli-Gondolo} show where axions constitute the whole of the DM according to  some of these calculations (see~\cite{Visinelli:2014twa} for details) for $m_a \simeq 100$ $\mu$eV,  $f_a \simeq 10^{11}$ GeV. These  masses (see Fig.~\ref{axion-limits}) are not within the reach of current searches with cavities in ADMX (1.9  to 10 $\mu$eV) but could be tested by ADMX-HF, with smaller cavities~\cite{vanBibber:2013ssa}.

Finally, let us mention that  axions are also relics of the pre-BBN era  and its characteristics could change in non-standard cosmologies~\cite{Visinelli:2009kt}.

\begin{figure}[t]
\centerline{\includegraphics[width=0.73\textwidth]{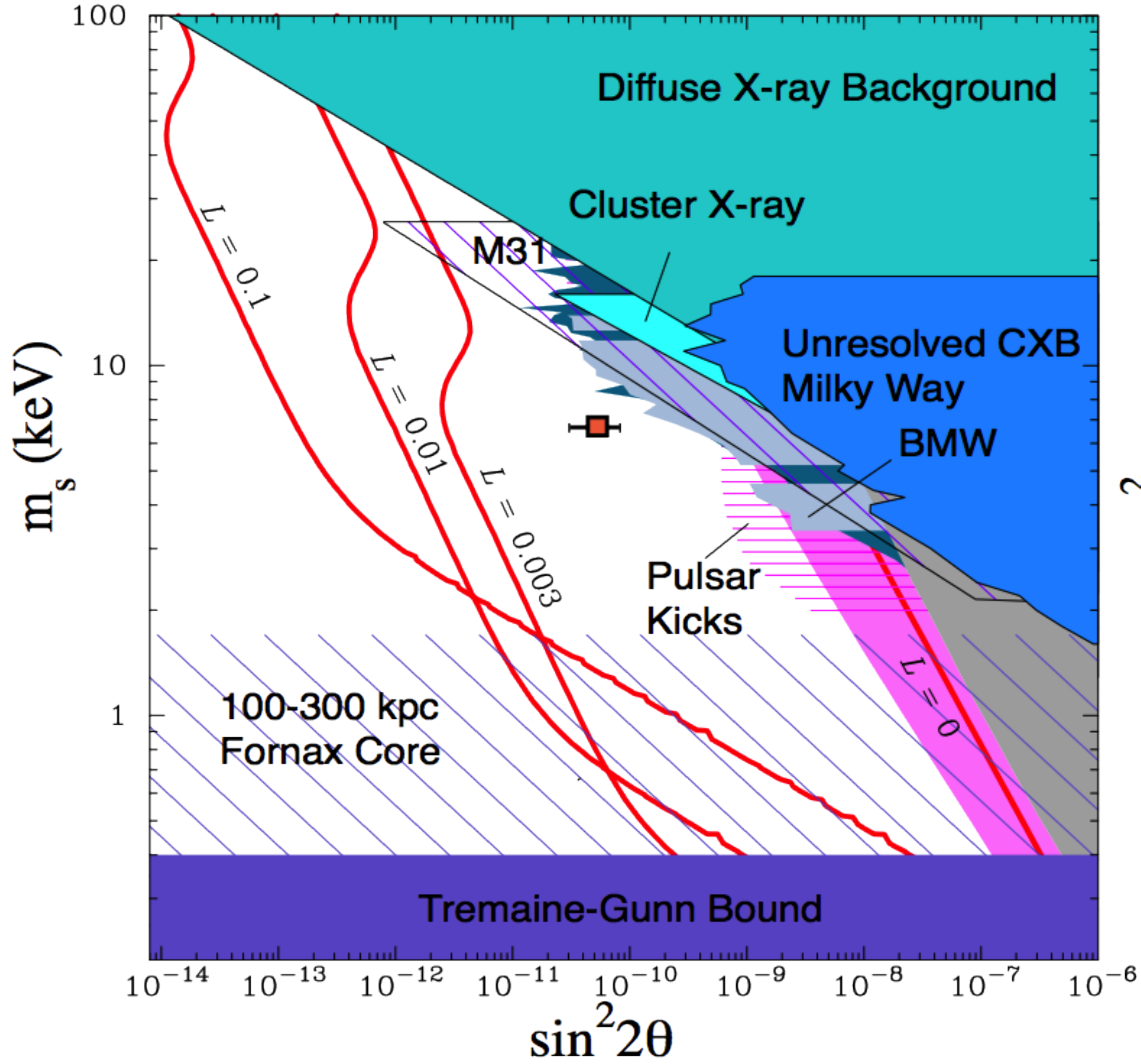}}
\caption{Prediction for sterile neutrinos which constitute the whole of the DM (red lines) produced 
non-resonantly~\cite{Dodelson:1993je}  ($L=0$ - the pink band is due to uncertainties in this production 
mechanism~\cite{Asaka:2006nq}) or resonantly~\cite{Shi:1998km} (for large lepton asymmetries $L=$ 0.003, 0.001 and 0.1).  Also shown are limits from phase-space considerations in different structures (Tremaine-Gunn and Fornax Core) and  non-observation of X-rays generated in their decay.  ``Pulsar-Kicks" indicates a particular region of interest 
(see e.g~\cite{Kusenko:2013saa}). The red symbol corresponds to the sterile neutrino which could produce a recent potential line signal  reported at 3.5 keV. Fig. from~\cite{Bulbul:2014sua}.
See~\cite{Bulbul:2014sua} for explanations and references.
}
\label{Sterile}
\end{figure}

\section{Sterile neutrinos as  DM candidates} 

The SM has three ``active neutrinos" $\nu_a$, i.e. neutrinos with  weak interactions, but others without weak interactions (called ``sterile" by  Bruno Pontecorvo) $\nu_s$  can be easily added (one or more, of any mass). These are mixed to the active neutrinos through their common mass matrix. Considering only one active and one sterile neutrinos, the mass eigenstates $\nu_1$ and $\nu_2$ are admixtures of both: $\nu_a = \cos \theta~ \nu_1 + 
\sin\theta ~\nu_2 $ and  $\nu_s = -\sin \theta~ \nu_1  + \cos \theta  ~\nu_2 $.
Here $\nu_{a,s}$ are interaction eigenstates and $\nu_{1,2}$ are mass eigenstates, with $m_1 << m_2 \equiv m_s$. The mostly sterile neutrino mass eigenstate is a good candidate to be WDM if its mass is $m_s$=O(keV)~\cite{Abazajian:2001nj}. This neutrino can be created via active-sterile oscillations, without~\cite{Dodelson:1993je} or with~\cite{Shi:1998km} a large Lepton Asymmetry $L$ (much larger than the baryon asymmetry in the Universe),  and respectively
 be WDM or ``Cool DM" (i.e. a cooler form of WDM, with a spectrum peaked at smaller momenta). These two types of sterile neutrinos are called ``Dodelson-Widrow (DW)"~\cite{Dodelson:1993je} (or  ``non-resonantly produced") and ``resonantly produced", respectively (because a large $L$  produces an $L$-driven resonant  conversion of active into sterile neutrinos). DW sterile neutrinos have an almost thermal spectrum, with average momentum over temperature $\left<p/T\right> \simeq 3.15$, but  resonantly produced ones can  easily have $\left<p/T\right> \simeq 1$ to 2. 
 
 A combination of lower bounds on DW  sterile neutrino masses, coming from early structures  in the Universe (called Ly-$\alpha$ clouds)  and  sub-halo counts  in simulations of galaxy formation, leads to $m_s >$ 8 keV.   Upper limits on the same mass  coming from the non-observation of X-rays due to the decay $\nu_s \rightarrow \nu \gamma$ in galaxies and clusters of galaxies lead to $m_s < 3$ keV.  These together rule out DW sterile neutrinos as the major component of the DM (see e.g.~\cite{Abazajian:2012ys} and  \cite{Horiuchi:2013noa} and references therein).
The same limits are less restrictive when applied to resonantly produced sterile neutrinos, which remain valid candidates to account for the whole of the DM.  Also,  sterile neutrinos can be produced in other ways besides oscillations (e.g. in the decay of new scalar fields or heavier sterile neutrinos) which yield sterile neutrinos that can also constitute the bulk of the DM (see e.g.~\cite{Kusenko:2013saa} and references therein).

If they exist, sterile neutrinos are remnants of the pre-BBN era and the aforementioned constraints assume a standard cosmology.  Their relic abundance could be very different in non-standard cosmologies (see e.g.~\cite{Gelmini:2004ah}).

Recently a potential weak line signal was reported at 3.5 keV~\cite{Bulbul:2014sua, Boyarsky:2014jta} which could be due to the two-body decay $\nu_s \rightarrow \nu \gamma$ of a sterile neutrino with $m_s=7$ keV and $\sin^2 2 \theta \simeq  10 ^{-10}$ (indicated by the red spot in Fig.~\ref{Sterile}).  The line was found  by one group in stacked observations of 73 galaxy clusters and in the Perseus cluster~\cite{Bulbul:2014sua}, and by another group in the Andromeda galaxy,  Perseus cluster and  the Milky Way center~\cite{Boyarsky:2014jta}. This could be a resonantly produced sterile neutrino~\cite{Abazajian:2014gza}, but other explanations in terms of atomic line emissions or backgrounds cannot be excluded. Further observations and analyses are necessary to confirm or reject this potential DM signal.

\section{Outlook}

There is no compelling observational or experimental evidence in favor of any of our  DM candidates, such as WIMPs,  axions, sterile neutrinos, primordial black holes or any other of the many  that have been proposed. It is only through experiments and observations that we will elucidate the nature of the DM. The next decades will be a very exciting time for DM research. Direct detection experiments will move to the ton-scale, indirect detection efforts will continue developing very rapidly, the LHC with its full capabilities (and possibly the next particle collider) will reframe what we know about physics beyond the SM and thus DM candidates. 

The importance of the possible payoff of these searches is enormous. A confirmed detection of a DM candidate would open the doors to the age of precision DM studies to determine its properties and to DM astronomy.

\section*{Acknowledgments}

I thank Lance Dixon and Frank Petriello for organizing TASI 2014, and Tom Degrand and K.~T. Mahanthappa for their hospitality at Boulder during the school. This work was supported in part by the US Department of Energy under Award Number DE-SC0009937 and also by the National Science Foundation under Grant No. NSF PHY11-25915 (through the Kavli Institute for Theoretical Physics, KITP, of the University of California, Santa Barbara, where most of these these lectures were written).
  


\begin{thebibliography}{999}

\bibitem{Hooper:2009zm} 
  D.~Hooper,
  arXiv:0901.4090 [hep-ph].

\bibitem{Gorenstein:2014iba} 
  P.~Gorenstein and W.~Tucker,
  Adv.\ High Energy Phys. {\bf 2014}, 878203.

\bibitem{zwicky}
F. Zwicky, Helv. Phys. Acta 6 (1933) 110;  Astrophys. J. {\bf 86}  217 (1937). 

\bibitem{clusters}
N. Bahcall and X. Fan, Astrophys. J. {\bf 504} (1998) 1;
A. Kashlinsky, Phys. Rep. {\bf 307}  67 (1998);
R.~G.~Carlberg et al., 
Astrophys. J. {\bf 516}  552 (1999);
J.~A.~Tyson, G.~P.~Kochanski and I.~P.~Dell'Antonio,  
 Astrophys.\ J.\  {\bf 498}, L107 (1998)
H.~Dahle, 
arXiv:astro-ph/0701598.
\bibitem{Hoekstra:2013via} 
  H.~Hoekstra {\it{et al.}}
  Space Sci.\ Rev.\  {\bf 177}, 75 (2013)
  [arXiv:1303.3274]

\bibitem{rotationcurves}
A.~Borriello and P.~Salucci, 
Mon.\ Not.\ Roy.\ Astron.\ Soc.\  {\bf 323}, 285 (2001) 
[arXiv:astro-ph/0001082].  

\bibitem{weaklensing}
  E. van Uitert {\it{et al}} 
  Astron.\ Astroph.\ {\bf 545} A71(2012)
  [arXiv:1206.4304].
  
\bibitem{strong}
L.~A.~Moustakas and R.~B.~Metcalf,
Mon.\ Not.\ Roy.\ Astron.\ Soc.\  {\bf 339}, 607 (2003)
[arXiv:astro-ph/0206176].

\bibitem{Planck} 
  P.~A.~R.~Ade {\it et al.}  [Planck Coll.],
  Astron.\ Astrophys.\  {\bf 571}, A1 (2014)
  [arXiv:1303.5062 [astro-ph.CO]] 
 
\bibitem{Hannestad:2004px} 
  S.~Hannestad,
  Phys.\ Rev.\ D {\bf 70}, 043506 (2004)
  [astro-ph/0403291].
  
\bibitem{Ade:2015fwj} 
  P.~A.~R.~Ade {\it et al.}  [BICEP2 and Keck Array Collaborations],
  arXiv:1502.00643 [astro-ph.CO].
 
\bibitem{Ade:2014xna} 
  P.~A.~R.~Ade {\it et al.} [BICEP2 Coll.],
  Phys.\ Rev.\ Lett. {\bf 112}, 241101 (2014)
  [arXiv:1403.3985 [astro-ph.CO]].

\bibitem{Dai:2014jja} 
  L.~Dai, M.~Kamionkowski and J.~Wang,
  Phys.\ Rev.\ Lett.\  {\bf 113}, 041302 (2014)
  [arXiv:1404.6704 [astro-ph.CO]].

\bibitem{mond}
  M.~Milgrom,
  Astrophys.\ J.\  {\bf 270}, 365 (1983).
 
\bibitem{Bekenstein:2004ne}
  J.~D.~Bekenstein,
  Phys.\ Rev.\  D {\bf 70}, 083509 (2004)
  [Erratum-ibid.\  D {\bf 71}, 069901 (2005)]
  [arXiv:astro-ph/0403694].
  
\bibitem{McGaugh:2014nsa} 
  B.~Famaey and S.~McGaugh,
  Living Rev.\ Rel.\  {\bf 15}, 10 (2012)
  [arXiv:1112.3960]; 
  S.~S.~McGaugh,
  arXiv:1404.7525. 
  
\bibitem{Clowe:2006eq} 
 D.~Clowe {\it et al.} 
  Astrophys.\ J.\  {\bf 648}, L109 (2006)
  [astro-ph/0608407].

\bibitem{Angus:2007mn} 
  G.~W.~Angus, B.~Famaey and D.~A.~Buote,
  Mon.\ Not.\ Roy.\ Astron.\ Soc.\  {\bf 387}, 1470 (2008)
  [arXiv:0709.0108 [astro-ph]].
  
\bibitem{Milgrom:2008rv} 
  M.~Milgrom,
  arXiv:0801.3133 [astro-ph].
  
\bibitem{Overduin:2004sz} 
  J.~M.~Overduin and P.~S.~Wesson,
  Phys.\ Rept.\  {\bf 402}, 267 (2004)
  [astro-ph/0407207].
  
  \bibitem{Pospelov:2000bq}
 M.~Pospelov and T.~ter Veldhuis,
 Phys.\ Lett.\  B {\bf 480} (2000) 181
 [arXiv:hep-ph/0003010].
 
\bibitem{Sigurdson:2004zp}
 K.~Sigurdson {\it et al.}
 Phys.\ Rev.\  D {\bf 70} (2004) 083501
 [Erratum-ibid.\  D {\bf 73} (2006) 089903]
 [arXiv:astro-ph/0406355].
 
\bibitem{Barger:2010gv}
 V.~Barger, W.~Y.~Keung and D.~Marfatia,
 Phys.\ Lett.\  B {\bf 696} (2011) 74
 [arXiv:1007.4345 [hep-ph]].

\bibitem{Ho:2012bg} 
  C.~Ho and R.~Scherrer,
  Phys.\ Lett.\ B{\bf 722} 341 (2013)
  [arXiv:1211.0503].
  
\bibitem{Feldman:2007wj} 
  D.~Feldman, Z.~Liu and P.~Nath,
  Phys.\ Rev.\ D {\bf 75}, 115001 (2007)
  [hep-ph/0702123].

\bibitem{CHAMPs} 
  L.~Chuzhoy and E.~W.~Kolb,
  JCAP {\bf 0907}, 014 (2009)
  
  
\bibitem{Jacobs:2014yca} 
  D.~M.~Jacobs, G.~D.~Starkman and B.~W.~Lynn,
ÊÊ
ÊÊ[arXiv:1410.2236].


\bibitem{Cline:2012is} 
  J.~M.~Cline, Z.~Liu and W.~Xue,
  Phys.\ Rev.\ D {\bf 85}, 101302 (2012)
  [arXiv:1201.4858 [hep-ph]]. 
  
\bibitem{Holdom:1985ag} 
  B.~Holdom,
  Phys.\ Lett.\ B {\bf 166}, 196 (1986).
 

\bibitem{Burrage:2009yz} 
  C.~Burrage, J.~Jaeckel, J.~Redondo and A.~Ringwald,
  JCAP {\bf 0911}, 002 (2009)
  [arXiv:0909.0649 [astro-ph.CO]].
 
\bibitem{McDermott:2010pa} 
  S.~D.~McDermott, H.~B.~Yu and K.~M.~Zurek,
  Phys.\ Rev.\ D {\bf 83}, 063509 (2011)
  [arXiv:1011.2907 [hep-ph]].

\bibitem{Goldberg:1986nk} 
  H.~Goldberg, L.~J.~Hall,
  Phys.\ Lett.\ B {\bf 174}, 151 (1986).
  
\bibitem{Feng:2009mn} 
  J.~L.~Feng, M.~Kaplinghat, H.~Tu and H.~-B.~Yu,
  JCAP {\bf 0907}, 004 (2009)
  [arXiv:0905.3039 [hep-ph]].

\bibitem{Kaplan:2009de} 
  D.~E.~Kaplan, G.~Z.~Krnjaic, K.~R.~Rehermann and C.~M.~Wells,
  JCAP {\bf 1005}, 021 (2010)
  [arXiv:0909.0753].
  
\bibitem{Foot:1991bp} 
  R.~Foot, H.~Lew and R.~R.~Volkas,
  Phys.\ Lett.\ B {\bf 272}, 67 (1991).
  
\bibitem{Pospelov:2007mp} 
  M.~Pospelov, A.~Ritz and M.~B.~Voloshin,
  Phys.\ Lett.\ B {\bf 662}, 53 (2008)
  [arXiv:0711.4866 [hep-ph]].
 
\bibitem{ArkaniHamed:2008qn} 
  N.~Arkani-Hamed, D.~P.~Finkbeiner, T.~R.~Slatyer and N.~Weiner,
  Phys.\ Rev.\ D {\bf 79}, 015014 (2009)
 [arXiv:0810.0713].
 
\bibitem{Boehm:2014vja} 
  C.~Boehm {\it et al.}
  Mon.\ Not.\ Roy.\ Astron.\ Soc.\  {\bf 445}, L31 (2014)
  [arXiv:1404.7012 [astro-ph.CO]].

\bibitem{DDDM} 
  J.~Fan, A.~Katz, L.~Randall and M.~Reece,
  Phys. Dark Univ. {\bf 2} 139 (2013)
  [arXiv:1303.1521] and 
  Phys.\ Rev.\ Lett.\  {\bf 110} 211302 (2013)
  [arXiv:1303.3271].
  
 \bibitem{DD}
  J.~I.~Read {\it et al.}
  Mon.\ Not.\ Roy.\ Astron.\ Soc.\  {\bf 397}, 44 (2009)
  [arXiv:0902.0009].



\bibitem{Griest:2014cqa} 
  K. Griest, A. Cieplak and M. Lehner,
  Astrophys.\ J.\  {\bf 786}, 158 (2014).

\bibitem{Alcock:1998fx} 
  C.~Alcock {\it et al.}  [MACHO and EROS Colls.],
  Astrophys.\ J.\  {\bf 499}, L9 (1998)
  [astro-ph/9803082];
 
  \bibitem{granularity} 
  J.~Yoo, J.~Chaname and A.~Gould,
  Astrophys.\ J.\  {\bf 601}, 311 (2004)
  [astro-ph/0307437].
  
\bibitem{Carr:1974nx} 
  B.~Carr and S.~Hawking,
  Mon.\ Not.\ Roy.\ Astron.\ Soc.\  {\bf 168}, 399 (1974).
 
\bibitem{Kusenko:2013saa} 
  A.~Kusenko and L.~J.~Rosenberg,
ÊÊarXiv:1310.8642 [hep-ph].

\bibitem{Green:2014faa} 
  A.~M.~Green,
  arXiv:1403.1198 [gr-qc].
 
\bibitem{fuzzy-DM}
  W.~Hu, R.~Barkana and A.~Gruzinov,
  Phys.\ Rev.\ Lett.\  {\bf 85}, 1158 (2000)
  [astro-ph/0003365].
 
 \bibitem{Tremaine-Gunn}
  S.~Tremaine and J.~E.~Gunn,
  Phys.\ Rev.\ Lett.\  {\bf 42}, 407 (1979); 
  J.~Madsen,
  Phys.\ Rev.\ D {\bf 64}, 027301 (2001)
  [astro-ph/0006074].


\bibitem{Zavala:2012us} 
  J.~Zavala, M.~Vogelsberger and M.~Walker,
  Mon.\ Not.\ Roy.\ Astron.\ Soc.\.: Letters {\bf 431}, L20 (2013)
  [arXiv:1211.6426].

\bibitem{SIDM}
  D.~N.~Spergel and P.~J.~Steinhardt,
  Phys.\ Rev.\ Lett.\  {\bf 84}, 3760 (2000)
  [astro-ph/9909386].
   
\bibitem{Vogelsberger:2012ku} 
  M.~Vogelsberger, J.~Zavala and A.~Loeb,
  Mon.\ Not.\ Roy.\ Astron.\ Soc.\  {\bf 423}, 3740 (2012)
  [arXiv:1201.5892 [astro-ph.CO]].
 
\bibitem{too-big-to-fail} 
  M.~Boylan-Kolchin, J.~Bullock and M.~Kaplinghat,
  Mon.\ Not.\ Roy.\ Astron.\ Soc.\  {\bf 415}, L40 (2011)
  [arXiv:1103.0007].
  
\bibitem{Kolb:1990vq} 
  E.~W.~Kolb and M.~S.~Turner,
  Front.\ Phys.\  {\bf 69}, 1 (1990).

\bibitem{Kirby:2014sya} 
  E.~N.~Kirby {\it et al.}
  Mon.\ Not.\ Roy.\ Astron.\ Soc.\  {\bf 439}, 1015 (2014)
  [arXiv:1401.1208 [astro-ph.GA]].
 
\bibitem{too-small-to-succeed}  
  R.~Kennedy, C.~Frenk, S.~Cole and A.~Benson,
  Mon.\ Not.\ Roy.\ Astron.\ Soc.\  {\bf 442}, 2487 (2014)
  [arXiv:1310.7739 [astro-ph.CO]].
  


\bibitem{Gelmini:2010zh} 
  G.~Gelmini and P.~Gondolo,
  In *Bertone, G. (ed.): Particle dark matter* 121-141
  [arXiv:1009.3690 [astro-ph.CO]].

\bibitem{Barrow:1982ei} 
  J.~D.~Barrow,
  Nucl.\ Phys.\ B {\bf 208}, 501 (1982).
  
  
\bibitem{Gondolo:1990dk} 
  P.~Gondolo and G.~Gelmini,
  Nucl.\ Phys.\ B {\bf 360}, 145 (1991).
  
\bibitem{BBC-pdg} 
K. Olive and J. Peacock ``Big-Bang Cosmology",
J.~Beringer {\it et al.}  [Part. Data Gr.],
  Phys.\ Rev.\ D {\bf 86}, 010001 (2012) and updates.
  
\bibitem{Baltz:2004tj} 
  E.~A.~Baltz,
  eConf C {\bf 040802}, L002 (2004)
  [astro-ph/0412170].
 
\bibitem{Lee:1977ua} 
B.~W.~Lee and S.~Weinberg,
  Phys.\ Rev.\ Lett.\  {\bf 39}, 165 (1977);
  P.~Hut,
  Phys.\ Lett.\ B {\bf 69} 85 (1977).
  
\bibitem{Griest:1989wd} 
  K.~Griest and M.~Kamionkowski,
  Phys.\ Rev.\ Lett.\  {\bf 64}, 615 (1990).
  
\bibitem{Feng:2003xh}
J.~L.~Feng, A.~Rajaraman and F.~Takayama,
Phys.\ Rev.\ Lett.\  {\bf 91}  011302 (2003).
[arXiv:hep-ph/0302215].
  

\bibitem{Nussinov:1985xr} 
  S.~Nussinov,
  Phys.\ Lett.\ B {\bf 165}  55 (1985).
  
\bibitem{Gelmini:1986zz} 
  G.~Gelmini, L.~Hall and M.~Lin,
 {\it  Nucl.\ Phys.\ B} {\bf 281} (1987) 726.
   
\bibitem{Petraki:2013wwa} 
  K.~Petraki and R.~Volkas,
  Int.\ J.\ Mod.\ Phys.\ A{\bf 28} 1330028 (2013)
  [arX iv:1305.4939]; 
  K.~Zurek
  Phys.\ Rept.\  {\bf 537} 91 (2014)
  [arXiv:1308.03 38]. 

\bibitem{Griest:1986yu} 
  K.~Griest and D.~Seckel,
  Nucl.\ Phys.\ B {\bf 283}  681 (1987)
  [Erratum-ibid.\ B {\bf 296}  1034 (1988)].
  
\bibitem{Iminniyaz:2011cd}
  H.~Iminniyaz, M.~Drees and X.~Chen,
  JCAP {\bf 1107}  003 (2011)
  [astro-ph/1104.5548].

\bibitem{Gelmini:2013awa} 
  G.~B.~Gelmini, J.~H.~Huh and T.~Rehagen,
  JCAP {\bf 1308}, 003 (2013)
  [arXiv:1304.3679 [hep-ph]].
  
  
  
\bibitem{Gelmini:2009yh} 
  G.~Gelmini,
  Nucl.\ Phys.\ Proc.\ Suppl.{\bf 194} 63 (2009)
  [arXiv:0907.1694]; 
  G.~Gelmini and P.~Gondolo,
  Phys. Rev. D {\bf 74} 023510 (2006)
  [hep-ph/06022 30].

  
\bibitem{Lin:2000qq}
  W.~Lin {\it et al.}
  Phys.\ Rev.\ Lett.\  {\bf 86}, 954 (2001)
  [arXiv:astro-ph/0009003];
  J.~Hisano, K.~Kohri and M.~Nojiri,
  Phys.\ Lett.\  B {\bf 505}, 169 (2001)
  [arXiv:hep-ph/0011216]; 
  G.~Gelmini and C.Yaguna,
  Phys.\ Lett.\  B {\bf 643}, 241 (2006)
 [arXiv:hep-ph/0607012].

\bibitem{Gelmini:2008sh}
  G. Gelmini and P. Gondolo,
  JCAP {\bf 0810} 002 (2008)
  [arXiv:0803.2349]; 
  L.~Visinelli and P.~Gondolo,
  arXiv:1501.02233. 

\bibitem{Duda:2001ae} 
  G.~Duda, G.~Gelmini and P.~Gondolo,
  Phys.\ Lett.\ B {\bf 529}, 187 (2002)
  [hep-ph/0102200].
  
\bibitem{Cushman:2013zza} 
  P.~Cushman {\it et al.}
  arXiv:1310.8327 [hep-ex]. 
  
\bibitem{Drukier:1986tm} 
  A.~Drukier, K.~Freese and D.~Spergel,
  Phys.\ Rev.\ D {\bf 33}, 3495 (1986).  
  

\bibitem{Read:2014qva} 
  J.~I.~Read,
  J.\ Phys.\ G {\bf 41}, 063101 (2014)
  [arXiv:1404.1938].
  
 
\bibitem{Savage:2008er} 
  C.~Savage, G.~Gelmini, P.~Gondolo and K.~Freese,
  JCAP {\bf 0904}, 010 (2009)
  [arXiv:0808.3607 [astro-ph]].
  
\bibitem{Bozorgnia:2012eg} 
  N.~Bozorgnia, G.~B.~Gelmini and P.~Gondolo,
  JCAP {\bf 1208}, 011 (2012)
  [arXiv:1205.2333 [astro-ph.CO]].

  
\bibitem{Smith:2006ym} 
  M.~C.~Smith
  {\it et al.},
  Mon.\ Not.\ Roy.\ Astron.\ Soc.\  {\bf 379}, 755 (2007)
  [astro-ph/0611671].
  
\bibitem{Piffl:2013mla} 
  T.~Piffl 
   {\it et al.},
  Astron. \& Astroph.{\bf 562}, A91 (2014)
ÊÊ[arXiv:1309.4293].


  \bibitem{Vogelsberger} 
  M.~Vogelsberger {\it et al.}
  Mon.\ Not.\ Roy.\ Astron.\ Soc.\  {\bf 395}, 797 (2009)
  [arXiv:0812.0362 [astro-ph]].
  
  
\bibitem{Kuhlen:2008aw} 
  M.~Kuhlen, J.~Diemand and P.~Madau,
  Astrophys.\ J.\  {\bf 686}, 262 (2008)
  [arXiv:0805.4416 [astro-ph]].

  
  \bibitem{stream} 
  C.~W.~Purcell, A.~R.~Zentner and M.~Y.~Wang,
  JCAP {\bf 1208}, 027 (2012)
  [arXiv:1203.6617 [astro-ph.GA]].
  
  \bibitem{debris-flows}
  M.~Lisanti and D.~N.~Spergel,
  Phys.\ Dark Univ.\  {\bf 1}, 155 (2012)
  [arXiv:1105.4166]; 
  M.~Kuhlen, M.~Lisanti and D.~N.~Spergel,
  Phys.\ Rev.\ D {\bf 86}, 063505 (2012)
  [arXiv:1202.0007].
  
 
  
  \bibitem{LDM} 
  R.~Bernabei {\it et al}
  Phys.\ Rev.\ D {\bf 77}, 023506 (2008)
  [arXiv:0712.0562]; 
  J.~Kopp, V.~Niro, T.~Schwetz and J.~Zupan,
  Phys.\ Rev.\ D {\bf 80}, 083502 (2009)
  [arXiv:0907.3159]. 
  R.~Essig, J.~Mardon and T.~Volansky,
  Phys.\ Rev.\ D {\bf 85}, 076007 (2012)
  [arXiv:1108.5383]. 
  R.~Essig {\it et al.}
  Phys.\ Rev.\ Lett.\  {\bf 109}, 021301 (2012)
  [arXiv:1206.2644]. 
  B.~Batell, R.~Essig and Z.~Surujon,
  Phys.\ Rev.\ Lett.\  {\bf 113} 171802 (2014)
  [arXiv:1406.2698].
  
\bibitem{Angle:2011th} 
  J.~Angle {\it et al.}  [XENON10 Coll.],
  Phys.\ Rev.\ Lett.\  {\bf 107}, 051301 (2011)
  [Erratum-ibid.\  {\bf 110}, 249901 (2013)]
  [arXiv:1104.3088].


\bibitem{Helm:1956zz} 
  R.~H.~Helm,
  Phys.\ Rev.\ {\bf 104}   1466 (1956).
  
  
  \bibitem{TuckerSmith:2001hy}
  D.~Tucker-Smith and N.~Weiner,
  Phys.\ Rev.\ D {\bf 64}, 043502 (2001)
  [hep-ph/0101138].
  
\bibitem{Graham:2010ca} 
  P.~W.~Graham, R.~Harnik, S.~Rajendran and P.~Saraswat,
  Phys.\ Rev.\ D {\bf 82}, 063512 (2010)
  [arXiv:1004.0937 [hep-ph]].
  
 
\bibitem{Jungman:1995df}
  G.~Jungman, M.~Kamionkowski and K.~Griest,
  Phys.\ Rept.\  {\bf 267}, 195 (1996).
  [arXiv:hep-ph/9506380], and references therein.

  

\bibitem{IsospinViolating}
  A.~Kurylov and M.~Kamionkowski,
  Phys.\ Rev.\ D {\bf 69} 063503 (2004);
  [hep-ph/0307185]; 
  J.~L.~Feng, J.~Kumar, D.~Marfatia and D.~Sanford,
  Phys.\ Lett.\ B {\bf 703} 124  (2011)
  [arXiv:1102.4331 [hep-ph]].


\bibitem{Gelmini:2014psa} 
  G.~B.~Gelmini, A.~Georgescu and J.~H.~Huh,
  JCAP {\bf 1407}, 028 (2014)
  [arXiv:1404.7484 [hep-ph]].

\bibitem{Bednyakov:2004xq} 
  V.~A.~Bednyakov and F.~Simkovic,
  Phys.\ Part.\ Nucl.\  {\bf 36}, 131 (2005)
  [Fiz.\ Elem.\ Chast.\ Atom.\ Yadra {\bf 36}, 257 (2005)]
  [hep-ph/0406218].
  
  \bibitem{interactions} 
  A.~L.~Fitzpatrick {\it et al},
  JCAP {\bf 1302}, 004 (2013)
  [arXiv:1203.3542]. 
  
\bibitem{Barello:2014uda} 
  G.~Barello, S.~Chang and C.~A.~Newby,
  Phys.\ Rev.\ D {\bf 90}, no. 9, 094027 (2014)
  [arXiv:1409.0536 [hep-ph]].
  
 
  
   \bibitem{Bernabei:2010mq}
 R.~Bernabei {\it et al.}  [DAMA/LIBRA Coll.],
 Eur.\ Phys.\ J.\  C {\bf 67}, 39 (2010)
 [arXiv:1002.1028 [astro-ph.GA]];
  R.~Bernabei {\it et al.}
  Eur.\ Phys.\ J.\ C {\bf 73}, 2648 (2013)
  [arXiv:1308.5109 [astro-ph.GA]].


\bibitem{Aalseth:2010vx}
 C.~E.~Aalseth {\it et al.}  [CoGeNT collaboration],
 Phys.\ Rev.\ Lett.\  {\bf 106}, 131301 (2011)
 [arXiv:1002.4703 [astro-ph.CO]];
%
 Phys.\ Rev.\ Lett.\  {\bf 107}, 141301 (2011)
 [arXiv:1106.0650 [astro-ph.CO]];
  arXiv:1401.3295 [astro-ph.CO];
  arXiv:1401.6234 [astro-ph.CO].

\bibitem{Agnese:2013rvf} 
  R.~Agnese {\it et al.}  [CDMS Coll.],
  Phys.\ Rev.\ Lett.\  {\bf 111}, 251301 (2013)
  [arXiv:1304.4279 [hep-ex]].

\bibitem{Angloher:2014myn} 
  G.~Angloher {\it et al.}  [CRESST-II Coll.,
  Eur.\ Phys.\ J.\ C {\bf 74}, no. 12, 3184 (2014)
  [arXiv:1407.3146 [astro-ph.CO]].

\bibitem{Angloher:2011uu}
 G.~Angloher {\it et al.},
 Eur.\ Phys.\ J.\  C {\bf 72}, 1971 (2012)
 [arXiv:1109.0702].

\bibitem{DelNobile:2014sja} 
  E.~Del Nobile, G.~B.~Gelmini, P.~Gondolo and J.~H.~Huh,
  arXiv:1405.5582 [hep-ph]. TAUP 2013 Proceedings.

\bibitem{Ahlen:1987mn} 
  S.~P.~Ahlen {\it et al.}
  Phys.\ Lett.\ B {\bf 195}, 603 (1987).
  
\bibitem{Smith:1988kw} 
  P.~F.~Smith and J.~D.~Lewin,
  Phys.\ Rept.\  {\bf 187}, 203 (1990).
  
  
    \bibitem{Fox:2010bz}
 P.~J.~Fox, J.~Liu and N.~Weiner,
 Phys.\ Rev.\  D {\bf 83}, 103514 (2011)
 [arXiv:1011.1915 [hep-ph]].

\bibitem{Frandsen:2011gi}
 M.~T.~Frandsen {\it et al.}  
 JCAP {\bf 1201}, 024 (2012)
 [arXiv:1111.0292].

\bibitem{Gondolo:2012rs}
 P.~Gondolo and G.~Gelmini,
 JCAP{\bf 1212}, 015 (2012)
 [arXiv:1202.6359].

\bibitem{Gelmini:2014boa} 
  G.~B.~Gelmini,
  arXiv:1411.0787 [hep-ph].
 
\bibitem{DelNobile:2013cva} 
  E.~Del Nobile, G.~Gelmini, P.~Gondolo and J.~H.~Huh,
  JCAP {\bf 1310}, 048 (2013)
  [arXiv:1306.5273 [hep-ph]].
  
  
\bibitem{Arina:2014yna} 
  C.~Arina, E.~Del Nobile and P.~Panci,
  Phys.\ Rev.\ Lett.\  {\bf 114}, 011301 (2015)
  [arXiv:1406.5542 [hep-ph]].


\bibitem{Gould:1987ir} 
  A.~Gould,
  Astrophys.\ J.\  {\bf 321}, 571 (1987).
  
 
\bibitem{Aartsen:2012kia} 
  M.~G.~Aartsen {\it et al.}  [IceCube Coll.],
  Phys.\ Rev.\ Lett.\  {\bf 110}, no. 13, 131302 (2013)
  [arXiv:1212.4097 [astro-ph.HE]].

\bibitem{Zornoza:2014dma} 
  J.~D.~Zornoza {\it et al.}  [ANTARES Coll.],
  Nucl.\ Instrum.\ Meth.\ A {\bf 742}, 173 (2014)
  [arXiv:1404.0148 [astro-ph.HE]].
  
  
\bibitem{Zentner:2009is} 
  A.~R.~Zentner,
  Phys.\ Rev.\ D {\bf 80}, 063501 (2009) 
  [arXiv:0907.3448]; 
  I.~F.~M.~Albuquerque {\it et al}
  JCAP {\bf 1402}, 047 (2014)
  [arXiv:1312.0797].
  
  
\bibitem{Navarro:1996gj} 
    J.~Navarro, C.~Frenk and S.~White,
  Astrophys.\ J.\  {\bf 462} 563 (1996)
  [astro-ph/9508025];
  Astrophys.\ J.\  {\bf 490} 493 (1997)
  [astro-ph/ 9611107].

\bibitem{Weber:2009pt} 
  M.~Weber and W.~de Boer,
  Astron.\ Astrophys.\  {\bf 509}, A25 (2010)
  [arXiv:0910.4272 [astro-ph.CO]].
  
\bibitem{moore}
  B.~Moore {\it et al.}
  Astrophys.\ J.\  {\bf 524}, L19 (1999).
 
  
  
\bibitem{Weidenspointner:2008zz} 
  G.~Weidenspointner {\it et al.}
  Nature {\bf 451}, 159 (2008).

\bibitem{Boehm:2003bt} 
  C.~Boehm {\it et al}
  Phys.\ Rev.\ Lett.\  {\bf 92}, 101301 (2004)
  [astro-ph/0309686].

\bibitem{Finkbeiner:2007kk} 
  D.~P.~Finkbeiner and N.~Weiner,
  Phys.\ Rev.\ D {\bf 76}, 083519 (2007)
  [astro-ph/0702587].

  

\bibitem{Finkbeiner:2003im}
  D.~P.~Finkbeiner,
  Astrophys.\ J.\  {\bf 614}, 186 (2004);
  [arXiv:astro-ph/0311547].
  G.~Dobler and D.~P.~Finkbeiner,
  Astrophys.\ J.\  {\bf 680}, 1222 (2008);
  [arXiv:0712.1038 [astro-ph]].
   
\bibitem{Dobler:2009xz} 
  G.~Dobler {\it et al.}
  Astrophys.\ J.\  {\bf 717}, 825 (2010)
  [arXiv:0910.4583]; 
  M.~Su, T.~R.~Slatyer and D.~P.~Finkbeiner,
  Astrophys.\ J.\  {\bf 724}, 1044 (2010)
  [arXiv:1005.5480]. 
  
   
\bibitem{Hooper:2011ti} 
  D.~Hooper and T.~Linden
  Phys.\ Rev.\ D {\bf 84}, 123005 (2011)
  [arXiv: 1110.0006]. 
  D.~Hooper
  Phys.\ Dark Univ.\  {\bf 1}, 1 (2012)
  [arXiv:1201.1303].
  
\bibitem{Abazajian:2012pn} 
  K.~N.~Abazajian and M.~Kaplinghat,
  Phys.\ Rev.\ D {\bf 86}, 083511 (2012)
  [arXiv:1207.6047 [astro-ph.HE]].
  
\bibitem{Gordon:2013vta} 
  C.~Gordon and O.~Macias,
  Phys.\ Rev.\ D {\bf 88}, 083521 (2013)
  [Erratum-ibid.\ D {\bf 89}, no. 4, 049901 (2014)]
  [arXiv:1306.5725 [astro-ph.HE]].
  
\bibitem{Daylan:2014rsa}
  T.~Daylan {\it et al.}
  arXiv:1402.6703 [astro-ph.HE].
  
  
\bibitem{Abazajian:2014fta} 
  K.~N.~Abazajian, N.~Canac, S.~Horiuchi and M.~Kaplinghat,
  Phys.\ Rev.\ D {\bf 90}, 023526 (2014)
  [arXiv:1402.4090 [astro-ph.HE]].
  
\bibitem{Abazajian:2010zy} 
  K.~N.~Abazajian,
  JCAP {\bf 1103}, 010 (2011)
  [arXiv:1011.4275].
  
  
\bibitem{Bringmann:2012vr} 
  T. Bringmann {\it et al.}
  JCAP {\bf 1207}, 054 (2012)
  [arXiv:1203.1312]; 
  C.~Weniger,
  JCAP {\bf 1208}, 007 (2012)
  [arXiv:1204.2797]; 
  E.~Tempel, A.~Hektor and M.~Raidal,
  JCAP {\bf 1209}, 032 (2012)
  [Addendum-ibid.\  {\bf 1211}, A01 (2012)]
  [arXiv:1205.1045]; 
  A.~Boyarsky, D.~Malyshev and O.~Ruchayskiy,
  Phys.\ Dark Univ.\  {\bf 2}, 90 (2013)
  [arXiv:1205.4700]. 
  
\bibitem{Su:2012ft} 
  M.~Su and D.~P.~Finkbeiner,
  arXiv:1206.1616 [astro-ph.HE].


\bibitem{GeringerSameth:2012sr} 
  A.~Geringer-Sameth and S.~M.~Koushiappas,
  Phys.\ Rev.\ D {\bf 86}, 021302 (2012)
  [arXiv:1206.0796 [astro-ph.HE]].
  
  
\bibitem{Finkbeiner:2012ez} 
  D.~P.~Finkbeiner, M.~Su and C.~Weniger,
  JCAP {\bf 1301}, 029 (2013)
  [arXiv:1209.4562].
  A.~Hektor, M.~Raidal and E.~Tempel,
  Eur.\ Phys.\ J.\ C {\bf 73}, 2578 (2013)
  [arXiv:1209.4548].


\bibitem{Whiteson:2013cs} 
  D.~Whiteson,
  Phys.\ Rev.\ D {\bf 88}, no. 2, 023530 (2013)
  [arXiv:1302.0427].

\bibitem{Ackermann:2013uma} 
  M.~Ackermann {\it et al.}  [Fermi-LAT Collaboration],
  Phys.\ Rev.\ D {\bf 88}, 082002 (2013)
  [arXiv:1305.5597 [astro-ph.HE]].
  
\bibitem{Weniger:2013dya} 
  C.~Weniger,
  arXiv:1303.1798 [astro-ph.HE].
  
\bibitem{Weniger:2013tza} 
  C.~Weniger {\it et al.},
  arXiv:1305.4710 [astro-ph.HE].
  
  
\bibitem{Ackermann:2013yva} 
  M.~Ackermann {\it et al.}  [Fermi-LAT Coll.],
  Phys.\ Rev.\ D {\bf 89}, no. 4, 042001 (2014)
  [arXiv:1310.0828 [astro-ph.HE]].

  
\bibitem{Adriani:2008zr} 
  O.~Adriani {\it et al.}  [PAMELA Coll.],
  Nature {\bf 458}, 607 (2009)
  [arXiv:0810.4995 [astro-ph]].
  
\bibitem{FermiLAT:2011ab} 
  M.~Ackermann {\it et al.}  [Fermi LAT Coll.],
  Phys.\ Rev.\ Lett.\  {\bf 108}, 011103 (2012)
  [arXiv:1109.0521 [astro-ph.HE]].
  
\bibitem{Aguilar:2013qda} 
  M.~Aguilar {\it et al.}  [AMS Coll.],
  Phys.\ Rev.\ Lett.\  {\bf 110} 141102 (2013).
  
   
\bibitem{Cholis:2013lwa} 
  I.~Cholis and D.~Hooper,
  Phys. Rev. D {\bf 89} 043013 (2014)
  [arXiv:1312.2952 [astro-ph.HE]].
  
\bibitem{Adriani:2010rc} 
  O.~Adriani {\it et al.}  [PAMELA Coll.],
  Phys.\ Rev.\ Lett.\  {\bf 105}, 121101 (2010)
  [arXiv:1007.0821 [astro-ph.HE]].

\bibitem{Aharonian:2008aa} 
  F.~Aharonian {\it et al.}  [HESS Coll.],
  Phys.\ Rev.\ Lett.\  {\bf 101} 261104 (2008)
  [arXiv:0811.3894 [astro-ph]].

\bibitem{Cholis:2013psa} 
  I.~Cholis and D.~Hooper,
  Phys.\ Rev.\ D {\bf 88}, 023013 (2013)
  [arXiv:1304.1840 [astro-ph.HE]].
  
  \bibitem{kane} 
  G.~Kane, R.~Lu and S.~Watson,
  Phys.\ Lett.\  B {\bf 681}, 151 (2009).
 [arXiv:0906.4765 [astro-ph.HE]].
    
\bibitem{Meade}  P.~Meade, 
 M.~Papucci, A.~Strumia and T.~Volansky,
  Nucl.\ Phys.\  B {\bf 831}, 178 (2010)
  [arXiv:0905.0480 [hep-ph]].
  
  
  
\bibitem{Beltran:2010ww} 
  M.~Beltran {\it et al.},
  JHEP {\bf 1009}, 037 (2010)
  [arXiv:1002.4137 [hep-ph]].
  
\bibitem{Goodman:2010ku} 
  J.~Goodman {\it et al.},
  Phys.\ Rev.\ D {\bf 82}, 116010 (2010)
  [arXiv:1008.1783].
  
\bibitem{Fox:2012ee} 
  P.~J.~Fox {\it et al.}
  Phys.\ Rev.\ D {\bf 86}, 015010 (2012)
  [arXiv:1203.1662].
  
\bibitem{Bai:2012xg} 
  Y.~Bai and T.~Tait,
  Phys.\ Lett.\ B {\bf 723}, 384 (2013)
  [arXiv:1208.4361].
  
\bibitem{Askew:2014kqa} 
  A.~Askew {\it et al}
  Int.J.Mod.Phys.A{\bf 29} 1430041 (2014)
  [arXiv:1406.5662].
  
\bibitem{Kopp:2011eu} 
  J.~Kopp,
  arXiv:1105.3248 [hep-ph].
  
\bibitem{ATLAS:2012ky} 
  G.~Aad {\it et al.}[ATLAS Coll.]
  JHEP {\bf 1304} 075 (2013)
  [arXiv:1210.4491]. 
  
\bibitem{deSimone:2014pda} 
  A.~De Simone, G.~F.~Giudice and A.~Strumia,
  JHEP {\bf 1406}, 081 (2014)
  [arXiv:1402.6287 [hep-ph]].
 
\bibitem{Abdallah:2014hon} 
  J.~Abdallah
  {\it et al.},
  arXiv:1409.2893 [hep-ph].
  
  
\bibitem{Bauer:2013ihz} 
  D.~Bauer {\it et al.}
  arXiv:1305.1605 [hep-ph].
  
\bibitem{Cahill-Rowley:2014boa}
  M.~Cahill-Rowley {\it et al.}
  arXiv:1405.6716 [hep-ph].
  
  
\bibitem{Peccei:1977hh} 
  R.~D.~Peccei and H.~R.~Quinn,
  Phys.\ Rev.\ Lett.\  {\bf 38}, 1440 (1977).
  

\bibitem{Weinberg:1977ma} 
  S.~Weinberg,
  Phys.\ Rev.\ Lett.\  {\bf 40}, 223 (1978);
  F.~Wilczek,
  Phys.\ Rev.\ Lett.\  {\bf 40}, 279 (1978).
  
   
\bibitem{Peccei:2006as} 
  R.~D.~Peccei,
  Lect.\ Notes Phys.\  {\bf 741}, 3 (2008)
  [hep-ph/0607268].
  
\bibitem{Raffelt:2006cw} 
  G.~G.~Raffelt,
  Lect.\ Notes Phys.\  {\bf 741}, 51 (2008)
  [hep-ph/0611350].
  
\bibitem{Bae:2008ue} 
  K.~J.~Bae, J.~H.~Huh and J.~E.~Kim,
  JCAP {\bf 0809}, 005 (2008)
  [arXiv:0806.0497 [hep-ph]].

\bibitem{Beringer:1900zz} 
G. Raffelt and L. Rosenberg, ``Axions and other similar particles" in
  J.~Beringer {\it et al.}  [Part. Data Gr.],
  Phys.\ Rev.\ D {\bf 86}, 010001 (2012).

\bibitem{Ringwald:2012hr} 
  A.~Ringwald,
  Phys.\ Dark Univ.\  {\bf 1}, 116 (2012)
  [arXiv:1210.5081].

\bibitem{Visinelli:2014twa} 
  L.~Visinelli and P.~Gondolo,
  Phys.\ Rev.\ Lett.\  {\bf 113}, 011802 (2014)
  [arXiv:1403.4594 [hep-ph]]. 
  
\bibitem{Hindmarsh:1994re} 
  M.~B.~Hindmarsh and T.~W.~B.~Kibble,
  Rept.\ Prog.\ Phys.\  {\bf 58}, 477 (1995)
  [hep-ph/9411342].
  
  
\bibitem{Hiramatsu:2012gg} 
  T.~Hiramatsu, M. Kawasaki, K.~Saikawa and T.~Sekiguchi,
  Phys.\ Rev.\ D{\bf 85} 105020 (2012)
  [Erratum-ibid.\ D{\bf 86} 089902 (2012)]
  [arXiv: 1202.5851]
   and
  JCAP {\bf 1301}, 001 (2013)
  [arXiv:1207.3166 [hep-ph]].
  
\bibitem{vanBibber:2013ssa} 
  K.~van Bibber and G.~Carosi,
  arXiv:1304.7803 [physics.ins-det].
  
\bibitem{Visinelli:2009kt} 
  L.~Visinelli and P.~Gondolo,
  Phys.\ Rev.\ D {\bf 81} 063508 (2010)
  [arXiv:0912.0015]. 
  
  
\bibitem{Abazajian:2001nj} 
  K.~Abazajian, G.~M.~Fuller and M.~Patel,
  Phys.\ Rev.\ D {\bf 64}, 023501 (2001)
  [astro-ph/0101524].
  
\bibitem{Dodelson:1993je} 
  S.~Dodelson and L.~M.~Widrow,
   Phys.\ Rev.\ Lett.\  {\bf 72}, 17 (1994)
  [hep-ph/9303287];
  R.~Barbieri and A.~Dolgov,
  Phys.\ Lett.\ B {\bf 237}, 440 (1990);
  A.~D.~Dolgov,
 Phys.\ Rept.\ {\bf 370}, 333 (2002)
  [hep-ph/0202122].
  
\bibitem{Asaka:2006nq} 
  T.~Asaka, M.~Laine and M.~Shaposhnikov,
  JHEP {\bf 0701}, 091 (2007)
  [hep-ph/0612182].
  
\bibitem{Shi:1998km} 
  X. Shi and G. Fuller
 Phys. Rev. Lett.{\bf 82} 2832(1999)
  [astro-ph/9810076]
  
  
\bibitem{Abazajian:2012ys} 
  K.~N.~Abazajian {\it et al.}
  arXiv:1204.5379 [hep-ph].

\bibitem{Horiuchi:2013noa} 
  S.~Horiuchi {\it et al.}
  Phys.\ Rev.\ D {\bf 89}, 025017 (2014)
  [arXiv:1311.0282].


\bibitem{Gelmini:2004ah} 
  G.~Gelmini, S.~Palomares-Ruiz and S.~Pascoli,
  Phys.\ Rev.\ Lett.\  {\bf 93}, 081302 (2004)
  [astro-ph/0403323];
  T.~Rehagen and G.~B.~Gelmini,
  JCAP {\bf 1406}, 044 (2014)
  [arXiv:1402.0607 [hep-ph]].


\bibitem{Bulbul:2014sua} 
  E.~Bulbul {\it et al.}
  Astrophys.\ J.\  {\bf 789}, 13 (2014)
  [arXiv:1402.2301].

\bibitem{Boyarsky:2014jta} 
  A.~Boyarsky, O.~Ruchayskiy, D.~Iakubovskyi and J.~Franse,
  Phys.\ Rev.\ Lett.\  {\bf 113} 251301 (2014)
  [arXiv:1402.4119] 
and  arXiv:1408.2503.

\bibitem{Abazajian:2014gza} 
  K.~Abazajian,
  Phys.\ Rev.\ Lett.\  {\bf 112} 161303 (2014)
  [arXiv:1403.0954].
  
\end{thebibliography}
\end{document}